\documentclass{aa}  

\usepackage{graphicx}
\usepackage{grffile}
\usepackage{txfonts}
\usepackage{hyperref}
\hypersetup{
    colorlinks,
    citecolor=blue,
    linkcolor=blue
    }
\usepackage{placeins}
\usepackage{float}

\begin{document} 
\raggedbottom
   \title{Finding AGN remnant candidates based on radio morphology with machine learning}
   \author{Rafa\"el I.J. Mostert\thanks{E-mail: mostert@strw.leidenuniv.nl}\inst{1,2} \and
          Raffaella Morganti\inst{2,3} \and
  Marisa Brienza\inst{4,5,6}\and
          Kenneth J. Duncan\inst{7} \and
Martijn S.S.L. Oei\inst{1}\and
Huub J.A. R\"{o}ttgering\inst{1}\and
Lara Alegre\inst{7}\and
Martin J. Hardcastle\inst{8}\and
Nika Jurlin\inst{9}}

\institute{Leiden Observatory, Leiden University, P.O. Box 9513, 2300 RA Leiden, The Netherlands \and
ASTRON, the Netherlands Institute for Radio Astronomy, Oude Hoogeveensedijk 4, 7991 PD Dwingeloo, The Netherlands \and
Kapteyn Astronomical Institute, University of Groningen, P.O. Box 800, 9700 AV Groningen, The Netherlands \and
INAF - Osservatorio di Astrofisica e Scienza dello Spazio di Bologna, Via P. Gobetti 93/3, 40129 Bologna, Italy\and 
Dipartimento di Fisica e Astronomia, Università di Bologna, via P. Gobetti 93/2, I-40129, Bologna, Italy\and 
INAF - Istituto di Radioastronomia, Bologna Via Gobetti 101, I-40129 Bologna, Italy\and
Institute for Astronomy, Royal Observatory, Blackford Hill, Edinburgh, EH9 3HJ, UK \and
Centre for Astrophysics Research, Department of Physics, Astronomy and Mathematics, University of Hertfordshire, College Lane, Hatfield AL10 9AB, UK\and
Department of Astronomy, The University of Texas at Austin, 2515 Speedway, Stop C1400, Austin, TX 78712-1205, USA}

   \date{Received 31 January 2023 / Accepted 11 April 2023}

 
  \abstract
  {Remnant radio galaxies represent the dying phase of radio-loud active galactic nuclei (AGN). 
  Large samples of remnant radio galaxies are important for quantifying the radio galaxy life cycle. 
  The remnants of radio-loud AGN  can be identified in radio sky surveys based on their spectral index, or, complementary, through visual inspection based on their radio morphology.
  However, this is extremely time-consuming when applied to the new large and sensitive radio surveys.
  }
   {
   Here we aim to reduce the amount of visual inspection required to find AGN remnants based on their morphology,
   through supervised machine learning trained on an existing sample of remnant candidates.
   }
  { For a dataset of $4,107$ radio sources, with angular sizes larger than $60$ arcsec, from the LOw Frequency ARray (LOFAR) Two-Metre Sky Survey second data release (LoTSS-DR2), we started with $151$ radio sources that were visually classified as `AGN remnant candidate'.
	We derived a wide range of morphological features for all radio sources from their corresponding Stokes-I images: 
	from simple source catalogue-derived properties, to clustered Haralick-features, and self-organising map (SOM) derived morphological features. 
  We trained a random forest classifier to separate the `AGN remnant candidates' from the not yet inspected sources.
   }
   {The SOM-derived features and the total to peak flux ratio of a source are shown to be most salient to the classifier.
   For each source, our classifier outputs a positive prediction if it believes the source to be a likely `AGN remnant candidate', or else a negative prediction. 
	 The positive predictions of our model included all initially inspected `AGN remnant candidates', plus a number of not yet inspected sources.
   We estimate that $31\pm5\%$ of sources with positive predictions from our classifier will be labelled `AGN remnant candidates' upon visual inspection,
   while we estimate the upper bound of the $95\%$ confidence interval for `AGN remnant candidates' in the negative predictions at $8\%$.
     Visual inspection of just the positive predictions reduces the number of radio sources requiring visual inspection by $73\%$.
   }
   {
   This work shows the usefulness of SOM-derived morphological features and source catalogue-derived properties in capturing the morphology of AGN remnant candidates.
   The dataset and method outlined in this work bring us closer to the automatic identification of AGN remnant candidates based on radio morphology alone and can be carried out for similar projects that require automatic morphology-based classification based on small labelled sample sizes. }


\maketitle

\section{Introduction}
Mass accreting supermassive black holes, also known as active galactic nuclei (AGN), at the centres of galaxies can form collinear jets, producing synchrotron emission visible at radio frequencies, such jets are known to impact star-formation (SF) rates and thereby play a crucial role in the evolution of their host galaxy. 

Through detailed studies of single AGN and particular galaxy clusters, astronomers have identified a number of different stages of radio-loud AGN (see \citealt{morganti2017archaeology} for a review):
they can be active; dying or remnant; or restarting.
The time that AGN typically spend in each of these phases is a key factor in determining how SF rates in a galaxy typically evolve over cosmic time.

Furthermore, determining the ratio of remnant to active AGN will shed light on the rate and magnitude of the adiabatic and radiative energy losses of the radio lobes.
With time, the emission of the lobes of an AGN remnant will fall below the detection limit, as a switched-off AGN will not re-energise its lobes with electrons through its jets, while radiative and adiabatic losses continue \citep{godfrey2017population}.

AGN remnant candidates can be selected by visually inspecting radio galaxies at $\sim100$ MHz frequencies.
At $\sim100$ MHz frequencies, AGN that have recently switched-off should be traceable as the compact features (for example the flat spectrum core, jet or hotspot emission if applicable) have faded while the steep spectrum emission from the radio lobes can remain visible for a longer time. 
Due to the adiabatic expansion of the lobes, these AGN remnants are expected to have low surface brightness and to be amorphous in shape; bright hotspots in the lobes and jets are expected to disappear \citep{morganti2017archaeology}.

Remnants can also be selected based on their spectral index. 
The synchrotron-based emission from a radio source usually follows a power-law behaviour in its energy spectrum, such that we can describe the relation between the observed flux $S$ and frequency $\nu$ as $S\propto \nu^{-\alpha}$, where $\alpha$ is the spectral index.
Radio emission from an AGN core is usually flat, while emission from the lobes is often steeper and can become ultra-steep ($\alpha^{1.4 GHz}_{150 MHz}>1.2$) for AGN remnants.

AGN remnants are relatively rare objects, with estimates that they make up $<11\%$ of all radio sources \citep{Mahony2016,Brienza2017,Mahatma2018,jurlin2020life,Benjamin2021}.
This means that a lot of radio sources need to be visually inspected to find a remnant, so that gathering large samples of remnants is very time-consuming.
In the era of large radio sky surveys with unprecedented sensitivity, the number of sources to inspect becomes overwhelming.

Up till now, the research on AGN remnants has been carried out using individual sources \citep[e.g.][]{cordey1987ic,Brienza2018duty} or small samples of sources selected based on source-by-source visual inspection \citep[e.g.][]{Saripalli2012,Brienza2017,Mahatma2018,jurlin2020life,Benjamin2021} or small samples of sources based on spectral criteria \citep[e.g.][]{Parma2007,Murgia2011}. 
Small hand-curated samples of AGN remnant candidates are valuable as they can be complete and clean and therefore justify follow-up observations at higher frequencies to determine whether or not they are actually AGN remnants --- see for example the work by \citet{jurlin2020life}. 
However, this time-consuming selection method does not scale to the vast sky-areas needed to get large samples which allow stronger constraints on the population properties of AGN remnants and the time an AGN typically spends in the on, off or restarted phase.  
Large samples of AGN remnant candidates can already be selected based on their spectral index:
by creating spectral index maps of multi-frequency observations \citep[e.g.][]{Harwood2018}, individual (ultra-)steep sources can be visually inspected \citep[e.g.][]{Morganti2021}.
However, from Monte-Carlo simulations, \citet{godfrey2017population} and \citet{Brienza2017} deduce that sources with ultra-steep spectral indices ($\alpha^{1.4 GHz}_{150MHz}>1.2$) represent roughly half of the entire AGN remnant population.
Selecting all AGN remnants based on spectral criteria alone means we do not find the other half of the population.
Selection based on just spectral criteria would return the entire population only if observations at $>5$ GHz, where the spectral steepening occurs sooner, were available for the same large fractions of the sky as our lower frequency data, which is not the case \citep{Brienza2017}.
Finding automated ways to create our sample of AGN remnant candidates based on their morphology would thus complement automated spectral index selection and be a necessary step towards a complete census of AGN remnants.

Existing work on the automated sorting of radio galaxies based on morphology can be divided into (discrete) classification systems and (continuous) clustering systems.
The classification systems are best used when large numbers (hundreds) of labelled examples are available, for example, to classify Fanaroff--Riley \citep{Fanaroff1974a} class I (FRI) or class II (FRII) galaxies \citep[e.g.][]{Aniyan2017a,alhassan2018first,Ma2019machine,mingo2019revisiting,Bowles2021gated,Scaife2021equivariant,Mohan2022}.
Classification becomes more challenging but not impossible when labels are sparse. For example, \citet{Proctor2016} train oblique decision trees \citep{Murthy1993} using features of $48$ giant radio galaxies to find more giant radio galaxy (GRG) candidates in the NRAO VLA Sky Survey \citep[NVSS;][]{Condon1998}.
\citet{Proctor2016} compiled a list of the $1,600$ most likely GRG candidates in NVSS. A follow-up by \citet{Dabhade2020} which included a thorough visual inspection of all the candidates in the list, confirmed $151$ GRG.
For sparsely labelled datasets, pre-training on similar labelled datasets \citep[e.g. for optical galaxies:][]{Walmsley2022hybrid, Walmsley2022zoobot} or semi-supervised learning\footnote{Semi-supervised learning attempts to learn from a limited set of labels by propagating labels to examples where the model has a high prediction certainty and by generalizing the predictive capacity of the model by training on augmented versions of the labelled-data.} can, in theory, be applied \citep[e.g. for radio galaxies:][]{Slijepcevic2022shift}.
However, designing and implementing a radio galaxy-specific semi-supervised learning set-up is non-trivial.
For example, the set-up presented by \citet{Slijepcevic2022shift} does not yet outperform the simpler, supervised, convolutional neural network baseline by \cite{Tang2019transfer} on real unlabelled radio data. 
Notable clustering systems, based on unsupervised machine learning techniques, to find similar sources or perform outlier detection use self-organising maps \citep[SOM;][]{Kohonen1989,Kohonen2001} or Haralick features \citep{haralick1973textural}. 
For the application of these techniques to radio galaxies see \citet[][]{Galvin2019,Ralph2019,Galvin2020,Mostert2020} and \citet{Ntwaetsile2021} respectively.

In this work, we attempt to reduce the number of radio sources that require visual inspection to create a large sample of AGN remnant candidates based on their radio morphology from the ongoing Low Frequency Array \citep[LOFAR;][]{vanHaarlem2013} Two-metre Sky Survey \citep[LoTSS;][]{Shimwell2017,Shimwell2019}. 
Specifically, we will focus on sources within the HETDEX area of the sky within the second data release \citep[LoTSS-DR2;][]{Shimwell2022}, as this area of the sky is accompanied by a high-quality radio source catalogue (see Sec. \ref{sec:data}).

In the study presented here, we aim at a morphological separation of AGN remnant candidates in the set of well-resolved sources of LoTSS.
Visual inspection of all well-resolved radio sources in the HETDEX area of the LoTSS-DR2 catalogue ($4,107$ radio sources $>60$ arcsec) is extremely time-consuming.
Instead, we create a binary machine learning classifier that suggests more likely candidates, based on a small and incomplete set of already labelled AGN remnant candidates (Brienza et al. in prep.).
By random sampling from our positive and negative model predictions, we estimate the probability of finding more `AGN remnant candidates' in each set of predictions.
If, upon visual inspection, we succeed in having most `AGN remnant candidates' in the positive predictions, that means we have created a tool for people to create bigger samples of remnant candidates by visually inspecting only the positive predictions.
The methods in this work combine a supervised random forest (RF) classifier, which is a decision-tree type classifier as is used by \citet{Proctor2016}, with features, some of which come from unsupervised learning techniques as presented by \citet{Mostert2020} and \citet{Ntwaetsile2021}.

Section \ref{sec:data}, describes the LOFAR data that we used to find AGN remnant candidates and explains the initial visual inspection. 
Section \ref{sec:methods} describes the image pre-processing that we apply to subsequently, automatically, derive our various morphological features. 
It also describes the set-up and training of our random forest (RF).
In Sec. \ref{sec:results}, we show the resulting morphological features, the trained RF and an assessment of the corresponding feature importances.
Finally, we discuss the use and limitations of our method in Sec. \ref{sec:discussion} and summarise our conclusions in Sec. \ref{sec:conclusion}.

\section{Data}
\label{sec:data}
The data we use is confined to an area of $424$ square degrees in the HETDEX region of the sky (right ascension 10h45m00s to 15h30m00s and declination $45^{\circ}$ to $57^{\circ}$). 
The observations were taken using the high-band antennas of LOFAR at $120$--$168$ MHz. 
The LOFAR images with a resolution of $6\arcsec$ and a median sensitivity of $71\ \mu\mathrm{Jy\ beam}^{-1}$ are part of the LOFAR Two-metre Sky Survey second data release \citep[LoTSS-DR2;][]{LoTSSDR2}.
The data come with a value-added source catalogue (simply `source catalogue' hereafter) with manual radio source-component associations and a combination of automatic and manual identifications of the corresponding optical host galaxies \citep{Williams2019} and (Hardcastle et al. in prep.).
According to the source catalogue provided by \citet{Williams2019}, which is based on the earlier LoTSS-DR1 over the same sky area, the HETDEX region contains $318,520$ radio sources.

Visual inspection of radio emission with the aim to manually classify a source as `AGN remnant candidate' or not, requires one to judge whether a source has a low surface brightness, is amorphous in shape and lacks signs of core, jet or hotspot emission.
This is only sensible for very well-resolved sources, and in this work we therefore only consider the $4,107$ radio sources with an apparent angular size bigger than $60$ arcsec ($10$ times the $6$ arcsec synthesized beam).\footnote{The angular sizes, in arcsec, of the radio sources in the public LoTSS-DR1 catalogue are available through the `Maj' or the `LGZ\_Size' column. (The `Maj' column indicates that a radio source was created by source finder PyBDSF, while the `LGZ\_Size' column indicates that additional crowd-sourced information was used to check if the source finder's radio component association was correct.) By taking all sources with either `Maj' > $60$ arcsec ($31$ sources) or `LGZ\_Size' > $60$ arcsec ($4,076$ sources), we end up with $4,107$ sources.}

At present the number of AGN remnant candidates in the literature across all radio surveys is below one hundred \citep[e.g.][]{jurlin2020life,Benjamin2021}, which is a low number when compared to the hundreds of thousands of detected radio-loud AGN \citep[e.g.][]{Shimwell2022} and also when compared to the expected fraction of remnants \citep[e.g.][]{Brienza2017,godfrey2017population}.
It is clear that a larger training set of AGN remnant candidates needs to be built. 
Brienza et al. in prep. applied an automatic rule-based morphological selection building on the work of \citet{Brienza2017} and \citet{jurlin2020life} and after going through an extensive and time-consuming visual inspection of this rule-based selection, obtained $151$ radio sources, $>60$ arcsec, as likely AGN remnant candidates.
Specifically, these labels were based on the morphology of the radio source, also taking into account neighbouring radio emission, at $144$ MHz (LoTSS Stokes-I).
Optical images from the Panoramic Survey Telescope 
and Rapid Response System 1 (Pan-STARRS1) $3\pi$ steradian survey \citep{panstarrs} centred on the location of the radio source were also inspected to be able to classify sources with a corresponding nearby star-forming host galaxy as noncandidate.
More details about the labelling procedure can be found in the accompanying paper (Brienza et al. in prep.).\footnote{All data, including the list of $151$ AGN remnant candidates are at \url{https://lofar-surveys.org/finding_agn_remnants.html}.}

\begin{figure}\begin{center}
\includegraphics[width=0.9\columnwidth]{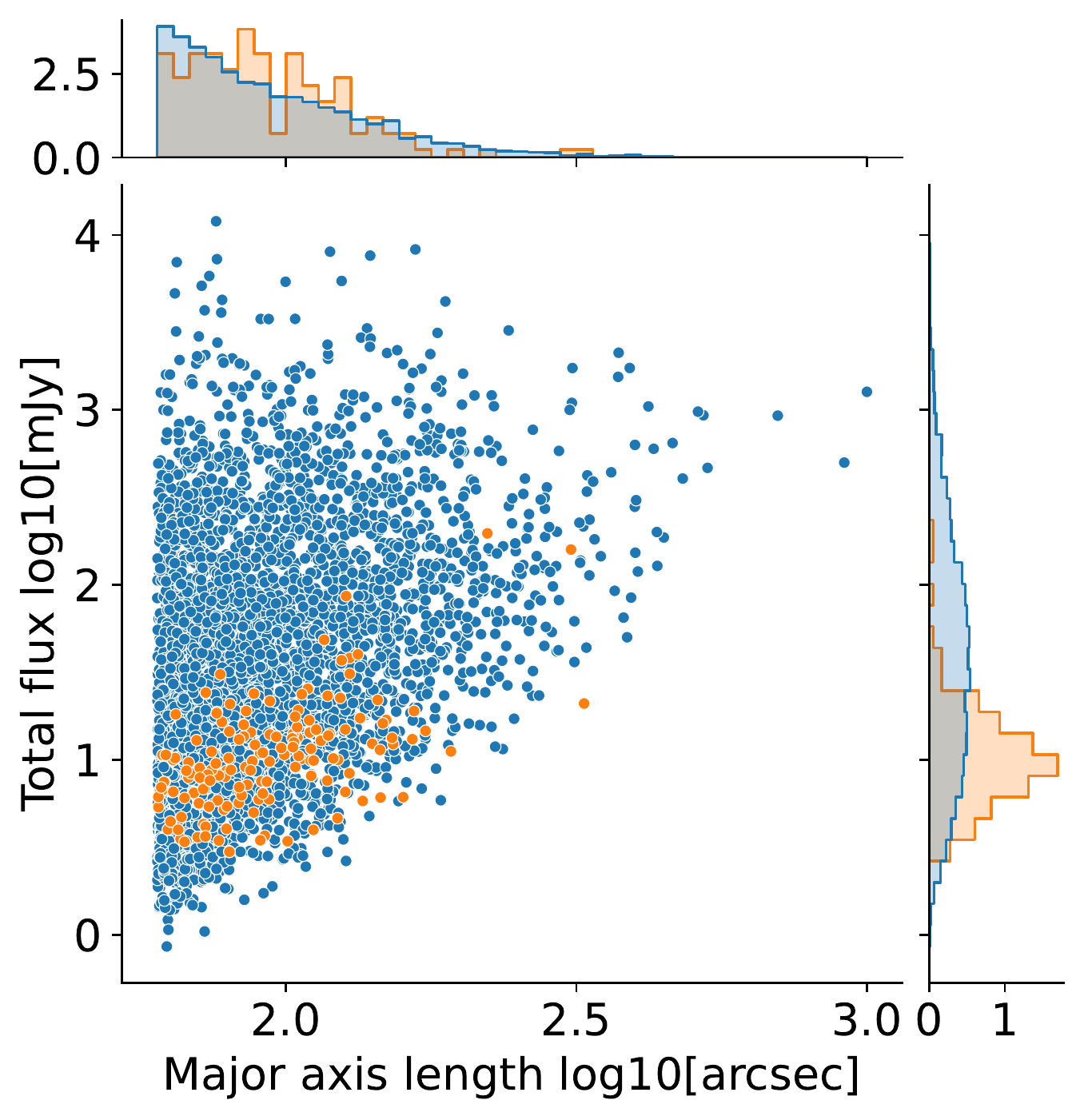}
\caption{Source size versus total flux density for the sources in HETDEX, $>60$ arcsec, labelled as `AGN remnant candidate' following initial visual inspection by Brienza et al. in prep. (orange dots) and `noncandidates' (blue dots).} \label{fig:size_vs_flux}
\end{center}\end{figure}

In this manuscript, we took the $151$ radio sources labelled as AGN remnant candidates by Brienza et al. in prep., and compare them to the sources $>60$ arcsec in the HETDEX area that have not yet been visually inspected.
Until Sec. \ref{sec:discussion} we will refer to all these not yet inspected sources other than the $151$ candidates as `noncandidate'.
To get a sense of the labels, see Fig. \ref{fig:size_vs_flux} for our respective size and flux distributions of sources in each class.

\section{Methods}
\label{sec:methods}

\begin{figure*}\begin{center}
\includegraphics[width=0.9\textwidth]{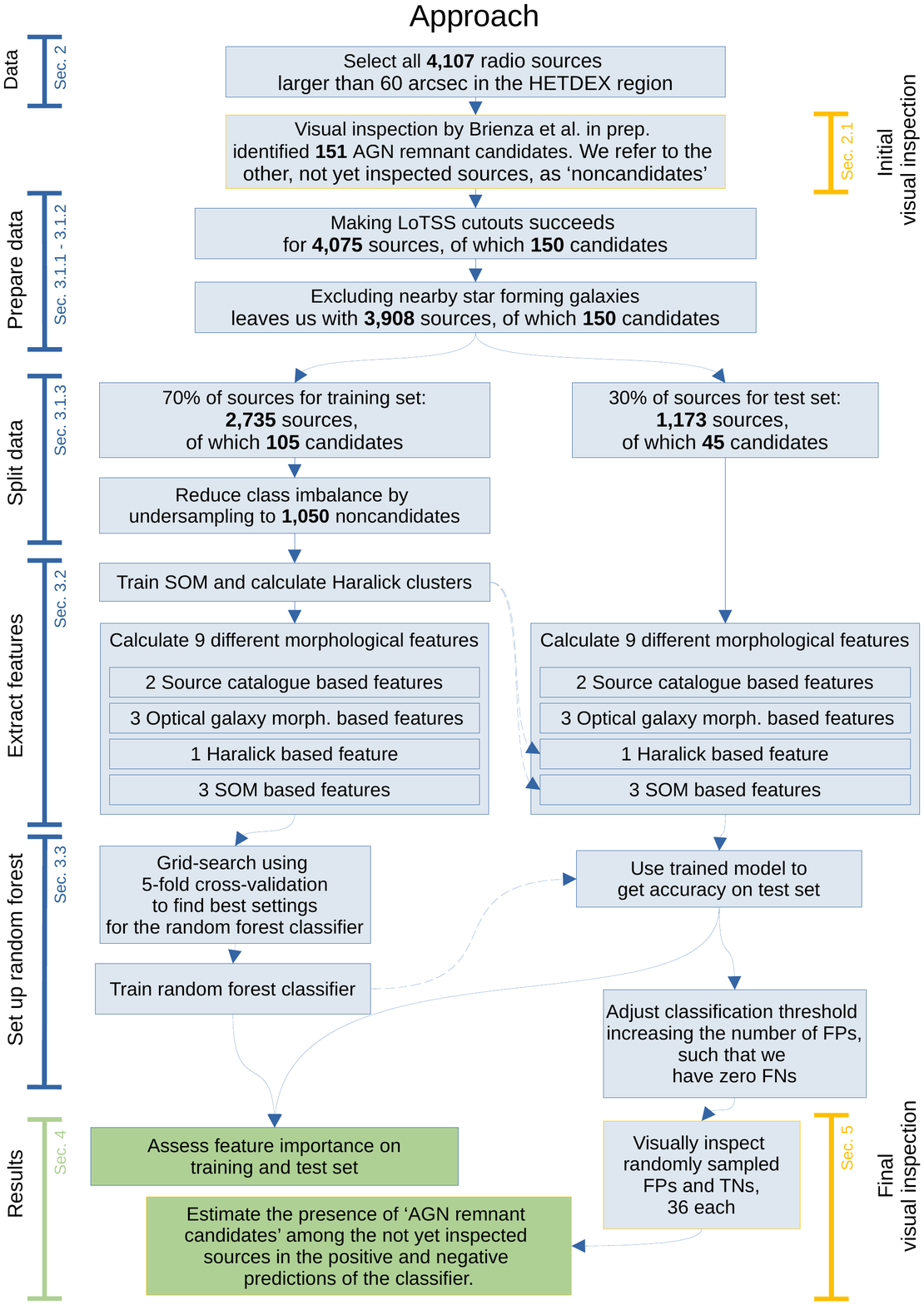}
\caption{Diagram of our approach to create a model that classifies radio sources as either `AGN remnant candidate' or `noncandidate' based on a number of morphological features. The solid arrows show the flow of our data, while the dashed arrows indicate the use of various trained models, specifically a SOM,  HDBSCAN* clustered Haralick features, and a random forest, all trained using only the data from the (undersampled) training set.
The brackets indicate the different stages of our method and mention the sections in which corresponding details can be found.
} \label{fig:approach}
\end{center}\end{figure*}

AGN remnant morphological definitions are fuzzy and not readily identifiable by a single metric. 
Therefore, we extract and test a wide range of morphological features in a reproducible manner: from simple source properties, to morphology metrics normally used in optical astronomy, to Haralick features, to SOM-derived features \citep{Alegre2022}.
The SOM-derived features are created by building on the rotation and flipping invariant SOM algorithm implemented by \citet{Polsterer2015} and built upon by \citet{Mostert2022}.
In this work, we add source-size invariance to the rotation and flipping invariant SOM.
We also introduce a `compressed-SOM' technique that allows for better generalization of the trained neurons. 
We then train a random forest (RF) to use the morphological features to classify radio sources as `AGN remnant candidate' or `noncandidate'.
To establish how well the classifier is able to leverage the features to predict these labels we use label-permutations.
Furthermore, we quantify which morphological features are most salient to detect AGN remnant candidates using feature-permutations.

Our hypothesis is that the positive predictions from the trained RF are more likely to be labelled `AGN remnant candidate' upon visual inspection than the negative predictions.
This is irrespective of what a source turns out to be after visual inspection, a `positive' prediction simply means that the model predicts the label `AGN remnant candidate' and a `negative' prediction means that the model predicts the label `noncandidate'.
By visually inspecting random samples of positive and negative predictions, we can test this hypothesis.
If this hypothesis is strong enough, we will be able to reduce the number of radio sources that require visual inspection when creating a large sample of AGN remnant candidates based on morphology, by only having to inspect the positive predictions of our model.
An overview of our approach is visualized in Fig. \ref{fig:approach}.

\subsection{Data pre-processing}
\label{sec:prepro}
The first step in our automated AGN candidate suggestion method is to reduce the observed Stokes-I image maps to a single, pre-processed image cutout per radio source.
These cutouts can later be used to automatically extract morphological features for each radio source.
The second step in our methods is to automatically exclude most nearby star-forming radio galaxies (SFGs) for which the radio emission is associated with star formation rather than to (past) AGN activity.

\subsubsection{Creating image cutouts}
\label{sec:cutouts}
Apart from the source catalogue--based features, all morphological features are based on the image of the considered radio source.
Thus we extracted image cutouts for each of the $4,107$ sources.
Specifically, we created square cutouts from the Stokes-I LoTSS-DR2 image pointings centred on the corresponding sky coordinates in the source catalogue with a size $1.5$ times the source size as reported by the source catalogue. 
These variable-sized images were then all resized to fixed size $174\times 174$ arcsec\textsuperscript{2} images using bilinear interpolation.
This ensures that the various morphological features derived downstream are always comparing sources with roughly the same pixel-extent, thereby approximating source size invariance.
As the angular pixel resolution of the LoTSS images is $1.5$ arcsec per pixel, this results in images of $116 \times 116$ pixels\textsuperscript{2}.
To prevent radio emission in the corner of these square cutouts to influence the morphology measures, we mask all pixels outside the circular aperture with a diameter of $116$ pixels centred on the middle of the cutout.
The cutout-extraction process fails for sources for which the Stokes-I image contains NaNs.
The $32$ sources for which cutout-extraction failed will be discarded from our dataset.
Thus, after cutout-extraction, we are left with $4,075$ sources of which $150$ AGN remnant candidates.

\citet{Mostert2020} reduced the effect of noise in the radio source image cutouts by clipping away signal that is $<1.5$ times the local noise. 
In this work, we do not apply such clipping as AGN remnants are likely to have a surface brightness close to the noise.
To prevent the morphological features that we derive later on from providing different outcomes for sources with similar morphologies, but different apparent brightness, we rescale all pixel values $p_{i}$ in all images to $p_{i, \text{norm}}$ according to Eq. \ref{eq:minmax}:
\begin{equation}
\label{eq:minmax}
p_{i, \text{norm}} = \frac{p_{i} - min(p_{i})}{max(p_{i})}.
\end{equation} 

Unlike the work presented by \citet{Mostert2020}, we use a radio catalogue that includes manual and likelihood ratio--based source-component association \citep{Williams2019} as the starting point for creating cutouts, and thus assume that the source-component association is mostly correct.
This means that we are less prone to mistake a spatially separated single radio lobe for an amorphous AGN remnant candidate.
Moreover, this allowed us to remove neighbouring unrelated radio components that fall within our square image cutouts.
We did so by subtracting the Gaussian components that, according to the Python Blob Detection and Source Finder \citep[\textit{PyBDSF};][]{Mohan2015} source-model, constitute the neighbouring sources from the Stokes-I image cutout.

\subsubsection{Exclusion of nearby star-forming radio galaxies}
\label{sec:sfg}
\begin{figure}\begin{center}
\includegraphics[width=0.9\columnwidth]{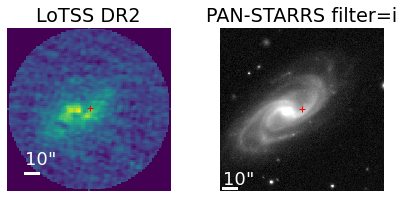}
\caption{Radio source that can be rejected as AGN remnant candidate as the optical image, taken by Pan-STARRS1, reveals that the corresponding optical host of the radio source is a nearby star-forming spiral galaxy.} \label{fig:nearby_spiral}
\end{center}\end{figure}

A single amorphous radio blob at $144$ MHz might indicate the presence of an AGN remnant, but the same morphology can also signify synchrotron radiation from the supernovas in the star-forming regions from a nearby galaxy without any (past) AGN activity, see Fig. \ref{fig:nearby_spiral} for an example. 
The extent of the radio emission we observe must be far larger than the extent of optical/infrared emission to ensure that the radio emission originates from (past) AGN activity, and not from a nearby SFG.
After trial and error, we adopted the criterion that the LoTSS source major axis must be at least 10 times larger than the optical size given by the extended source catalogue \citep[2MASX;][]{jarrett2000a} from the Two Micron All Sky Survey \citep[2MASS;][]{skrutskie2006a}.
From the $4,075$ successfully extracted sources $> 60$ arcsec, $167$ sources were labelled as potential nearby SFGs using this method.
After visual inspection comparing the LoTSS and the Pan-STARRS1 images,
we found that this label is correct for $86.2\%$ 
($144/167$) of the sources.

When classifying radio sources as AGN remnant candidates in even larger datasets in the 
future, we will not visually inspect these potential nearby SFGs but simply discard them.
In this work too, we discard all $167$ sources labelled as potential 
nearby SFGs, leaving us with $3,908$ radio sources of which $150$ candidates.
This means that we do not consider the $(167-144)/4075=0.6\%$ of sources $>60$ arcsec that could 
have been AGN remnant candidates. 

An alternative method to exclude nearby SFGs is to use the optical/near-IR information to derive SF rates and to compare the corresponding expected radio continuum flux density through the radio--SF rate correlation \citep{Gurkan2018} with the observed flux density.
However, we decide not to adopt this method as it is only reliable for sources with sufficient multi-wavelength information \citep[e.g.][]{Smith2021}.

\subsubsection{Splitting data into a training and a test set}
\label{sec:split}
Statistics derived from applying a model to the same data that was used to train the model are potentially unrealistically optimistic due to (possible) overfitting.
Thus, we split $70\%$ of our data into a training set and used the remaining $30\%$ of the data for our test set.
We used a stratified split, which means we randomly assign each radio source to either the training or test set while roughly maintaining the same class balance (`candidate' or `noncandidate') as in the full dataset, leaving us with a training set that contains $2,630$ `AGN remnant candidates' and $105$ `noncandidates'.

Like many classifiers, an RF achieves worse performance on an imbalanced dataset. 
An imbalanced dataset is a dataset for which the number of examples is not evenly spread over the different classes.
In our case, for each AGN remnant candidate in the data we work with, there are more than $26$ sources labelled `noncandidate', thereby making it easy for a classifier to achieve high accuracy by always predicting that a given radio source is a `noncandidate'.
There are broadly three ways to minimize the effect of an imbalanced dataset.
One can oversample the minority class, undersample the majority class or adjust the weight of each sample during training based on its class \citep{murphy2012machine,Vanderplas2019}.
What works best depends on the dataset used.
We combined undersampling the majority class with class weight adjustment (for details on the latter, see Sec. \ref{sec:rf}).
Specifically, we undersampled the majority class in the training set from $2,630$ to $1,050$ sources, 
increasing the ratio of AGN remnant candidate versus `noncandidate' in the training set from $1$ in $26$, to $1$ in $10$, thereby helping our classifier by reducing the hay (`noncandidates') that hides the needles (`candidates').
Hereafter, `training set' will refer to the $1,155$ sources in our undersampled training dataset and the `test set' to all $1,173$ sources in our test dataset.

\subsection{Deriving morphological features}
\label{sec:morpho}
In this section, we detail how we automatically derive $9$ morphological features for the sources in the training and test set.
These $9$ features will feed into our classifier on a per-source-basis, to predict whether or not the source is a likely AGN remnant candidate. 

\subsubsection{Source catalogue-derived properties}
\label{sec:simple}
From the radio source catalogue produced by \citet{Williams2019}, we first calculated the ratio of the length of the major axis to the length of the minor axis of the source (as projected on the sky-plane), as we expect that AGN remnants have a lower ratio due to adiabatic expansion of their lobes in all directions combined with an absence of a longitudinally outward pushing jet.
Second, we calculated the ratio of the total flux density to the peak flux density of the sources, as we expect AGN remnants to have a low and uniform surface brightness due to the adiabatic expansion and the absence of jet-formed hotspots in the lobes.
We do not use the total flux density from the catalogue as a feature, since this metric does not inform us about intrinsic brightness and will bias our classifier towards sources with a low apparent flux.

\subsubsection{Applying optical galaxy morphology metrics to radio sources}
\label{sec:alternative_metrics}
We included three morphological metrics that are normally used in quantifying optical galaxy morphology to see whether they are able to help separate AGN remnant candidates from the not yet inspected sources when applied to radio morphology.
Specifically, we implemented the concentration index, clumpiness index and Gini coefficient as described by \citet{Conselice2014}. 

The concentration index ($C$) is defined as the radius at which $80\%$ of the flux of a source is enclosed ($r_{80}$), divided by the radius at which $20\%$ of the flux is enclosed ($r_{20}$):
\begin{equation}
\label{eq:concentration}
C \coloneqq 5\log{\left(\frac{r_{80}}{r_{20}}\right)}.
\end{equation}
As expected, we see (Appendix \ref{app:optical_morpho}) that low concentrations match with sources that have little flux in their geometrical centres, so either FRIIs or strongly bent FRIs. High concentration indices match with sources that have more flux in their geometrical centres, so FRIs and diffuse objects.
Using the concentration index, a qualitative inspection shows that it (reassuringly) separates sources where the brightness is core-dominant from those where the brightness is lobe-dominant.

The clumpiness index ($S$) is defined as:
\begin{equation}
\label{eq:clumpiness}
S \coloneqq 10\frac{\sum(I-I^{\sigma})}{\sum I},
\end{equation}
where we have a pixel-wise subtraction of the radio source image $I$ and the same image smoothed by a Gaussian kernel $I^{\sigma}$ with a standard deviation that is $12$ arcsec (two times the synthesized beam).
Sources with smooth surface brightness should have low clumpiness values, while sources with small-scale structures should have high clumpiness values.
The smaller the angular extent of the sources, the less structure they can exhibit (we should not be able to discern features smaller than our 6 arcsec beam). We expect AGN remnant candidates to have a smooth brightness distribution due to the lack of compact structures, leading to low clumpiness values. 
We also expect our remnant candidates to be faint as the particles in the lobes radiate their energy away while they are, by definition, not replenished in an AGN remnant.
However, for sources that barely surpass the noise level, the noise pattern might be picked up by the clumpiness index, leading to higher clumpiness index values. 

To determine the Gini coefficient of an image, we sort the pixel values ($f$) in ascending order within a circular aperture that contains $95\%$ of the total flux density within the cutout. 
The Gini coefficient ($G$) is then given by:
\begin{equation}
\label{eq:gini}
G \coloneqq \frac{1}{|\bar{f}|n(n-1)}\sum_{i}^{n} (2i-n-1) |f_i|,
\end{equation}
where $\bar{f}$ is the average pixel value within the circular aperture of the image. 
The Gini coefficient is $1$ if all flux is concentrated in one pixel, and zero if every pixel shares an equal amount of flux.
The Gini coefficient behaves similarly to the clumpiness index: 
sources with highly concentrated surface brightness areas, like sources with bright core or lobe emission, will receive high Gini coefficient values, while diffuse sources and sources close to the noise level will receive low Gini coefficient values (see Appendix \ref{app:optical_morpho}).

\subsubsection{Haralick-based morphological features}
\label{sec:haralick}
Haralick texture features, Haralick's Coefficients or simply Haralick features are statistics that describe an image or image patch based on the spatial correlation of different pixel-intensities within the image \citep{haralick1973textural}.
\citet{Ntwaetsile2021} used Haralick features to cluster radio sources and detected outlying sources based on their radio morphologies, motivating us to include Haralick-based morphological features in our classifier.
\citet{haralick1973textural} define $14$ statistics, the first $13$ of which are computationally stable, so we will only use those, all based on a grey-level co-occurrence matrix (GLCM).
The GLCM is a square matrix with a width and length equal to the number of discrete grey levels in the image ($256$ in our case).
Each value $p_{i,j}$ in this matrix, with $i$ the row-index and $j$ the column-index, is the probability of a transition from a pixel having grey-value $i$ to a pixel having grey-value $j$ in the direction $\theta$ at a distance of $d$ pixels.
We refer the reader to \citet{haralick1973textural} for the $13$ statistics subsequently extracted from the GLCM.

We calculated the Haralick features using the Python implementation by \citet{coelho2013mahotas}.\footnote{\url{http://pypi.Python.org/pypi/mahotas}}
The images we use are the pre-processed Stokes-I cutouts described in Sec. \ref{sec:prepro}, 
with the additional step that we reduce each pixel value to 8-bit as in the work of \citet{Ntwaetsile2021}, resulting in pixel values ranging from 0 to 256 in integer steps only, as a meaningful GLCM can only be constructed using integer values.
Following \citet{Ntwaetsile2021}, we adopt a distance $d=1$ and use the offset directions up, down, right and left for $\theta$, thereby ensuring that the subsequently derived statistics are at least equivalent for any integer number of $90$ degree rotations of the input image.
We thus ended up with Haralick features: $13$ continuous values per source.

If more data labelled as AGN remnants would have been available, we could have incorporated all $13$ Haralick features.
As is, adding all of them would massively enlarge the feature-space, leading to overfitting.
Thus instead, we proceeded to cluster the Haralick features akin to the method laid out by \citet{Ntwaetsile2021} using an HDBSCAN* algorithm.

HDBSCAN* \citep{campello2015} is a clustering method based on density, it groups data points that have many nearby neighbours to a cluster. 
The number of clusters is not pre-set as with a k-nearest neighbours algorithm but is an emergent property based on the data and the chosen hyperparameters.
Apart from these (high-density) clusters, HDBSCAN* also creates a single outlier or `noise' cluster to which data points without many nearby neighbours are all assigned. 

Specifically, we used the Python HDBSCAN* implementation created by \citet{McInnes2017}.
We adopted the HDBSCAN* implementation default parameters except for the `minimum cluster size' and the `minimum sample size', which, together with the data input dimensions influences how many clusters are formed.
According to \citet{Ntwaetsile2021}, the desired number of clusters as influenced by these parameters is somewhat arbitrary and mostly up to the user, they suggest setting both values to $64$.
With our data that leads to only two clusters apart from the noise cluster; thus we ended up halving both parameters to $32$, leading to the formation of six clusters plus the noise cluster.

The feature that we fed into our classifier will be the ratio of AGN remnant candidates in the Haralick cluster of a particular source.
It is important to note that this feature changes as a function of the dataset for which it is evaluated.
Calculating this feature in advance would thus lead to data leakage, which leads to over-optimistic results and reproducibility issues \citep{Kapoor2022}.
Therefore, we adapted the classifier training, validation and inference such that this feature is calculated using only the data available within the specific training, validation or test sets that we employ.

\subsubsection{SOM-derived morphological features}
\label{sec:som}
A SOM is a single-layer neural network, made up of neurons arranged on a lattice.
It has the property that, during training, each neuron will become more and more similar to a subset of the training inputs.
After training, each neuron can thus be said to represent a certain set of inputs.
Within a well-trained SOM, neurons with similar properties will be located close to each other on the lattice \citep[][]{Kohonen2001}.

After training, each input from the training set can be `mapped to the SOM': it can be assigned to the neuron to which it is most similar.
The same can be done with images outside of the training set, as long as they go through the same preprocessing steps and are not out of distribution.
We can, for example, take images from the same survey, but in a different area of the sky and map these to the trained SOM. 
By looking at the number (and ratio) of `AGN remnant candidate' mapped to each neuron of a trained SOM, we can find to what extent which neurons are more indicative of remnant-morphology than others. 

\paragraph{Training the self-organising map}
\label{sec:training}
\begin{table*}[]
\centering
\caption{hyperparameters used for training the SOM}
\label{Run_parameters}
\begin{tabular}{l|l}
\hline\hline
SOM lattice dimensions (w x h x d)                      & $9\times9\times1$ \\
Number of channels or layers	& $1$ \\ 
Training image dimensions	& $116 \times 116$ pixels\textsuperscript{2}, or  $82 \times 82$ pixels\textsuperscript{2} after rotation\\
Neuron image dimensions	& $82 \times 82$ pixels\textsuperscript{2} \\
Neighbourhood radius start $\theta_0$ | decrease $\theta_d$ & $5$ | $0.9$       \\
Learning rate start $\alpha_0$ | decrease $\alpha_d$ | end $\alpha_e$ & $1$ | $0.8 $ | $0.3$\\
Periodic boundary conditions	& False \\
Number of training epochs & $25$ \\
Initialisation	 & Zeros \\
\hline
\end{tabular}
\end{table*}
To use a SOM, we must first train the SOM using a set of inputs from our training set.
In our case, the inputs are the Stokes-I image cutouts centred on the $1,155$ radio sources in our training set, see Sec. \ref{sec:rf}, for which we detailed the selection and preprocessing in Sec. \ref{sec:data}.
That means that the SOM does not use the source catalogue, it only takes the pixel-information of each source into account.
Training establishes the properties of a SOM: each neuron will come to represent a subset of images from the training set, and similar neurons are located close to each other on the lattice.
At each step in the SOM training procedure, a training image is compared to all the neurons in the map using the Euclidean norm between the training image and the neuron. 
The pixel values of the neuron with the lowest Euclidean norm to the image are then slightly (controllable with the `learning rate' hyperparameter $\alpha$) updated to match the training image. 
The neighbouring neurons are also updated to match the training image by an amount that decreases with increasing distance from the best-matching neuron (controllable with the `neighbourhood radius' hyperparameter $\theta$). 
The procedure we used to train the SOM is to a large extent analogous to the procedure detailed by \citet{Mostert2020}, see Table \ref{Run_parameters} for the hyperparameter values we used for training.

The original SOM training and mapping algorithms are not equivariant under rotation or flipping, while this equivariance is crucial for clustering radio galaxy morphologies.
\citet{Polsterer2015} achieve rotational and flipping invariance with respect to the training images by comparing each neuron with many rotated and flipped copies of the training image. 
The neuron for which one of the rotated or flipped copies has the lowest Euclidean norm to each other is updated using this copy.
We used the Parallelized rotation and flipping INvariant Kohonen maps (PINK) code developed by \citet{Polsterer2015} and introduced an approximate form of angular size invariance by resizing the central source in each image to the same fixed size using bilinear interpolation (as explained in Sec. \ref{sec:prepro}), before feeding the image to the SOM.
We discuss the effect of resizing sources on the SOM in Sec. \ref{sec:discussion}.

The size of a SOM lattice is a user-set hyperparameter, and it is arbitrary to a certain extent.
For our SOM, we adopted a 2D lattice consisting of $N=9\times 9$ neurons.
To allow neurons to emerge that represent a wide range of common morphologies from the training images, one can set the neighbourhood radius to be very small, such that updates do mostly affect a single neuron.
However, it is easy to lose coherence in this way, which is to say that multiple far-away neurons may independently start to represent the same common morphology. 
A larger lattice (a lattice with more neurons) with a higher neighbourhood radius preserves the coherence and allows a wide range of morphologies to emerge in the neurons.
However, this will also increase the number of redundant neurons, neurons that are close by and roughly represent the same set of sources.
Additionally, if the number of sources in our calibration set is of the same magnitude as $N$, the best-matching neurons to the sources will not be very robust: minor morphological changes to an input might cause the source to be mapped to a neighbouring neuron.
This is likely to negatively affect the informativeness of the morphological features that we derive from our SOM.
These features are, the absolute and relative number of calibration sources that are also mapped to the same best-matching neuron of a particular source.

To get the best of both worlds, we trained our SOMs using a $9\times 9$ lattice, preserving coherence and allowing a wide range of common morphologies to be represented, after which we prune the SOM by removing every other row and column.
This leaves us with a $5\times 5$ lattice, each of the neurons in this `compressed' SOM being relatively more unique than the ones in the  $9\times 9$ SOM.
Moreover, the smaller lattice allows the morphological features that we derive from the SOM to be more robust as each neuron contains more sources.

\paragraph{Mapping to the compressed SOM}
\label{sec:mapping}
After training, we mapped all sources to the trained, compressed, $5\times5$ SOM.
`Mapping' means that we compare an input to each of the trained neurons using the Euclidean norm, and identify the lattice coordinates of the neuron to which it has the smallest Euclidean norm.
The features that we used in our RF are the absolute number of AGN remnant candidates mapped to the same neuron best-matching the considered source, and the ratio of AGN remnant candidates to all sources mapped to that neuron.

Similar to the Haralick cluster ratios, the number of AGN remnant candidates per mapped neuron and the ratio features change as a function of the dataset for which these features are evaluated.
Here again, we adapted our classifier training, validation and inference to prevent data leakage.

\paragraph{Using the SOM similarity metric as a morphological feature}
\label{sec:euclidean_norm}
We used the Euclidean norm as the similarity metric to decide which SOM-neuron best represents a certain input, both during the SOM training and during the SOM mapping stage.
The Euclidean norm leads to a continuous value that is larger if the difference between the pixels of the input and the pixels of the neurons increases.
By assigning an input to its best-matching neuron, this Euclidean norm is reduced from a continuous value to the discrete coordinate of the neuron on the SOM-lattice. 
However, the continuous value could still provide us with more information.
Even within the set of sources mapped to a single neuron, there are differences in the Euclidean norms (in other words, there are differences in how well they morphologically match this neuron).
If the Euclidean norms of the calibration inputs lie in a specific range, the RF could use this to classify our sources as AGN remnant candidates.
Thus we extracted the Euclidean norms of all our sources as a feature that the RF can use.

\subsection{Training the random forest classifier}
\label{sec:rf}
After deriving the different morphological features, we wanted to test the relationship between these features and our initial labels.
Specifically, we wanted to train a binary classifier that uses our derived features to predict whether a radio source is likely to be an AGN remnant candidate, we wanted to test the predictive power of this classifier (how much better than chance is it), we wanted to see how it performs on unseen data, and we wanted to establish which features are most salient for this classifier.

A decision tree is one of the few classifiers that can handle both discrete feature-inputs (like the absolute number of AGN remnant candidates for a specific neuron) and continuous feature-inputs (like the total to peak flux ratio of a source).
Furthermore, a decision tree is robust to outlying feature-input values, and is insensitive to monotone transformations of its feature-inputs, thus allowing our features to be vastly different in absolute magnitude and the value-ranges \citep{murphy2012machine,Vanderplas2019}.
The disadvantage of a decision tree is that it has a high variance, as features that are high up in the tree have a disproportionate effect on its outcome.
This disadvantage can be reduced by using an ensemble of decision trees. 
An RF classifier returns the average over the outcome of many decision trees, each of which randomly samples only a fraction of the available features, thereby reducing the variance introduced by a single decision tree \citep{Breiman2001}.
For these reasons, we choose to use an RF classifier; specifically, we use the RF as implemented by the sci-kit learn Python library \citep{scikit-learn}.

An RF has a number of hyperparameters to tune, these are user-set parameters that impact the performance of a classifier but are not directly learnt when training a classifier (or any other machine-learning estimator). The optimal hyperparameter values depend on the dataset used and the metric one wishes to optimise for. Optimal hyperparameter values can thus be approximated by testing a range of reasonable values for the desired metric on an independent subset of the data (the validation set).
The most important hyperparameter for an RF is the number of trees in the forest, where a larger number leads to longer computation times but also leads to lower variance up to a certain point. 
Using $5,000$ trees allows us to tune the other hyperparameters of the tree in a time on the order of minutes.
With our relatively small number of data samples (radio sources) and many features, we are prone to overfit our data.
To reduce the tendency to overfit, the maximum depth of each tree in the RF can be limited and the maximum number of features to use per tree can be altered \citep{Breiman2001}. 
(Notably, RF does not overfit as a function of the number of trees \citep{Breiman2001}.)
To further address the class imbalance in our dataset, we can vary the weight of each sample with respect to the weight of samples from the majority class.
We employed a grid search on our training set to find good values for these hyperparameters.
Specifically, we checked the following maximum depth values: $[4,8,16,32,64,128,256,512,1024]$, the following maximum feature-ratios $[0.2,0.3,0.4,0.5,0.6]$,
and the following class weight ratios $[0.01,0.04,0.16,0.32,0.5]$.\footnote{A class weight ratio $x$ indicates that samples from the majority class are weighted by $x$, while samples from the minority class are weighted by $1-x$. The low values of $x$ that we explore in our grid search imply that we down-weight the examples from the majority class to increase the exposure of the minority class (for which we have fewer samples) to the classifier.}
We set the grid search to optimize for $F_2$-score, as finding most AGN remnant candidates in large samples is our main objective.
A $F_1$-score strives for a balance between precision and recall, but as we are more interested in recall than precision, we used the $F_2$-score, which is a special case of the $F_\beta$-score (with $\beta=2$) defined as:
\begin{equation}
\label{eq:f2}
F_\beta = (1+\beta^2) \frac{precision \cdot recall}{(\beta^2 \cdot precision) + recall}.
\end{equation}
We found the best hyperparameter values, by using a five-fold stratified cross-validation on the training set, whereby `stratified' means that the class balance in each cross-validation fold is similar to the class balance in the training set.
For all other hyperparameters, we adopted the default values of the \citet{scikit-learn} RF implementation.\footnote{Tuning the maximum size of decision trees in the RF is a way to trade-off performance versus overfitting. We arbitrarily chose to do so through the maximum RF depth, but a similar effect can (alternatively or additionally) be achieved by tuning the maximum number of leaf nodes, or tuning the minimum number of samples required for a node split in the RF.}
Finally, we trained an RF with these optimal hyperparameter values using all $1,155$ sources in the training set (as opposed to using $1/5$th of the training sources as we did during cross-validation).

\section{Results}
\label{sec:results}
Following feature extraction and the RF set-up, we can first explore the position of the AGN remnant sample within our feature space, 
and then look at the RF training results.

\subsection{Resulting morphological features}
\label{sec:results_features}

\begin{figure}\begin{center}
\includegraphics[width=0.99\columnwidth]{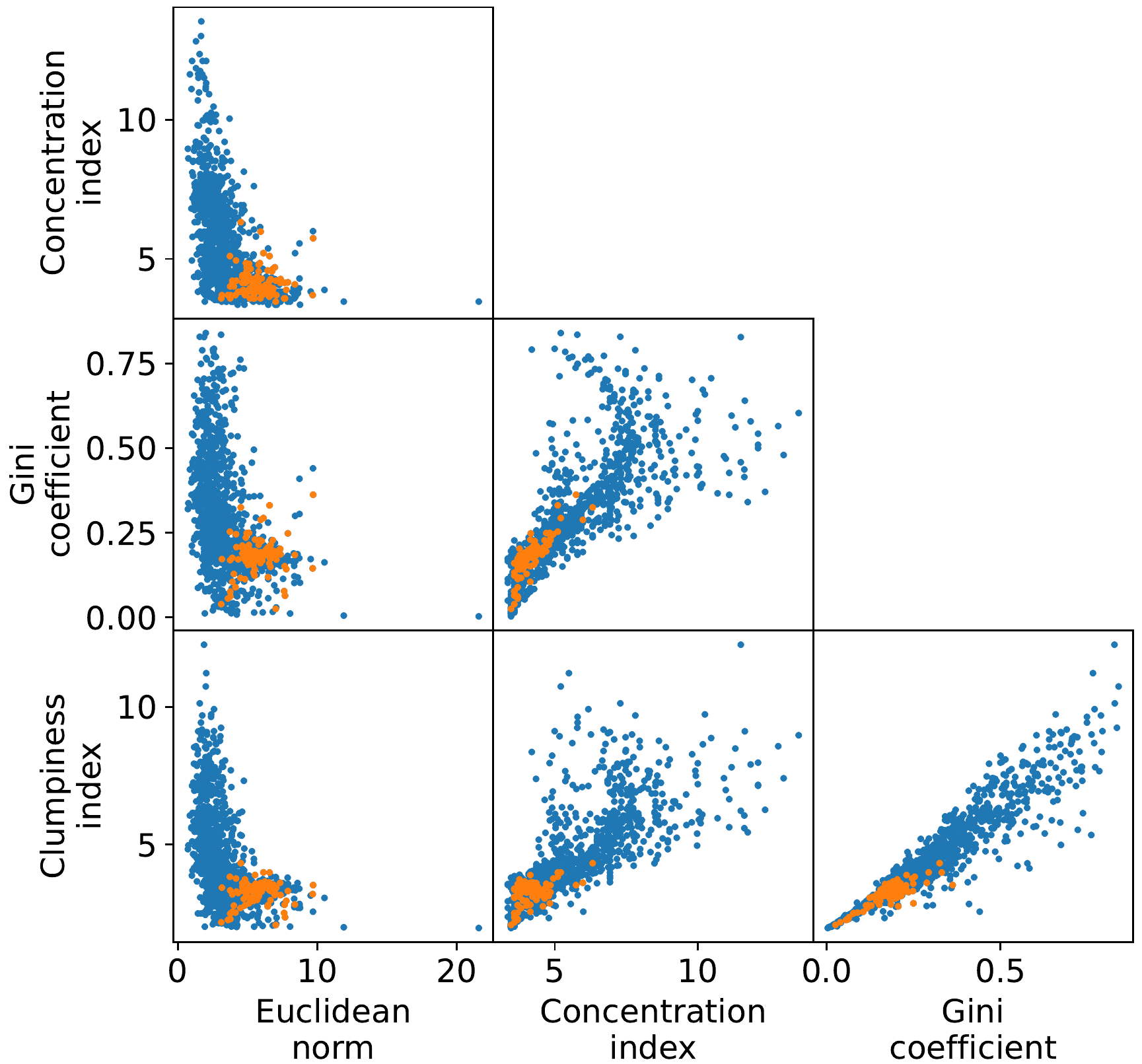}
\caption{Morphology metrics for the AGN remnant candidates in the training set (orange) and the other sources in the training set (blue).}
\label{fig:metrics}
\end{center}\end{figure}

Figure \ref{fig:metrics} shows the distribution of values we obtained for the Euclidean norm, Gini coefficient, clumpiness index and concentration index, for sources in the training set.
Generally, the AGN remnant candidates (orange) overlap with the noncandidates. 
However, the AGN remnant candidates seem to have higher Euclidean norm values, and lower Gini coefficients, clumpiness indices and concentration indices.
The high Euclidean norm values are not unexpected since AGN remnants are rare and hence not well represented by the SOM (which leads to high Euclidean norm values).
The lower Gini-coefficients and clumpiness indices are expected as remnants have a smoother surface brightness than regular AGN.
The lower clumpiness indices can be expected due to the absence of significant core-emission and hotspots for AGN remnants.
The figures in App. \ref{app:optical_morpho} and \ref{app:hara} show examples of radio source cutouts in different ranges of Gini coefficient, clumpiness index, concentration index values, and different Haralick clusters.
The figures show that the features are able to separate sources with different morphological properties, but they also show that the surrounding noise(-pattern) and residual emission from (artefacts of) neighbouring sources have a strong impact on the feature values, making classification based on these features harder.

\subsection{Radio morphology in the trained SOM}
\label{sec:results_SOM}

\begin{figure*}\begin{center}
\includegraphics[width=0.99\textwidth]{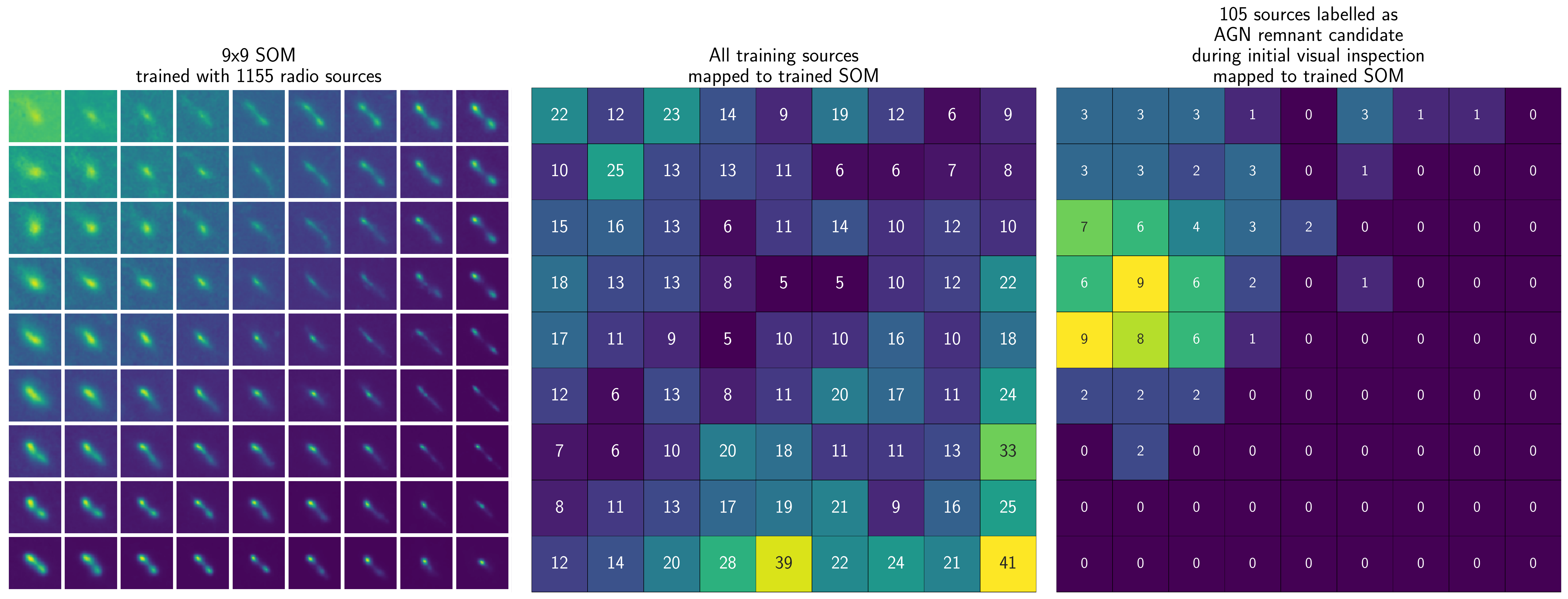}
\caption{First panel shows a $9\times9$ SOM trained with our training set. The second panel shows how all sources in the training set map to the SOM, and the third panel shows the subset of `AGN remnant candidates' from the training set mapped to the SOM.}
\label{fig:som_uncompressed}
\end{center}\end{figure*}

\begin{figure*}\begin{center}
\includegraphics[width=0.99\textwidth]{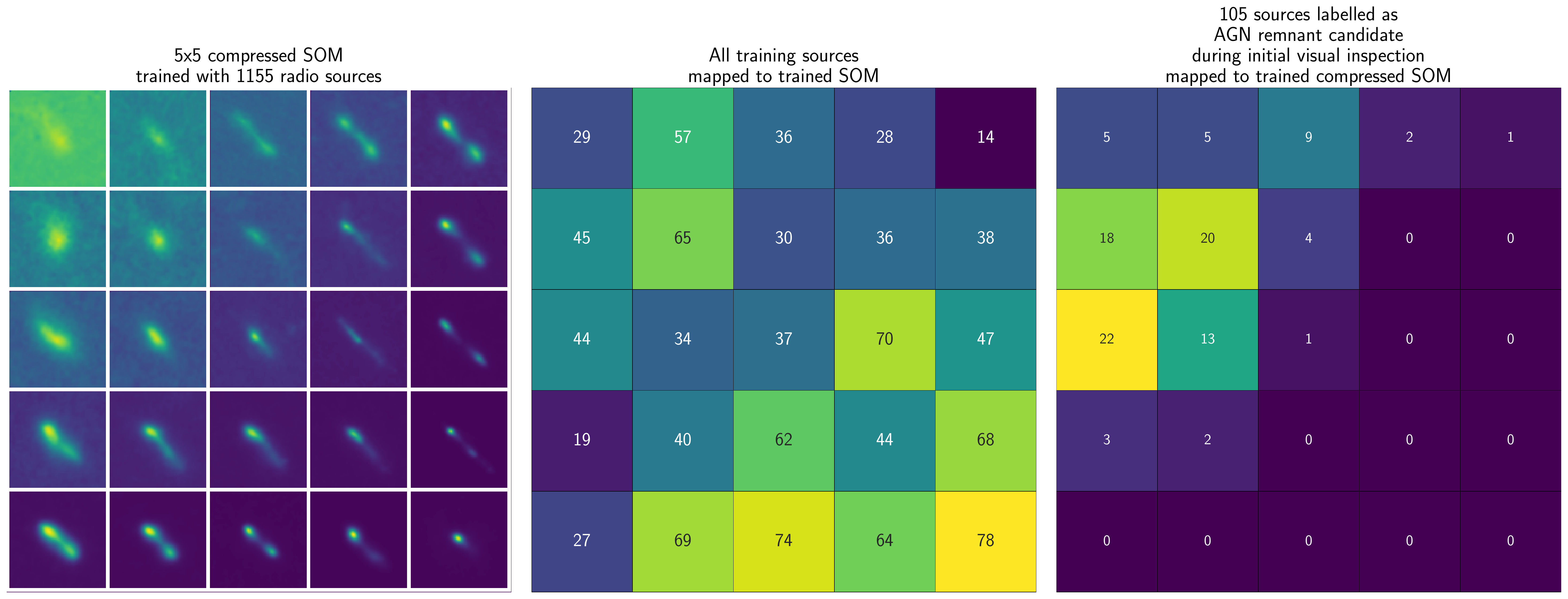}
\caption{Same as Fig. \ref{fig:som_uncompressed} except that we compressed the trained SOM to a $5\times5$ lattice. The second and third panels respectively show all training sources and the subset of `AGN remnant candidates' mapped to this compressed SOM.} \label{fig:som_compressed}
\end{center}\end{figure*}

Figure \ref{fig:som_uncompressed} shows the results from our SOM-derived features.
By inspecting the SOM --- the leftmost panel of Fig. \ref{fig:som_uncompressed} --- we observe a key property of SOMs: neurons of the SOM that are close together on the $9\times9$ map, are also similar in the morphologies that they represent and the transition between different morphologies is smooth. 
As a result, we can see smooth gradients of changes in morphologies. 
The neurons close to the top-left to bottom-right diagonal show morphologies representing FRI--like (core-dominated) sources, while away from this diagonal we see neurons representing FRII--like sources with edge-brightened lobes.
Going from the top-left to the bottom-right, we see neurons representing morphologies that gradually go from less to more elongated (i.e. the neurons represent sources going from a higher to a lower width-to-length ratio).
Additionally, we note that in the bottom-right, the neurons represent sources that are far above the local noise, while in the top-left, the neurons represent sources with surface brightness levels closer to the local noise.

The middle and right panels of  Fig. \ref{fig:som_uncompressed} show the mapping of all training sources to the SOM and the mapping of the AGN remnant candidates to the SOM respectively.
The middle panel of Fig. \ref{fig:som_uncompressed} shows that the sources are distributed over the entire SOM, with the highest concentrations in the bottom right area of the SOM-lattice.
The right panel of Fig. \ref{fig:som_uncompressed} shows that the AGN remnant candidates are mostly located in the top left region of the SOM, even though this pattern is somewhat scattered due to the low bin counts throughout. 
This low bin count effect will be worsened during training, cross-validation and testing of our classifier as that will entail mapping smaller samples of sources to the SOM.

The left panel of Fig. \ref{fig:som_compressed} shows our compressed SOM, for which the same observations as for the SOM in Fig. \ref{fig:som_uncompressed} hold.
However, additionally, we see that the mapping of the radio sources within this compressed SOM is more robust: the smaller number of neurons causes the distribution over the SOM of the low number of AGN remnant candidates to be less affected by small bin counts, especially during cross-validation.
The AGN remnant candidate mappings to the compressed SOM, in the middle panel of Fig. \ref{fig:som_compressed}, shows a clearer pattern: the AGN remnant candidates are mostly located in the diagonal band going from the top left to the bottom right with most mapped to the centre of the SOM, while the noncandidates are uniformly located, with most mapped to the bottom right area of the SOM.

In our earlier work \citep{Mostert2020}, multiple neurons were needed to represent sources with similar morphology but with different apparent sizes.
Figure \ref{fig:som_compressed} shows that the source size rescaling that we introduced at the data pre-processing stage (Sec. \ref{sec:cutouts}) of this work, reduces the size degeneracy in the resulting SOMs.

We note that there is no intrinsic significance to the absolute location of a neuron on the lattice.
If one trains multiple SOMs with the same training data, similar neurons will always end up relatively close together on the lattice, but the absolute location of these neurons may change.

\begin{figure}\begin{center}
\includegraphics[width=0.99\columnwidth]{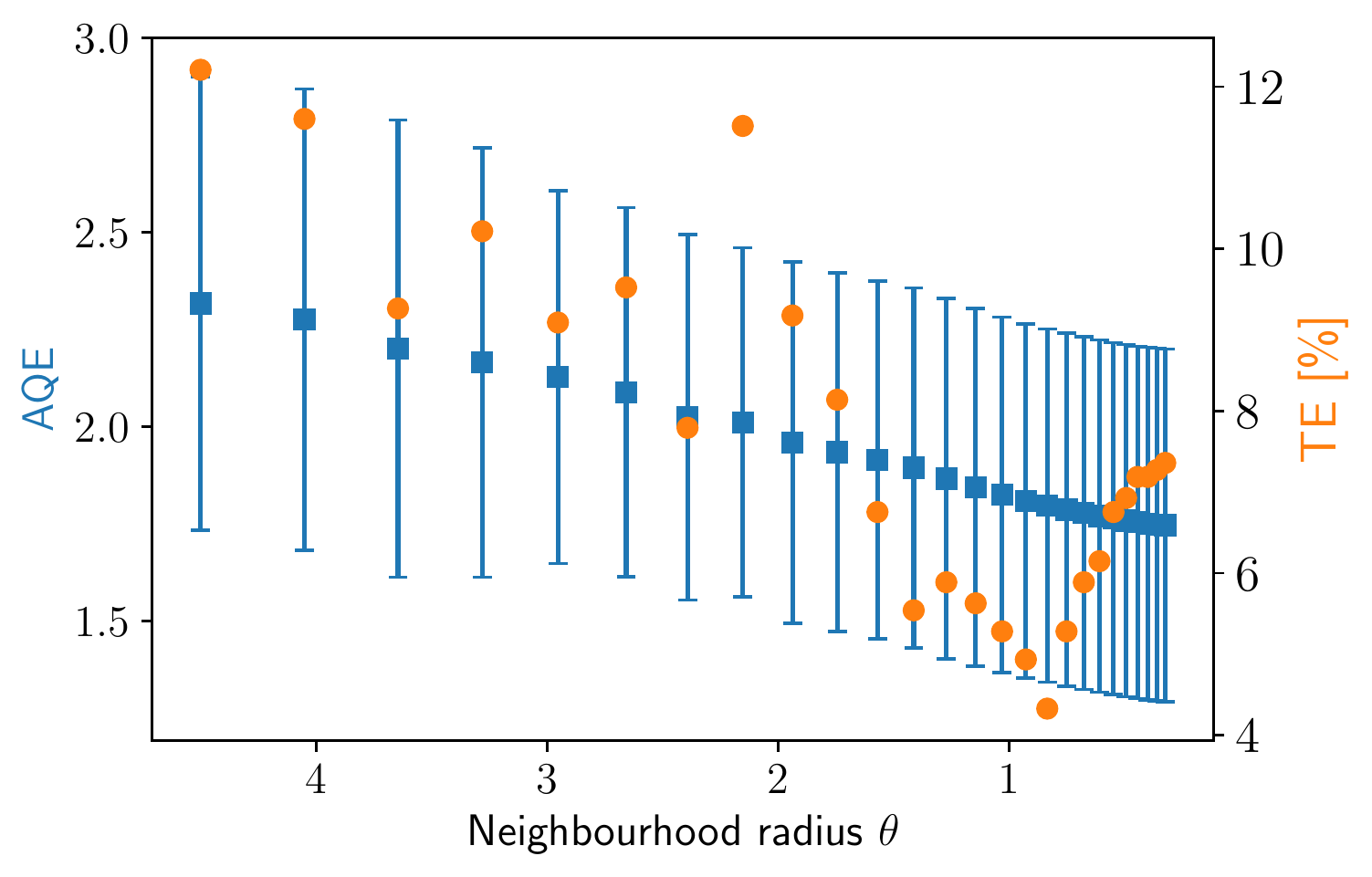}
\caption{Quantitative training metrics for the SOM in Fig. \ref{fig:som_uncompressed} showing the steadily decreasing average quantisation error (AQE) on the left y-axis and the topological error (TE) on the right axis.} \label{fig:som_metrics}
\end{center}\end{figure}

Quantitatively, the SOM training was monitored by logging the average quantisation error (AQE) and the topological error (TE).
The AQE \citep{Kohonen2001} is defined as taking the average over the summation of all the values of the Euclidean norm from each image in the dataset to its corresponding best-matching neuron.
A lower AQE indicates that the SOM is a better representation of the data.
The TE \citep{Villmann1994} is defined as the percentage of image cutouts in our data
for which the second best-matching neuron is not a direct neighbour of the best-matching neuron,
where `direct neighbour' is defined as all eight neighbours for neurons on a rectangular SOM lattice.
TE is a measure for the coherence of the SOM where lower TE equals better coherency.
With each SOM training update, one tries to lower AQE while keeping TE at acceptable values (arbitrary thresholds below $10\%$ are common).
Similar to the trends of the metrics in \citet{Mostert2020}, Fig. \ref{fig:som_metrics} shows a steadily declining AQE and a TE that declines initially, reaches a minimum, and then climbs up again.
In \citet{Mostert2020} the rule of thumb to stop training when the per epoch decline of AQE dropped below $1\%$ was applied, while in the current work, with the particular training set that we use, we noticed that there was room to continue training up to 25 epochs to achieve a slightly lower $AQE$ with a $TE$ that is still solidly below $10\%$.
	
\subsection{Resulting hyperparameters for the trained random forest classifier}
\label{sec:results_RF} 
In the cross-validation phase of training our RF, 
the grid search reveals that a maximum tree depth of $4$,
a maximum feature-ratio of $0.4$ 
and a class weight of $0.16$ for our majority class 
(and $0.84$ for our minority class),
are good hyperparameter choices for our RF when optimizing for $F_2$-score. 

\begin{figure}\begin{center}
\includegraphics[width=0.99\columnwidth]{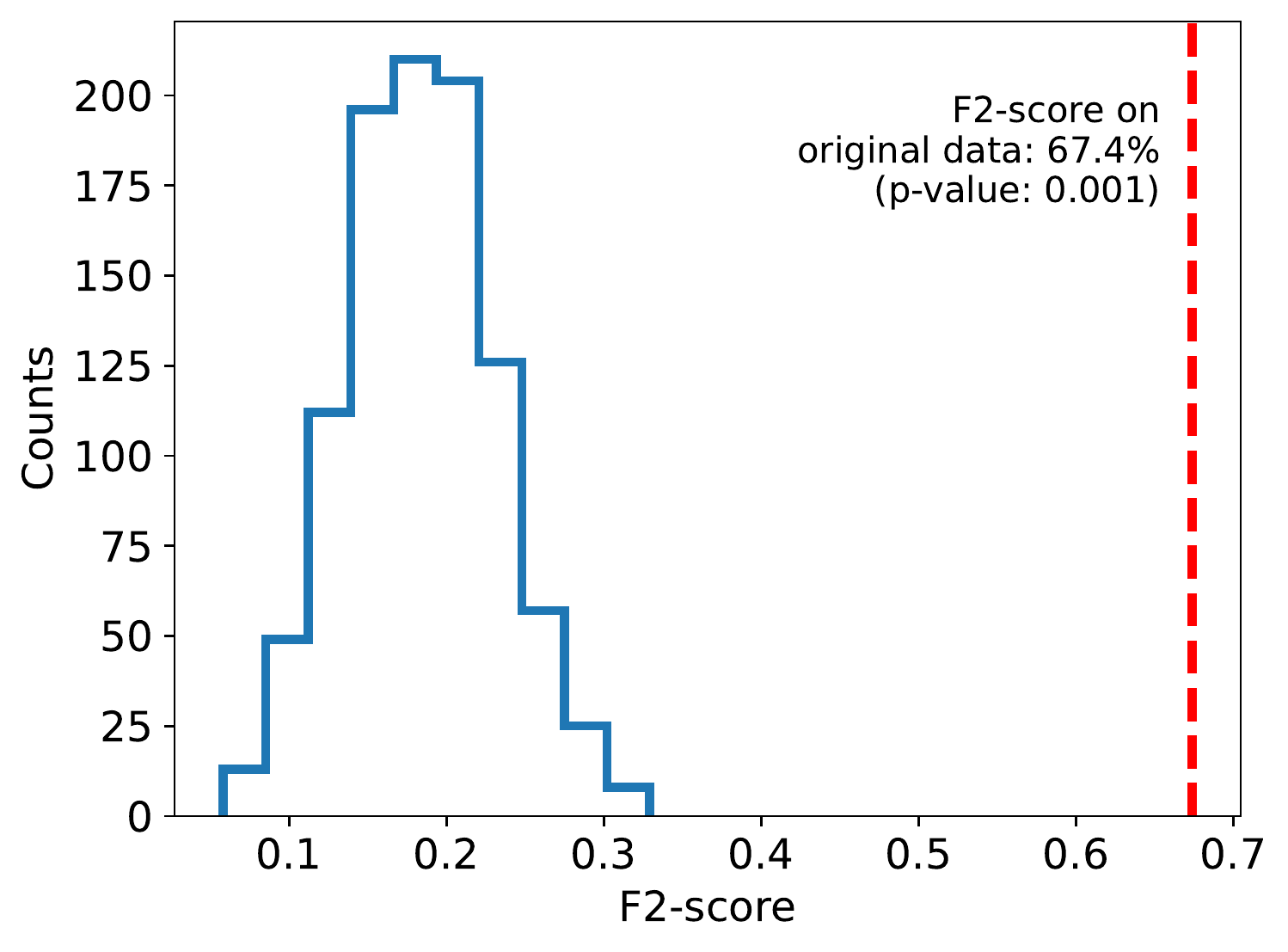}
\caption{Permutation test scores. The red dashed line indicates the cross-validated $F_2$-score of our RF classifier on the training data.
The blue histogram shows the cross-validated $F_2$-score of our RF classifier on the training data when the corresponding labels are randomly permuted.
We repeated the shuffling of the labels and cross-validated score assessment $1000$ times with different random seeds.
The difference between the scores for the permuted data and those for the original data indicates the significance of our trained model.} 
\label{fig:test_score}
\end{center}\end{figure}

As the size of our training set is modest, we tested if the predictions of our classifier are significantly better than chance.
We did so by randomly permuting the labels in a five-fold stratified cross-validation process, and assessed how well our classifier is able to fit the features it receives to predict the noisy (permuted) labels.
This allowed us to calculate the empirical p-value against the null hypothesis that our features and labels are independent.
As the p-value represents the fraction of the permuted datasets where our trained model performed as well or better than in the original data,
a high p-value ($>0.05$) indicates a lack of dependency between the features and our labels or a badly trained model, and conversely a low p-value indicates the existence of a dependency and a well trained-model. 
Figure \ref{fig:test_score} shows the results of this permutation test.
With a p-value of $0.001$ for the $F_2$-score of our model, we conclude that our features and labels are not independent and that our classifier is able to model a significant part of the mapping from our features to our labels.

\subsection{Feature importance}
\label{sec:results_importance}
We took two approaches to assess which (morphological) features contribute most to the predictive power of our classifier.
The first is `Mean Decrease Impurity feature importance' \citep{Breiman2001,louppe2014understanding}, also known as the Gini importance (not to be confused with the Gini coefficient). 
Features used at the base of a decision tree contribute to the predicted label of a larger number of radio sources than features near the leaves of a decision tree. 
The relative depth of a feature in a tree of the RF can thus be used as a proxy for its importance in predicting our class labels. 
We have to be cautious in interpreting these values, as impurity-based feature importance can be misleading for features with a high cardinality, favouring (continuous) features that can take on a high number of unique values over (discrete/ordinal) features with a low number of unique values \citep{strobl2007bias}.
In our case, the only ordinal feature is the `remnants per SOM neuron'.
Furthermore, the impurity-based importance is based on the training set and features that the RF uses to overfit will also show up as important.

\begin{figure*}\begin{center}
\includegraphics[width=0.99\textwidth]{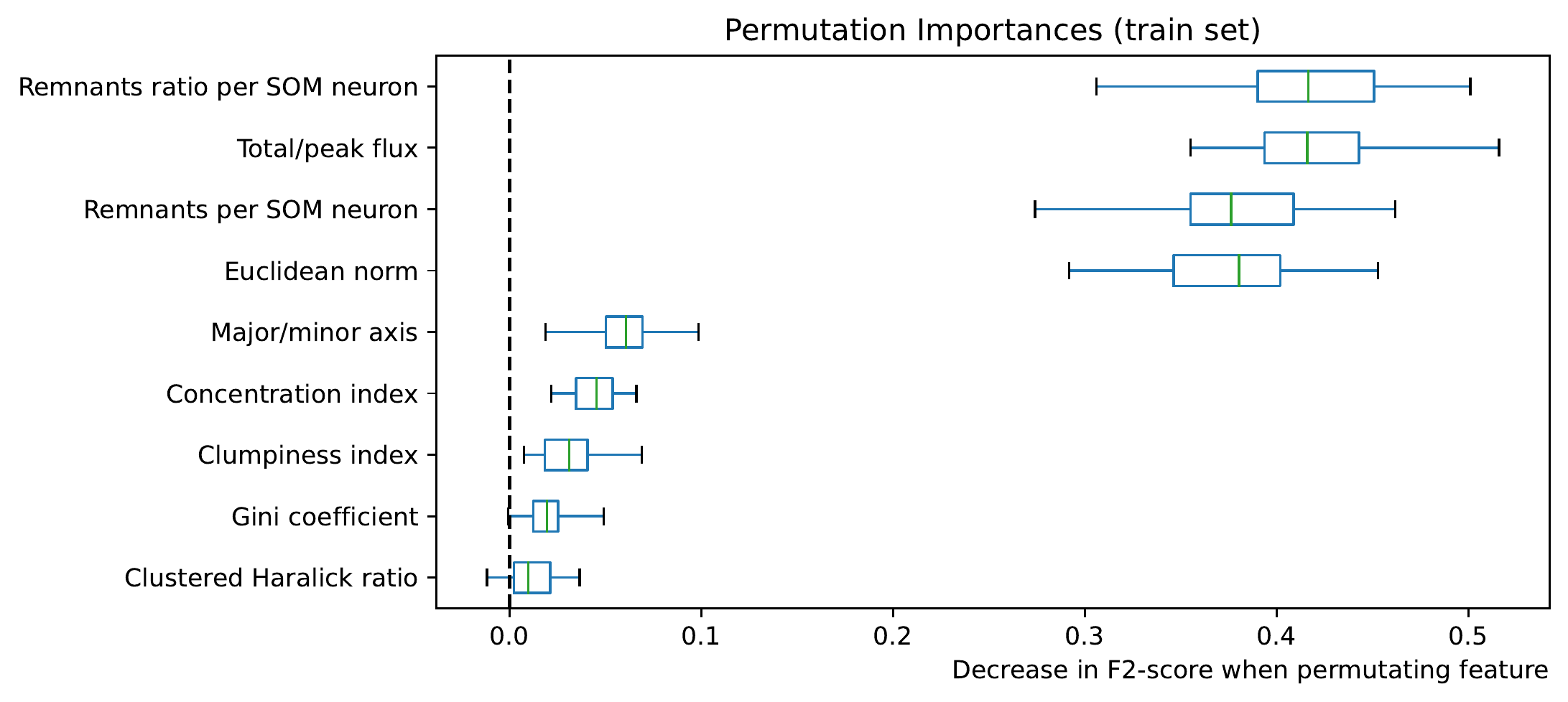}
\includegraphics[width=0.99\textwidth]{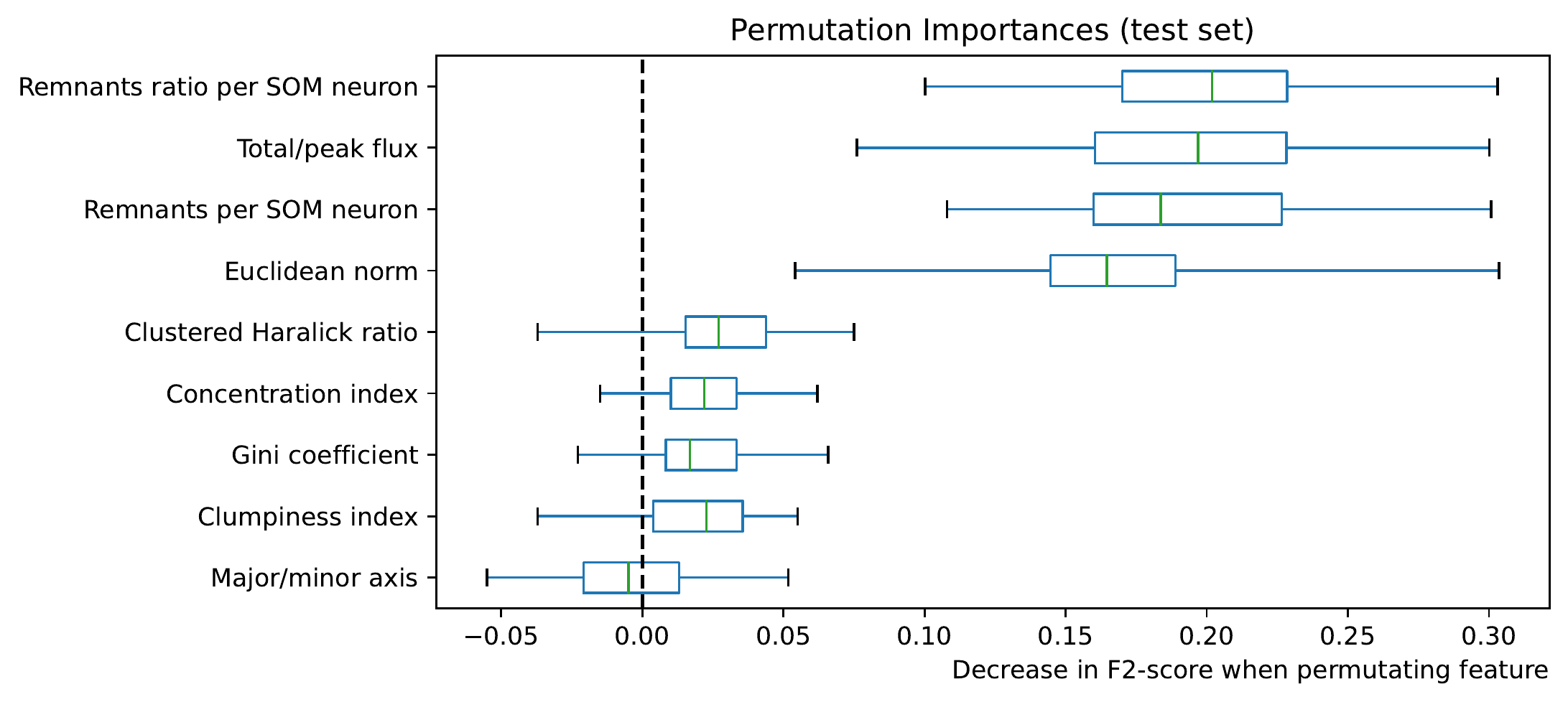}
\caption{Feature permutation importances for our training set (top panel) and our test set (bottom panel) with respect to the $F_2$-score. 
A higher $F_2$-score indicates better predictive performance.
The more important a feature is to the RF, the more the $F_2$-score decreases after permuting this feature.} 
\label{fig:feature_perm}
\end{center}\end{figure*}

\begin{table}
\centering
\caption{Normalized impurity-based feature importance of the training set according to RF. Higher values indicate higher importance.}
\label{tab:impurity_importance}
\begin{tabular}{ll}
\hline\hline
Feature & Importance \\
\hline
Remnants ratio per SOM neuron & $0.19$ \\
Remnants per SOM neuron & $0.18$ \\
Total/peak flux      & $0.17$ \\
Euclidean norm       & $0.16$ \\
Clustered Haralick ratio & $0.1$ \\
Clumpiness index     & $0.08$ \\
Concentration index  & $0.05$ \\
Major/minor axis     & $0.04$ \\
Gini coefficient     & $0.04$ \\
\hline
\end{tabular}
\end{table}

In our second approach, we test the saliency of the different features that we use by means of feature permutation.
Feature permutation importance is calculated by comparing the accuracy of our classifier with the original train or test data to the accuracy attained when the values of one of the features in the data are randomly shuffled.
This process is repeated for all features and the results are normalized.
We have to keep in mind that, using feature permutation, correlated features are shown as less important, because part of the performance loss of one permuted correlated feature is negated by the corresponding unpermuted correlated feature.
In our case, the SOM-features are correlated, and from Fig. \ref{sec:results_features}, the Gini coefficient, Clumpiness index and Concentration index are also correlated.

Table \ref{tab:impurity_importance} shows the results of the impurity-based feature importance and Fig. \ref{fig:feature_perm} the results of the feature permutation importance.
It is reassuring to see that the two different feature importance methods point towards the same conclusion.
Both approaches indicate that the importance of our features can be roughly divided into two groups: the `major to minor axis ratio', the `Haralick-derived features', and the `features borrowed from optical galaxy morphology' account for only a few per cent of the predictive power of the model, while the `SOM-derived features' and the `total to peak flux ratio' are most salient.
Taking a closer look at Fig. \ref{fig:feature_perm}, we see the feature permutation importance for both the training and the test set. 
The test set values indicate which features are actually important in predicting the right class label.
High positive values for features in the training set that are less positive or even negative for the test set indicate that the classifier uses these features to overfit on the training data.
Looking at the difference between the training and test set values, most features are used by the RF to improve the model but also marginally overfit on the training set.
Removing the `major to minor axis ratio' feature might even marginally improve the resulting $F_2$-score on the test set.

\subsection{Random forest classifier performance}
\label{sec:results_performance}
\begin{figure}\begin{center}
\includegraphics[width=0.8\columnwidth]{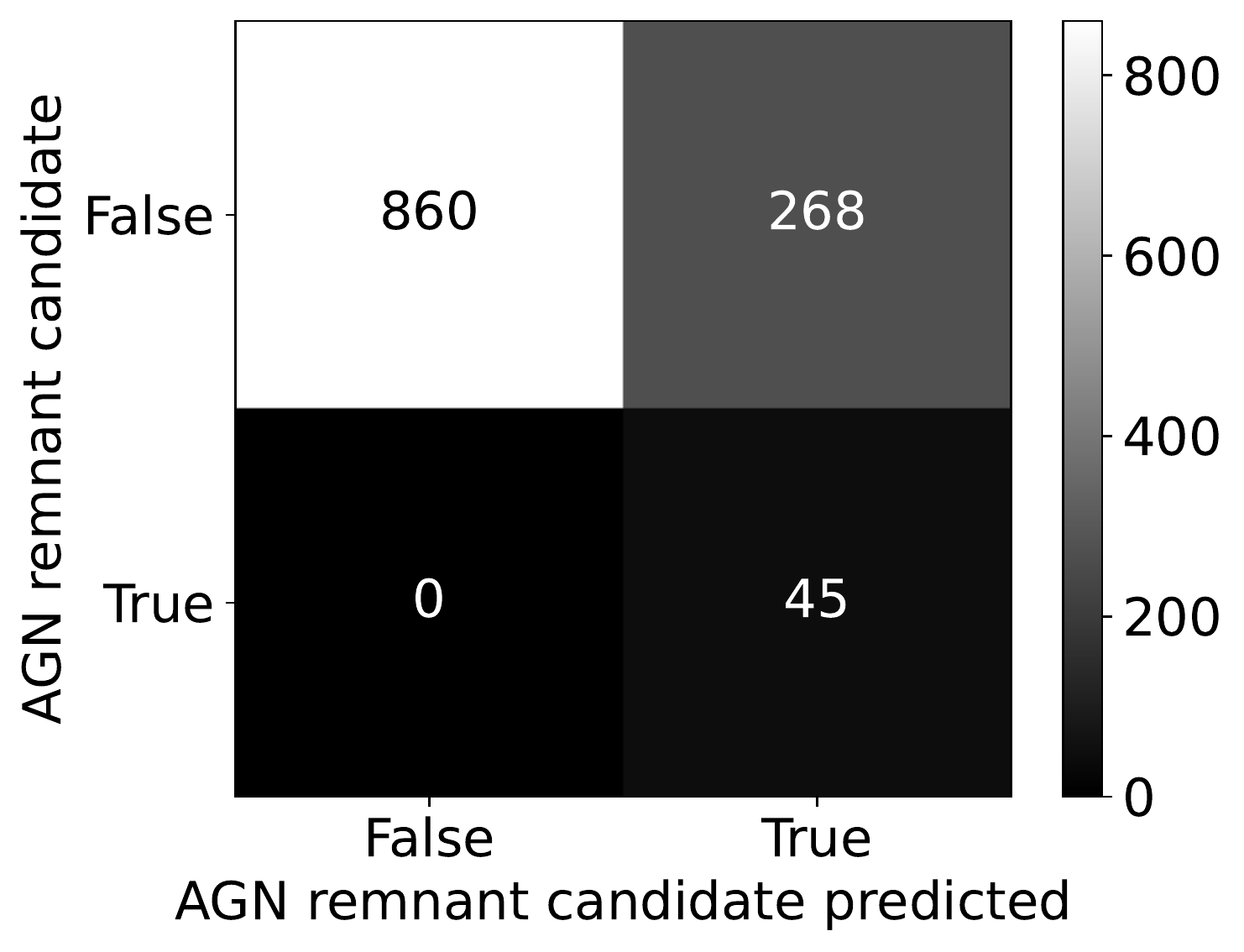}
\caption{Confusion matrix for our trained classifier applied to the test set, using a prediction threshold of $0.25$.} 
\label{fig:confusion}
\end{center}\end{figure}

As we look for a small number of AGN remnant candidates in a large sample of not yet inspected sources, 
isolating all candidates in the positive predictions is our most important objective for the classifier.
In classification terms, that means that a `false negative' -- predicting that a radio source is not a candidate while it is -- 
is less desirable than a `false positive' -- predicting that a radio source is a candidate while it is not.
We can change the prediction threshold of our trained classifier (which is set to $0.5$ by default) to generate fewer false negatives at the cost of more false positives.
We change the prediction threshold to $0.25$, as this is the point where the recall for our candidates is one hundred per cent.
Although we focussed on maximizing the $F_2$-score during training, resulting in a final $F_2$-score of $0.71$ on the train set and $0.46$ on the test set,
we also provide a summary of more common performance metrics in Table \ref{tab:perf} thereby enabling easy comparison to other classifiers.
The performance of the resulting classifier on our hold-out test can also be deduced from the corresponding confusion matrix shown in Fig. \ref{fig:confusion}.
The confusion matrix shows that for the $1,173$ radio sources from our test set,
we correctly discard $860$ sources (true negatives),
we correctly label $45$ sources as candidates (true positives),
we discard $0$ candidates (false negatives),
we label $268$ sources that have not yet been inspected as candidates (false positives).
Thus, if our hypothesis is true and we verify that upon visual inspection the (not yet inspected) false negatives do indeed turn out to contain way more AGN remnant candidates than the (not yet inspected) true negatives, we can reduce the number of to-be-inspected sources from $1,173$ 
to $268+45=313$, a reduction of $73\%$.
By accepting a non-zero number of false negatives, this reduction can be increased.

To put these numbers in perspective, we can estimate the absolute number of radio sources that would require visual inspection for the to-be-completed LoTSS.
Assuming that the completed LoTSS will cover $85\%$ of the northern hemisphere (as suggested by \citet{Shimwell2022}), and extrapolating the source numbers from the preliminary LoTSS-DR2 catalogue,\footnote{The preliminary LoTSS-DR2 catalogue, which covers $27\%$ of the northern sky, contains $14,287$ sources $>60$ arcsec.} we expect to find roughly $45$ thousand sources\footnote{Or fewer, as the noise levels for further data releases are expected to be higher due to observing closer to the galactic plane.} $>60$ arcsec in the complete LoTSS, for which our method would reduce visual inspection to roughly $12$ thousand sources.

\begin{table}
\centering
\caption{Performance of trained RF on  the test set with prediction threshold set to guarantee full recall of the AGN remnant candidates.}
\label{tab:perf}
\begin{tabular}{rrrrr}
\hline\hline
  & precision & recall & f1-score & support \\
\hline
noncandidate  & $1.00$ & $0.76$ & $0.87$ & $1128$ \\
remnant candidate  & $0.14$ & $1.00$ & $0.25$ & $45$ \\
  &  &  & &  \\
weighted average  & $0.97$ & $0.77$ & $0.84$ & $1173$ \\
\hline
accuracy & $0.77$ & & &  $1173$ \\
\hline
\end{tabular}
\end{table}

\begin{figure*}\begin{center}
\includegraphics[width=0.99\textwidth]{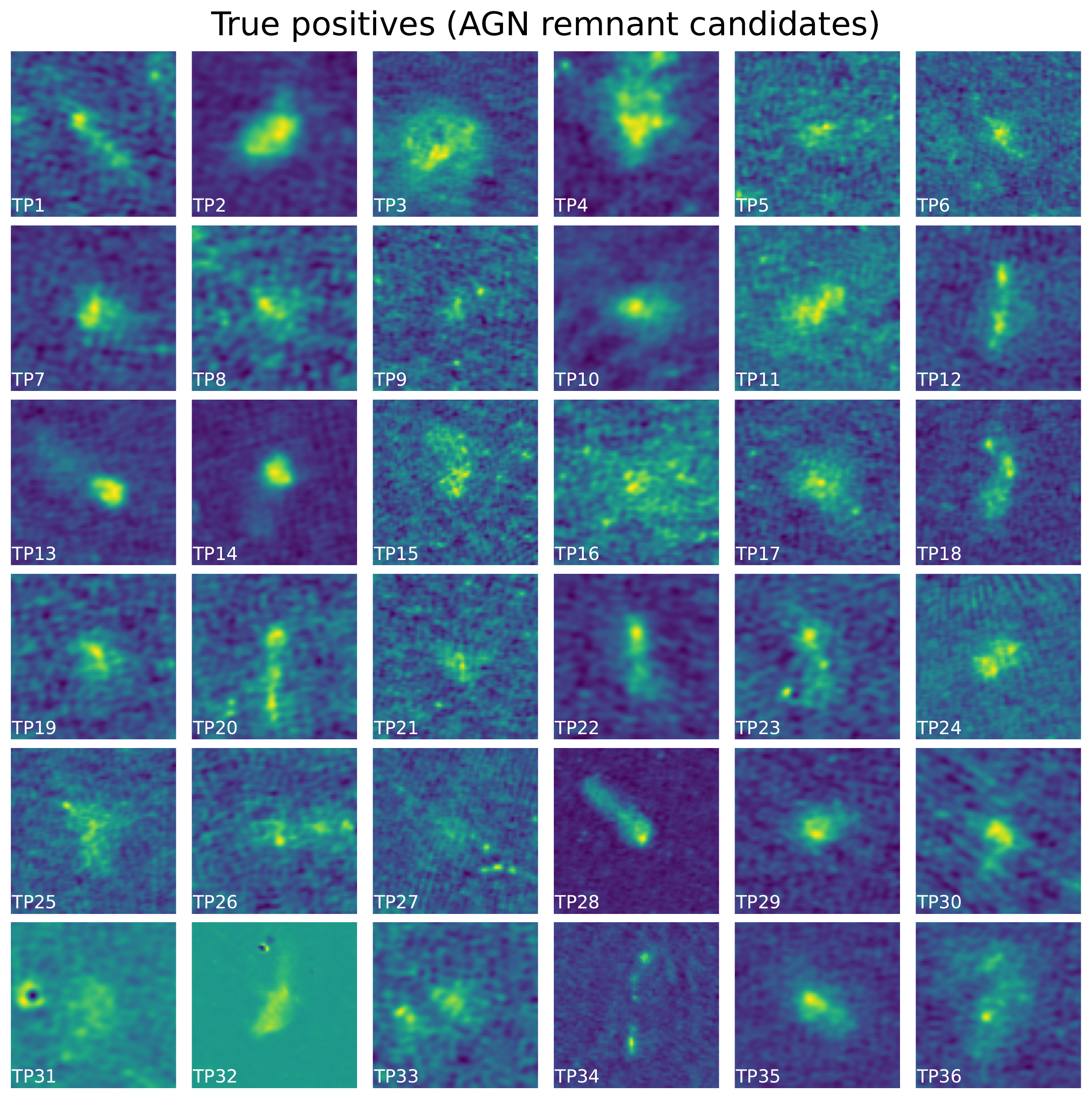}
\caption{Randomly sampled examples of true positives from the test set. 
True positives are radio sources for which our model-predicted label matches the `AGN remnant candidate' label from the initial visual inspection. } 
\label{fig:output_TP}
\end{center}\end{figure*}

\begin{figure*}\begin{center}
\includegraphics[width=0.99\textwidth]{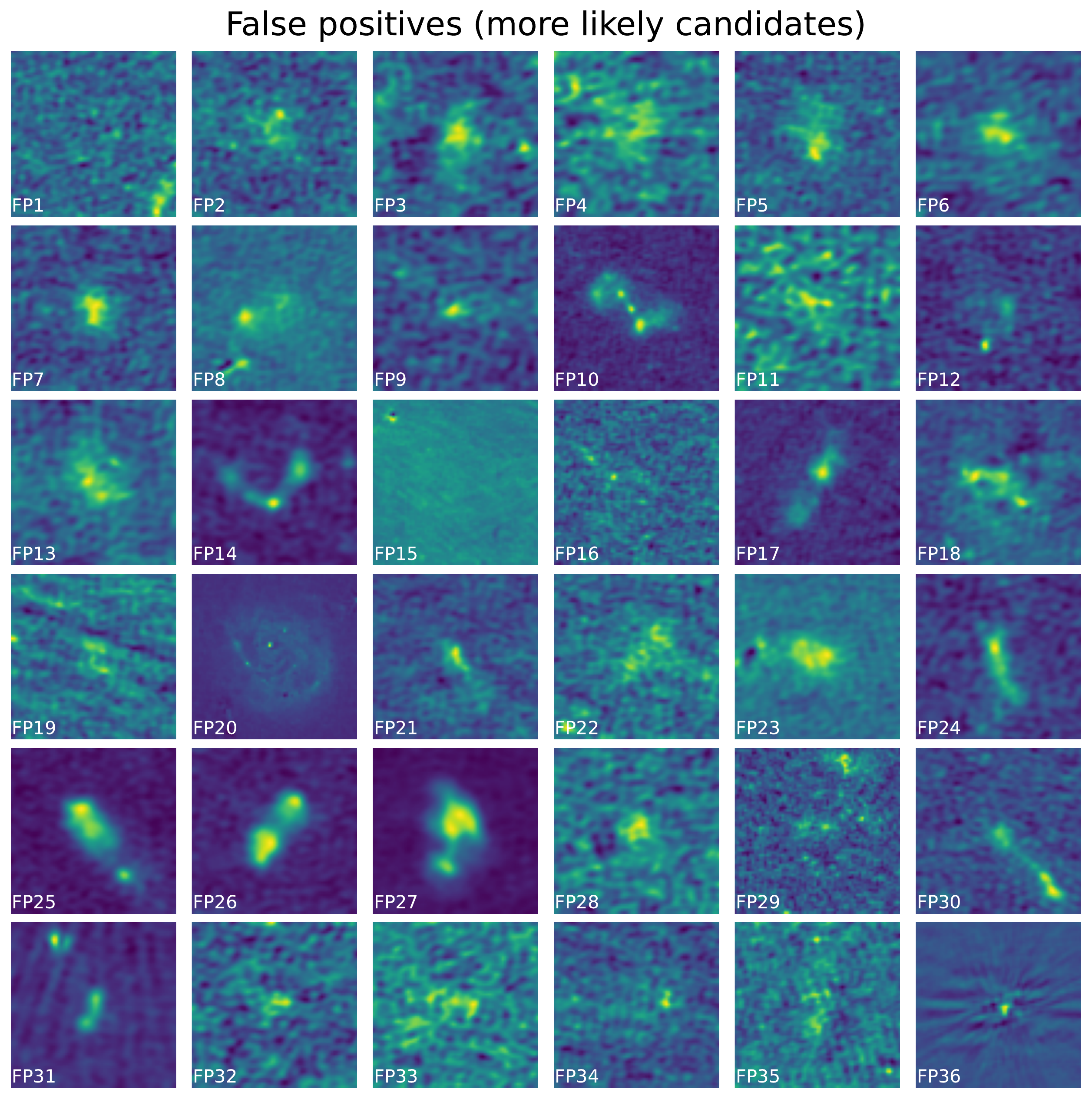}
\caption{Randomly sampled examples of false positives from the test set. 
False positives are not yet inspected radio sources for which the model-predicted label is `AGN remnant candidate'.
False positives indicate that our trained model cannot distinguish these sources from the sources labelled `AGN remnant candidate' during the initial visual inspection. 
Therefore, these false positives are \textit{more likely} to be labelled `AGN remnant candidate' upon subsequent visual inspection. 
} 
\label{fig:output_FP}
\end{center}\end{figure*}

\begin{figure*}\begin{center}
\includegraphics[width=0.99\textwidth]{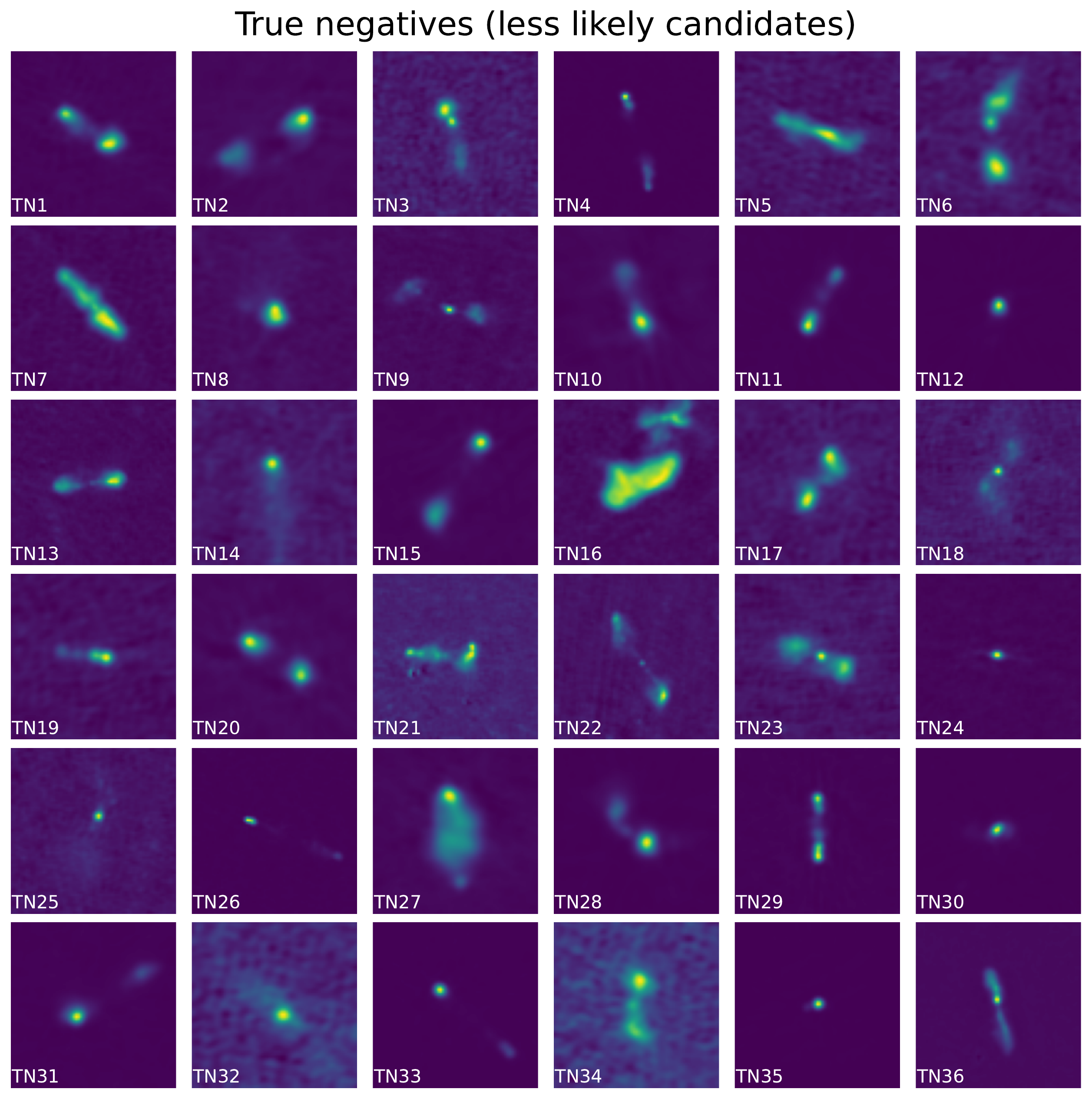}
\caption{Randomly picked examples of true negatives from the test set. 
True negatives are not yet inspected radio sources for which the model-predicted label is `noncandidate'.
As our trained model is able to distinguish these sources from the sources that were labelled `AGN remnant candidate' during initial visual inspection,
these sources are \textit{less likely} to be labelled `AGN remnant candidate' upon subsequent visual inspection. 
} 
\label{fig:output_TN}
\end{center}\end{figure*}

\section{Discussion}
\label{sec:discussion}
In this section, we discuss the saliency of the different features and estimate the percentage of `AGN remnant candidates' in the positive and negative predictions of our classifier and discuss the implications thereof. 
We also discuss the use and limitations of our methods, the added value of multi-frequency information, and future prospects.

\subsection{Feature saliency}
\label{sec:saliency}
\citet{Proctor2016}, who used a decision tree classifier in combination with six source catalogue-derived features to find giant radio galaxy candidates, suggested that including more features might allow for better isolation of the sought-after radio sources.
We observe that for AGN remnant candidates, SOM-derived morphological features can indeed complement source catalogue-derived features.
Table \ref{tab:impurity_importance} and Fig. \ref{fig:feature_perm} show that the SOM-derived features and the total to peak flux ratio are the most salient features.
Specifically the lower panel of Fig. \ref{fig:feature_perm} shows that the other features are roughly ten times less important to our RF classifier.

We suggest two explanations for the low saliency of the concentration index, the clumpiness index, the Gini coefficient, and
the Haralick clusters.
First, we can observe from Fig. \ref{fig:metrics}, that these features are correlated, which reduces their importance.
Second, an inspection of examples of sources and their feature value, see App. \ref{app:optical_morpho} and \ref{app:hara}, show that the noise around the central source and image artefacts from neighbouring sources strongly influence (or even dominate) the value of these extracted features.
For high signal-to-noise sources, we could have negated much of the noise using sigma-clipping during pre-processing.
Second, as we are concerned with sources that by definition have a relatively low surface brightness, we experimented with different (low) levels of sigma-clipping but we did not find a satisfactory compromise between loss of (diffuse) signal and noise removal, so we opted not to use sigma-clipping.
The high saliency of the SOM features, despite the presence of the noise and emission from neighbouring sources, might be explained by the fact that the neurons in the SOM (each of which can be regarded as a tiny cluster) are suspended in a two-dimensional lattice, allowing for a greater separation of both morphology and relative noise levels.

\subsection{Estimating the number of `AGN remnant candidates' after model prediction}
\label{sec:estimating}
Given that most sources in our training and test set have not yet been visually inspected,
we seek to answer the question: upon drawing a source from the subset of sources
for which our model prediction is positive (it predicts a `AGN remnant candidate'-label),
will visual inspection show that the source is indeed an `AGN remnant candidate'?
Furthermore, we also want to answer this question for sources for which the model 
prediction is negative (it predicts a `noncandidate'-label).
These answers allow us to estimate how many `AGN remnant candidates'
we expect to find upon visually inspecting all the positive model predictions,
and how many we miss by not visually inspecting the negative model predictions.

To answer these questions, we randomly sampled (without replacement) and 
visually inspected $36$
sources from the $268$ sources in our test set with a positive prediction that
have not yet been visually inspected (the `false positives' in our test set).
Similar to the initial visual inspection, we performed visual inspection by looking at the LoTSS-DR2 radio emission of the source, its surroundings and the overlapping optical data from Pan-STARRS1.
Upon visual inspection, 
targeted at confirming the diffuse amorphous emission and the absence of compact components 
in the radio, and the absence of SF in the optical, 
we found that $7$ of the $36$ draws
can be labelled as `AGN remnant candidate'.
From this experiment, plus the knowledge that $45$ sources in the set of positive predictions
were already assigned the label `AGN remnant candidate' during initial 
visual inspection (Sec. \ref{sec:data}), we know that the random variable,
\begin{equation}
p*_1 := \frac{45}{313} + \frac{7}{36} \cdot  \frac{268}{313}  = 31\%,
\end{equation}
is an unbiased estimator of the probability to draw a source among positive predictions that turns out to
have the label `AGN remnant candidate' after visual inspection.
Furthermore, we estimate the variance on $p*$ to be:
\begin{equation}
W := \frac{268^2}{(313^2 \cdot 36^2)} \cdot 36 \cdot \frac{7}{36} \cdot (1-\frac{7}{36}) \cdot \frac{232}{268-1}=0.003.
\end{equation}
As the standard deviation is the square root of the variance, we retrieve the result:
$p*_1 = 31\pm5\%$.
(See App. \ref{app:derivation} for the full derivation.)

Analogously, we randomly sampled (without replacement) and visually inspected $36$
sources from the $860$ sources in our test set with a negative prediction that
have not yet been visually inspected (the `true negatives' in our test set).
After visual inspection, $0$ of the draws is labelled as `AGN remnant candidate'.
That would lead to the unsatisfying estimator $p*_2=0$, 
so instead we can use the upper bound of a (conservative) confidence interval.
We apply the rule of three and estimate with $95\%$ confidence that 
fewer than $1$ in $36/3=12$ sources in our negative predictions 
will be an `AGN remnant candidate'.
These findings reinforce our hypothesis that we can speed up the process of finding more AGN 
remnant candidates by visually inspecting only the positive predictions of our model that 
was trained to separate a relatively small sample of radio sources with given 
`AGN remnant candidate' label from a larger sample of not yet inspected radio sources.

As the fraction of positive predictions is $f_p=\frac{45+268}{45+268+860}=0.27$; 
if we were to base a census on just the positive predictions of our classifier,
we are likely to conclude that $f_{p}\cdot 31\pm5\%=10\pm2\%$ 
of all the sources with a projected linear size $>60$ arcsec in LoTSS are AGN remnant candidates.
This percentage is in agreement with the AGN remnant fractions $<9\%$,$<11\%$, and $4\%$--$10\%$, within samples of $>1$ arcmin sources,
reported respectively by \citet{Mahatma2018}, \citet{jurlin2020life}, and \citet{Benjamin2021}.

\subsection{Use and limitations}
\label{sec:limitations}
Our methodology can also be repeated for the classification of other morphological groups for which only a few hundred labelled samples are available, like core-dominated FRIs, wide-angle tailed or narrow-angle tailed (bent) AGN sources \citep{mingo2019revisiting}. 
Likewise, in the case of a new calibration set of AGN remnant candidates (because of observational follow-up or new insights) the method explained in this work could simply be re-applied.
Compared to a deep neural network, which might also be considered for finding similar sources but is more akin to a `black-box' approach, our methods allow for more insight into which features are most salient to the classifier.

Below, we discuss two main limitations of our method.
First, our classifier is trained using expert labels in a supervised fashion.
As a result, human error in the labelling (known as `label noise') will unavoidably negatively affect the results of our classifier.
Label noise decreases classification performance and forces a model to be more complex than necessary to be able to fit the noisy labels \citep{frenay2013classification}.
Label noise negatively affects any type of classification system, but
RFs have at least demonstrated \citep{Breiman2001,folleco2008software} to be more robust for datasets with asymmetric label noise than the commonly used Naive Bayes \citep{murphy2012machine}, or C4.5 \citep{Quinlan1993} classifiers.

Second, the strength of our method depends on the quality of the radio source catalogue.
For example, a catalogue entry centred on a single radio-lobe, that is erroneously not associated with its other radio-lobe, will appear as a rather amorphous oblong shape and might erroneously be classified as an AGN remnant candidate.

\subsection{Future prospects}
\label{sec:future}
To improve future results using the methodology presented in this manuscript, as discussed in Sec. \ref{sec:saliency}, it is essential to mitigate the effect of the noise around the radio sources.
A sensible improvement would be to switch the source-detection software from PyBDSF, which returns a source-model build-up by multiple Gaussians and works best for unresolved or slightly resolved sources, towards a flood-fill type of source-detection software like ProFound \citep{robotham2018profound}, which returns a pixel-resolution mask for each source and works better for complex well-resolved sources \citep{hale2019radio}.
That would allow us to mask all pixels outside of the radio source's mask before extracting our morphological features and thereby reduce the effect of the surrounding noise.

Future research can also attempt to improve our classifier by additional training with sources that were selected for visual inspection (labelling) using active learning.
Active learning is a `human-in-the-loop' practice in machine learning that attempts to improve a model by alternating between model prediction and human labelling, with the concept that by labelling samples for which the model is least certain, the model requires fewer labels to acquire the same performance as a model that does not use active learning \citep{Settles2009a}.
\citet{Walmsley2020} show the benefits of using active learning in the context of classifying optical images of galaxies.
A personalised active learning framework like Astronomaly \citep{Lochner2021} could be used for this purpose.

Another path towards improving our classifier would be to build a `foundation model' \citep[e.g.][]{2021arXiv210105224A,bommasani2021opportunities,Walmsley2022hybrid}.
A foundation model is a very large neural network \citep[e.g.][]{lecun1989backpropagation,Goodfellow2016} that achieves improved performance on its predictions by (self-supervised) pre-training on a very large corpus of related data.\footnote{Self-supervised learning entails training a neural network using labels that are self-derived from the data \citep[e.g.][]{Ting2020,Jure2021}, in the image domain that can, for example, be achieved by masking a part of the image and training the network to predict the values of these masked pixels.}
The combined data from the Square Kilometre Array \citep[SKA;][]{Braun2015} pathfinder surveys\footnote{LoTSS and its low-frequency equivalent LoLSS \citep{Gasperin2022}, MeerKAT \citep{meerkat2016} international GHz tiered extragalactic exploration \citep[MIGHTEE;][]{mightee}, and the Evolutionary Map of the Universe survey \citep[EMU;][]{Norris2011} by the  Australian square-kilometre-array pathfinder \citep[ASKAP;][]{askap2008}.} is perfect to create a foundation model for these surveys and future large-scale sky surveys from the SKA.
Such a model would improve the performance on multiple downstream tasks like source morphology classification and source-component association \citep[e.g.][]{Mostert2022}.

\section{Conclusions}
\label{sec:conclusion}
Large samples of AGN remnants would progress the study of the radio galaxy life cycle.
The number of observed radio galaxies is high thanks to the current SKA-pathfinder surveys, however, current remnant samples are still small as AGN remnants are quite rare and identifying them based on their morphology through visual inspection is time-consuming.
In this work, we have presented an automated way to determine which radio sources in the LOFAR Two-metre Sky Survey are likely AGN remnant candidates based on the source's radio morphology. 
We train and test an RF classifier using features extracted from $4,075$ radio sources, $150$ of which were determined to be AGN remnant candidates through visual inspection by Brienza et al. in prep.

First, we automated the extraction of a large variety of morphological features, from simple statistics like a source's width-to-height ratio, the Gini coefficient, clumpiness and concentration indices normally used in optical astronomy, soft-clustered Haralick features, and SOM-derived features.
We introduced size-invariance and a compression technique that makes the SOM more robust to small sample sizes, a necessity given that we started from $150$ AGN remnant candidates, which is a small number of labels to build a classifier on.
Using a grid search and cross-validation, we evaluated the best hyperparameters to train our random forest classifier with.
As we prefer visually inspecting more radio sources than miss AGN remnant candidates, we chose our prediction threshold such that we sacrifice precision for a recall of $100\%$ on the already labelled AGN remnant candidates.

We took steps towards the quantification of the, so far, subjective nature of the radio morphology of AGN remnant candidates.
Through feature-permutation, we learned that with the chosen set-up, SOM-derived features and total to peak flux ratio are most salient in predicting the right class label, while clustered Haralick-features, Gini coefficient, concentration index, clumpiness index and major to minor axis ratio, are less relevant.

Running our model on a new sky region of the ongoing LoTSS survey,
we could visually inspect only the positive predictions, reducing the percentage of
sources requiring visual inspection with $73\%$.
Within this set of positive predictions, we expect 
$31\pm5\%$ to be `AGN remnant candidate'.
Using this approach we will miss the `AGN remnant candidates' that still reside in
the negative predictions, 
but we conservatively estimated that these constitute fewer than $8\%$ of the negative predictions.

The results from the method presented in this paper improve if a larger sample of labelled sources is available, as the predictive power of the features grows with a larger sample size.
We conclude that our method brings us closer to identifying complete samples of AGN remnant candidates in an automatic way.
However, given our current precision and the large volume of radio sources in LoTSS-DR2 and upcoming SKA and SKA-pathfinder data releases, there are further steps needed to reach the ultimate goal of fully automatic morphological selection of AGN remnant candidates.

 \begin{acknowledgements}
The data and code used to produce the results and figures of this paper are available at: \url{https://lofar-surveys.org/finding_agn_remnants.html}.
This research has made use of the Astropy \citep{astropy} and the scikit-learn \citep{scikit-learn} Python packages.
The authors also acknowledge the usefulness of the PINK (\url{https://github.com/HITS-AIN/PINK}) and HaralickFeatures (\url{https://github.com/KushathaNtwaetsile/HaralickFeatures}) implementations.
LOFAR data products were provided by the LOFAR Surveys Key Science project (LSKSP; https://lofar-surveys.org/) and were derived from observations with the International LOFAR Telescope (ILT). LOFAR (van Haarlem et al. 2013) is the Low Frequency Array designed and constructed by ASTRON. It has observing, data processing, and data storage facilities in several countries, which are owned by various parties (each with their own funding sources), and which are collectively operated by the ILT foundation under a joint scientific policy. The efforts of the LSKSP have benefited from funding from the European Research Council, NOVA, NWO, CNRS-INSU, the SURF Co-operative, the UK Science and Technology Funding Council and the Jülich Supercomputing Centre.
MB acknowledges financial support from INAF under the SKA/CTA PRIN “FORECaST”, from the agreement ASI-INAF n. 2017-14-H.O and from the PRIN MIUR 2017PH3WAT “Blackout”. M.S.S.L. Oei acknowledges support from the VIDI research programme with project number 639.042.729, which is financed by The Netherlands Organisation for Scientific Research (NWO). MJH acknowledges support from the UK STFC [ST/V000624/1]. LA is grateful for support from UK STFC via CDT studentship grant ST/P006809/1. 
\end{acknowledgements}

\bibliographystyle{aa} 
\bibliography{finding_AGN_remnants.bib}

\begin{thebibliography}{87}
\expandafter\ifx\csname natexlab\endcsname\relax\def\natexlab#1{#1}\fi

\bibitem[{{Alegre} {et~al.}(2022){Alegre}, {Sabater}, {Best}, {Mostert},
  {Williams}, {G{\"u}rkan}, {Hardcastle}, {Kondapally}, {Shimwell}, \&
  {Smith}}]{Alegre2022}
{Alegre}, L., {Sabater}, J., {Best}, P., {et~al.} 2022, \mnras, 516, 4716

\bibitem[{{Alhassan} {et~al.}(2018){Alhassan}, {Taylor}, \&
  {Vaccari}}]{alhassan2018first}
{Alhassan}, W., {Taylor}, A.~R., \& {Vaccari}, M. 2018, \mnras, 480, 2085

\bibitem[{{Aniyan} \& {Thorat}(2017)}]{Aniyan2017a}
{Aniyan}, A.~K. \& {Thorat}, K. 2017, \apjs, 230, 20

\bibitem[{{Azizi} {et~al.}(2021){Azizi}, {Mustafa}, {Ryan}, {Beaver},
  {Freyberg}, {Deaton}, {Loh}, {Karthikesalingam}, {Kornblith}, {Chen},
  {Natarajan}, \& {Norouzi}}]{2021arXiv210105224A}
{Azizi}, S., {Mustafa}, B., {Ryan}, F., {et~al.} 2021, arXiv e-prints,
  arXiv:2101.05224

\bibitem[{Bommasani {et~al.}(2021)Bommasani, Hudson, Adeli, Altman, Arora, von
  Arx, Bernstein, Bohg, Bosselut, Brunskill,
  {et~al.}}]{bommasani2021opportunities}
Bommasani, R., Hudson, D.~A., Adeli, E., {et~al.} 2021, arXiv preprint
  arXiv:2108.07258

\bibitem[{{Bowles} {et~al.}(2021){Bowles}, {Scaife}, {Porter}, {Tang}, \&
  {Bastien}}]{Bowles2021gated}
{Bowles}, M., {Scaife}, A. M.~M., {Porter}, F., {Tang}, H., \& {Bastien}, D.~J.
  2021, \mnras, 501, 4579

\bibitem[{{Braun} {et~al.}(2015){Braun}, {Bourke}, {Green}, {Keane}, \&
  {Wagg}}]{Braun2015}
{Braun}, R., {Bourke}, T., {Green}, J.~A., {Keane}, E., \& {Wagg}, J. 2015, in
  Advancing Astrophysics with the Square Kilometre Array (AASKA14), 174

\bibitem[{{Breiman}(2001)}]{Breiman2001}
{Breiman}, L. 2001, Machine Learning, 45, 5

\bibitem[{{Brienza} {et~al.}(2017){Brienza}, {Godfrey}, {Morganti}, {Prandoni},
  {Harwood}, {Mahony}, {Hardcastle}, {Murgia}, {R{\"o}ttgering}, {Shimwell}, \&
  {Shulevski}}]{Brienza2017}
{Brienza}, M., {Godfrey}, L., {Morganti}, R., {et~al.} 2017, \aap, 606, A98

\bibitem[{{Brienza} {et~al.}(2018){Brienza}, {Morganti}, {Murgia}, {Vilchez},
  {Adebahr}, {Carretti}, {Concu}, {Govoni}, {Harwood}, {Intema}, {Loi},
  {Melis}, {Paladino}, {Poppi}, {Shulevski}, {Vacca}, \&
  {Valente}}]{Brienza2018duty}
{Brienza}, M., {Morganti}, R., {Murgia}, M., {et~al.} 2018, \aap, 618, A45

\bibitem[{Campello {et~al.}(2015)Campello, Moulavi, Zimek, \&
  Sander}]{campello2015}
Campello, R.~J., Moulavi, D., Zimek, A., \& Sander, J. 2015, ACM Transactions
  on Knowledge Discovery from Data (TKDD), 10, 1

\bibitem[{{Chen} {et~al.}(2020){Chen}, {Kornblith}, {Norouzi}, \&
  {Hinton}}]{Ting2020}
{Chen}, T., {Kornblith}, S., {Norouzi}, M., \& {Hinton}, G. 2020, arXiv
  e-prints, arXiv:2002.05709

\bibitem[{Coelho(2013)}]{coelho2013mahotas}
Coelho, L.~P. 2013, Journal of Open Research Software, 1

\bibitem[{{Condon} {et~al.}(1998){Condon}, {Cotton}, {Greisen}, {Yin},
  {Perley}, {Taylor}, \& {Broderick}}]{Condon1998}
{Condon}, J.~J., {Cotton}, W.~D., {Greisen}, E.~W., {et~al.} 1998, \aj, 115,
  1693

\bibitem[{{Conselice}(2014)}]{Conselice2014}
{Conselice}, C.~J. 2014, \araa, 52, 291

\bibitem[{{Cordey}(1987)}]{cordey1987ic}
{Cordey}, R.~A. 1987, \mnras, 227, 695

\bibitem[{{Dabhade} {et~al.}(2020){Dabhade}, {Mahato}, {Bagchi}, {Saikia},
  {Combes}, {Sankhyayan}, {R{\"o}ttgering}, {Ho}, {Gaikwad}, {Raychaudhury},
  {Vaidya}, \& {Guiderdoni}}]{Dabhade2020}
{Dabhade}, P., {Mahato}, M., {Bagchi}, J., {et~al.} 2020, \aap, 642, A153

\bibitem[{de~Gasperin(2022)}]{Gasperin2022}
de~Gasperin, F. 2022, aap

\bibitem[{{Fanaroff} \& {Riley}(1974)}]{Fanaroff1974a}
{Fanaroff}, B.~L. \& {Riley}, J.~M. 1974, \mnras, 167, 31P

\bibitem[{Folleco {et~al.}(2008)Folleco, Khoshgoftaar, Van~Hulse, \&
  Bullard}]{folleco2008software}
Folleco, A., Khoshgoftaar, T.~M., Van~Hulse, J., \& Bullard, L. 2008, in 2008
  IEEE congress on evolutionary computation (IEEE world congress on
  computational intelligence), IEEE, 3853--3859

\bibitem[{Fr{\'e}nay \& Verleysen(2013)}]{frenay2013classification}
Fr{\'e}nay, B. \& Verleysen, M. 2013, IEEE transactions on neural networks and
  learning systems, 25, 845

\bibitem[{{Galvin} {et~al.}(2019){Galvin}, {Huynh}, {Norris}, {Wang},
  {Hopkins}, {Wong}, {Shabala}, {Rudnick}, {Alger}, \&
  {Polsterer}}]{Galvin2019}
{Galvin}, T.~J., {Huynh}, M., {Norris}, R.~P., {et~al.} 2019, \pasp, 131,
  108009

\bibitem[{{Galvin} {et~al.}(2020){Galvin}, {Huynh}, {Norris}, {Wang},
  {Hopkins}, {Polsterer}, {Ralph}, {O'Brien}, \& {Heald}}]{Galvin2020}
{Galvin}, T.~J., {Huynh}, M.~T., {Norris}, R.~P., {et~al.} 2020, \mnras, 497,
  2730

\bibitem[{{Godfrey} {et~al.}(2017){Godfrey}, {Morganti}, \&
  {Brienza}}]{godfrey2017population}
{Godfrey}, L.~E.~H., {Morganti}, R., \& {Brienza}, M. 2017, \mnras, 471, 891

\bibitem[{Goodfellow {et~al.}(2016)Goodfellow, Bengio, \&
  Courville}]{Goodfellow2016}
Goodfellow, I., Bengio, Y., \& Courville, A. 2016, Deep Learning (MIT Press),
  \url{http://www.deeplearningbook.org}

\bibitem[{{G{\"u}rkan} {et~al.}(2018){G{\"u}rkan}, {Hardcastle}, {Smith},
  {Best}, {Bourne}, {Calistro-Rivera}, {Heald}, {Jarvis}, {Prandoni},
  {R{\"o}ttgering}, {Sabater}, {Shimwell}, {Tasse}, \& {Williams}}]{Gurkan2018}
{G{\"u}rkan}, G., {Hardcastle}, M.~J., {Smith}, D.~J.~B., {et~al.} 2018,
  \mnras, 475, 3010

\bibitem[{{Hale} {et~al.}(2019){Hale}, {Robotham}, {Davies}, {Jarvis},
  {Driver}, \& {Heywood}}]{hale2019radio}
{Hale}, C.~L., {Robotham}, A.~S.~G., {Davies}, L.~J.~M., {et~al.} 2019, \mnras,
  487, 3971

\bibitem[{Haralick {et~al.}(1973)Haralick, Shanmugam, \&
  Dinstein}]{haralick1973textural}
Haralick, R.~M., Shanmugam, K., \& Dinstein, I.~H. 1973, IEEE Transactions on
  systems, man, and cybernetics, 610

\bibitem[{{Harwood} {et~al.}(2018){Harwood}, {Hardcastle}, {Croston}, \&
  {Goodger}}]{Harwood2018}
{Harwood}, J.~J., {Hardcastle}, M.~J., {Croston}, J.~H., \& {Goodger}, J.~L.
  2018, {BRATS: Broadband Radio Astronomy ToolS}, Astrophysics Source Code
  Library, record ascl:1806.025

\bibitem[{{Ivezi{\'c}} {et~al.}(2019){Ivezi{\'c}}, {Connelly}, {Vanderplas}, \&
  {Gray}}]{Vanderplas2019}
{Ivezi{\'c}}, {\v{Z}}., {Connelly}, A.~J., {Vanderplas}, J.~T., \& {Gray}, A.
  2019, {Statistics, Data Mining, and Machine Learning in Astronomy}

\bibitem[{{Jarrett} {et~al.}(2000){Jarrett}, {Chester}, {Cutri}, {Schneider},
  {Skrutskie}, \& {Huchra}}]{jarrett2000a}
{Jarrett}, T.~H., {Chester}, T., {Cutri}, R., {et~al.} 2000, \aj, 119, 2498

\bibitem[{{Jarvis} {et~al.}(2016){Jarvis}, {Taylor}, {Agudo}, {Allison},
  {Deane}, {Frank}, {Gupta}, {Heywood}, {Maddox}, {McAlpine}, {Santos},
  {Scaife}, {Vaccari}, {Zwart}, {Adams}, {Bacon}, {Baker}, {Bassett}, {Best},
  {Beswick}, {Blyth}, {Brown}, {Bruggen}, {Cluver}, {Colafrancesco}, {Cotter},
  {Cress}, {Dav{\'e}}, {Ferrari}, {Hardcastle}, {Hale}, {Harrison}, {Hatfield},
  {Klockner}, {Kolwa}, {Malefahlo}, {Marubini}, {Mauch}, {Moodley}, {Morganti},
  {Norris}, {Peters}, {Prand oni}, {Prescott}, {Oliver}, {Oozeer},
  {Rottgering}, {Seymour}, {Simpson}, {Smirnov}, \& {Smith}}]{mightee}
{Jarvis}, M., {Taylor}, R., {Agudo}, I., {et~al.} 2016, in MeerKAT Science: On
  the Pathway to the SKA, 6

\bibitem[{{Johnston} {et~al.}(2008){Johnston}, {Taylor}, {Bailes}, {Bartel},
  {Baugh}, {Bietenholz}, {Blake}, {Braun}, {Brown}, {Chatterjee}, {Darling},
  {Deller}, {Dodson}, {Edwards}, {Ekers}, {Ellingsen}, {Feain}, {Gaensler},
  {Haverkorn}, {Hobbs}, {Hopkins}, {Jackson}, {James}, {Joncas}, {Kaspi},
  {Kilborn}, {Koribalski}, {Kothes}, {Landecker}, {Lenc}, {Lovell}, {Macquart},
  {Manchester}, {Matthews}, {McClure-Griffiths}, {Norris}, {Pen}, {Phillips},
  {Power}, {Protheroe}, {Sadler}, {Schmidt}, {Stairs}, {Staveley-Smith},
  {Stil}, {Tingay}, {Tzioumis}, {Walker}, {Wall}, \& {Wolleben}}]{askap2008}
{Johnston}, S., {Taylor}, R., {Bailes}, M., {et~al.} 2008, Experimental
  Astronomy, 22, 151

\bibitem[{{Jonas} \& {MeerKAT Team}(2016)}]{meerkat2016}
{Jonas}, J. \& {MeerKAT Team}. 2016, in MeerKAT Science: On the Pathway to the
  SKA, 1

\bibitem[{{Jurlin} {et~al.}(2020){Jurlin}, {Morganti}, {Brienza}, {Mandal},
  {Maddox}, {Duncan}, {Shabala}, {Hardcastle}, {Prandoni}, {R{\"o}ttgering},
  {Mahatma}, {Best}, {Mingo}, {Sabater}, {Shimwell}, \&
  {Tasse}}]{jurlin2020life}
{Jurlin}, N., {Morganti}, R., {Brienza}, M., {et~al.} 2020, \aap, 638, A34

\bibitem[{Kaiser {et~al.}(2010)Kaiser, Burgett, Chambers, Denneau, Heasley,
  Jedicke, Magnier, Morgan, Onaka, \& Tonry}]{panstarrs}
Kaiser, N., Burgett, W., Chambers, K., {et~al.} 2010, in Ground-based and
  Airborne Telescopes III, ed. L.~M. Stepp, R.~Gilmozzi, \& H.~J. Hall, Vol.
  7733, International Society for Optics and Photonics (SPIE), 159 -- 172

\bibitem[{{Kapoor} \& {Narayanan}(2022)}]{Kapoor2022}
{Kapoor}, S. \& {Narayanan}, A. 2022, arXiv e-prints, arXiv:2207.07048

\bibitem[{Kohonen(1989)}]{Kohonen1989}
Kohonen, T. 1989, Self-Organization and Associative Memory (Springer Berlin
  Heidelberg)

\bibitem[{Kohonen(2001)}]{Kohonen2001}
Kohonen, T. 2001, Self-organizing maps (Berlin New York: Springer)

\bibitem[{LeCun {et~al.}(1989)LeCun, Boser, Denker, Henderson, Howard, Hubbard,
  \& Jackel}]{lecun1989backpropagation}
LeCun, Y., Boser, B., Denker, J.~S., {et~al.} 1989, Neural computation, 1, 541

\bibitem[{{Lochner} \& {Bassett}(2021)}]{Lochner2021}
{Lochner}, M. \& {Bassett}, B.~A. 2021, Astronomy and Computing, 36, 100481

\bibitem[{Louppe(2014)}]{louppe2014understanding}
Louppe, G. 2014, PhD thesis, University of Liege, Belgium, arXiv:1407.7502

\bibitem[{{Ma} {et~al.}(2019){Ma}, {Xu}, {Zhu}, {Hu}, {Li}, {Shan}, {Zhu},
  {Gu}, {Li}, {Liu}, \& {Wu}}]{Ma2019machine}
{Ma}, Z., {Xu}, H., {Zhu}, J., {et~al.} 2019, \apjs, 240, 34

\bibitem[{{Mahatma} {et~al.}(2018){Mahatma}, {Hardcastle}, {Williams},
  {Brienza}, {Br{\"u}ggen}, {Croston}, {Gurkan}, {Harwood},
  {Kunert-Bajraszewska}, {Morganti}, {R{\"o}ttgering}, {Shimwell}, \&
  {Tasse}}]{Mahatma2018}
{Mahatma}, V.~H., {Hardcastle}, M.~J., {Williams}, W.~L., {et~al.} 2018,
  \mnras, 475, 4557

\bibitem[{Mahony {et~al.}(2016)Mahony, Morganti, Prandoni, van Bemmel,
  Shimwell, Brienza, Best, Br{\"u}ggen, Calistro~Rivera, de~Gasperin,
  {et~al.}}]{Mahony2016}
Mahony, E., Morganti, R., Prandoni, I., {et~al.} 2016, Monthly Notices of the
  Royal Astronomical Society, 463, 2997

\bibitem[{McInnes {et~al.}(2017)McInnes, Healy, \& Astels}]{McInnes2017}
McInnes, L., Healy, J., \& Astels, S. 2017, The Journal of Open Source
  Software, 2

\bibitem[{{Mingo} {et~al.}(2019){Mingo}, {Croston}, {Hardcastle}, {Best},
  {Duncan}, {Morganti}, {Rottgering}, {Sabater}, {Shimwell}, {Williams},
  {Brienza}, {Gurkan}, {Mahatma}, {Morabito}, {Prandoni}, {Bondi}, {Ineson}, \&
  {Mooney}}]{mingo2019revisiting}
{Mingo}, B., {Croston}, J.~H., {Hardcastle}, M.~J., {et~al.} 2019, \mnras, 488,
  2701

\bibitem[{{Mohan} {et~al.}(2022){Mohan}, {Scaife}, {Porter}, {Walmsley}, \&
  {Bowles}}]{Mohan2022}
{Mohan}, D., {Scaife}, A. M.~M., {Porter}, F., {Walmsley}, M., \& {Bowles}, M.
  2022, \mnras, 511, 3722

\bibitem[{{Mohan} \& {Rafferty}(2015)}]{Mohan2015}
{Mohan}, N. \& {Rafferty}, D. 2015, {PyBDSF: Python Blob Detection and Source
  Finder}, Astrophysics Source Code Library

\bibitem[{Morganti(2017)}]{morganti2017archaeology}
Morganti, R. 2017, Nature Astronomy, 1, 39

\bibitem[{{Morganti} {et~al.}(2021){Morganti}, {Oosterloo}, {Brienza},
  {Jurlin}, {Prandoni}, {Orr{\`u}}, {Shabala}, {Adams}, {Adebahr}, {Best},
  {Coolen}, {Damstra}, {de Blok}, {de Gasperin}, {D{\'e}nes}, {Hardcastle},
  {Hess}, {Hut}, {Kondapally}, {Kutkin}, {Loose}, {Lucero}, {Maan}, {Maccagni},
  {Mingo}, {Moss}, {Mostert}, {Norden}, {Oostrum}, {R{\"o}ttgering}, {Ruiter},
  {Shimwell}, {Schulz}, {Vermaas}, {Vohl}, {van der Hulst}, {van Diepen}, {van
  Leeuwen}, \& {Ziemke}}]{Morganti2021}
{Morganti}, R., {Oosterloo}, T.~A., {Brienza}, M., {et~al.} 2021, \aap, 648, A9

\bibitem[{{Mostert} {et~al.}(2022){Mostert}, {Duncan}, {Alegre},
  {R{\"o}ttgering}, {Williams}, {Best}, {Hardcastle}, \&
  {Morganti}}]{Mostert2022}
{Mostert}, R. I.~J., {Duncan}, K.~J., {Alegre}, L., {et~al.} 2022, \aap, 668,
  A28

\bibitem[{{Mostert} {et~al.}(2021){Mostert}, {Duncan}, {R{\"o}ttgering},
  {Polsterer}, {Best}, {Brienza}, {Br{\"u}ggen}, {Hardcastle}, {Jurlin},
  {Mingo}, {Morganti}, {Shimwell}, {Smith}, \& {Williams}}]{Mostert2020}
{Mostert}, R. I.~J., {Duncan}, K.~J., {R{\"o}ttgering}, H. J.~A., {et~al.}
  2021, \aap, 645, A89

\bibitem[{{Murgia} {et~al.}(2011){Murgia}, {Parma}, {Mack}, {de Ruiter},
  {Fanti}, {Govoni}, {Tarchi}, {Giacintucci}, \& {Markevitch}}]{Murgia2011}
{Murgia}, M., {Parma}, P., {Mack}, K.~H., {et~al.} 2011, \aap, 526, A148

\bibitem[{Murphy(2012)}]{murphy2012machine}
Murphy, K.~P. 2012, Machine learning: a probabilistic perspective (MIT press)

\bibitem[{Murthy {et~al.}(1993)Murthy, Kasif, Salzberg, \& Beigel}]{Murthy1993}
Murthy, S.~K., Kasif, S., Salzberg, S., \& Beigel, R. 1993, in Proceedings of
  AAAI, Vol.~93, Citeseer, 322--327

\bibitem[{Norris {et~al.}(2011)Norris, Hopkins, Afonso, Brown, Condon, Dunne,
  Feain, Hollow, Jarvis, Johnston-Hollitt, {et~al.}}]{Norris2011}
Norris, R.~P., Hopkins, A.~M., Afonso, J., {et~al.} 2011, Publications of the
  Astronomical Society of Australia, 28, 215

\bibitem[{{Ntwaetsile} \& {Geach}(2021)}]{Ntwaetsile2021}
{Ntwaetsile}, K. \& {Geach}, J.~E. 2021, \mnras, 502, 3417

\bibitem[{{Parma} {et~al.}(2007){Parma}, {Murgia}, {de Ruiter}, {Fanti},
  {Mack}, \& {Govoni}}]{Parma2007}
{Parma}, P., {Murgia}, M., {de Ruiter}, H.~R., {et~al.} 2007, \aap, 470, 875

\bibitem[{Pedregosa {et~al.}(2011)Pedregosa, Varoquaux, Gramfort, Michel,
  Thirion, Grisel, Blondel, Prettenhofer, Weiss, Dubourg, Vanderplas, Passos,
  Cournapeau, Brucher, Perrot, \& Duchesnay}]{scikit-learn}
Pedregosa, F., Varoquaux, G., Gramfort, A., {et~al.} 2011, Journal of Machine
  Learning Research, 12, 2825

\bibitem[{{Polsterer} {et~al.}(2015){Polsterer}, {Gieseke}, \&
  {Igel}}]{Polsterer2015}
{Polsterer}, K.~L., {Gieseke}, F., \& {Igel}, C. 2015, in Astronomical Society
  of the Pacific Conference Series, Vol. 495, Astronomical Data Analysis
  Software an Systems XXIV (ADASS XXIV), ed. A.~R. {Taylor} \& E.~{Rosolowsky},
  81

\bibitem[{{Proctor}(2016)}]{Proctor2016}
{Proctor}, D.~D. 2016, \apjs, 224, 18

\bibitem[{{Quici} {et~al.}(2021){Quici}, {Hurley-Walker}, {Seymour}, {Turner},
  {Shabala}, {Huynh}, {Andernach}, {Kapi{\'n}ska}, {Collier},
  {Johnston-Hollitt}, {White}, {Prandoni}, {Galvin}, {Franzen},
  {Ishwara-Chandra}, {Bellstedt}, {Tingay}, {Gaensler}, {O'Brien}, {Rogers},
  {Chow}, {Driver}, \& {Robotham}}]{Benjamin2021}
{Quici}, B., {Hurley-Walker}, N., {Seymour}, N., {et~al.} 2021, \pasa, 38, e008

\bibitem[{Quinlan(1993)}]{Quinlan1993}
Quinlan, J.~R. 1993, C4.5: Programs for Machine Learning (Elsevier)

\bibitem[{{Ralph} {et~al.}(2019){Ralph}, {Norris}, {Fang}, {Park}, {Galvin},
  {Alger}, {Andernach}, {Lintott}, {Rudnick}, {Shabala}, \& {Wong}}]{Ralph2019}
{Ralph}, N.~O., {Norris}, R.~P., {Fang}, G., {et~al.} 2019, \pasp, 131, 108011

\bibitem[{{Robotham} {et~al.}(2018){Robotham}, {Davies}, {Driver}, {Koushan},
  {Taranu}, {Casura}, \& {Liske}}]{robotham2018profound}
{Robotham}, A.~S.~G., {Davies}, L.~J.~M., {Driver}, S.~P., {et~al.} 2018,
  \mnras, 476, 3137

\bibitem[{{Saripalli} {et~al.}(2012){Saripalli}, {Subrahmanyan}, {Thorat},
  {Ekers}, {Hunstead}, {Johnston}, \& {Sadler}}]{Saripalli2012}
{Saripalli}, L., {Subrahmanyan}, R., {Thorat}, K., {et~al.} 2012, \apjs, 199,
  27

\bibitem[{{Scaife} \& {Porter}(2021)}]{Scaife2021equivariant}
{Scaife}, A. M.~M. \& {Porter}, F. 2021, \mnras, 503, 2369

\bibitem[{{Schoenmakers} {et~al.}(2000){Schoenmakers}, {de Bruyn},
  {R{\"o}ttgering}, {van der Laan}, \& {Kaiser}}]{Schoenmakers2000}
{Schoenmakers}, A.~P., {de Bruyn}, A.~G., {R{\"o}ttgering}, H.~J.~A., {van der
  Laan}, H., \& {Kaiser}, C.~R. 2000, \mnras, 315, 371

\bibitem[{Settles(2009)}]{Settles2009a}
Settles, B. 2009, Active Learning Literature Survey, Tech. rep., University of
  Wisconsin-Madison Department of Computer Sciences

\bibitem[{{Shimwell} {et~al.}(2022{\natexlab{a}}){Shimwell}, {Hardcastle},
  {Tasse}, {Best}, {R{\"o}ttgering}, {Williams}, {Botteon}, {Drabent},
  {Mechev}, {Shulevski}, {van Weeren}, {Bester}, {Br{\"u}ggen}, {Brunetti},
  {Callingham}, {Chy{\.z}y}, {Conway}, {Dijkema}, {Duncan}, {de Gasperin},
  {Hale}, {Haverkorn}, {Hugo}, {Jackson}, {Mevius}, {Miley}, {Morabito},
  {Morganti}, {Offringa}, {Oonk}, {Rafferty}, {Sabater}, {Smith}, {Schwarz},
  {Smirnov}, {O'Sullivan}, {Vedantham}, {White}, {Albert}, {Alegre}, {Asabere},
  {Bacon}, {Bonafede}, {Bonnassieux}, {Brienza}, {Bilicki}, {Bonato}, {Calistro
  Rivera}, {Cassano}, {Cochrane}, {Croston}, {Cuciti}, {Dallacasa}, {Danezi},
  {Dettmar}, {Di Gennaro}, {Edler}, {En{\ss}lin}, {Emig}, {Franzen},
  {Garc{\'\i}a-Vergara}, {Grange}, {G{\"u}rkan}, {Hajduk}, {Heald}, {Heesen},
  {Hoang}, {Hoeft}, {Horellou}, {Iacobelli}, {Jamrozy}, {Jeli{\'c}},
  {Kondapally}, {Kukreti}, {Kunert-Bajraszewska}, {Magliocchetti}, {Mahatma},
  {Ma{\l}ek}, {Mandal}, {Massaro}, {Meyer-Zhao}, {Mingo}, {Mostert}, {Nair},
  {Nakoneczny}, {Nikiel-Wroczy{\'n}ski}, {Orr{\'u}}, {Pajdosz-{\'S}mierciak},
  {Pasini}, {Prandoni}, {van Piggelen}, {Rajpurohit}, {Retana-Montenegro},
  {Riseley}, {Rowlinson}, {Saxena}, {Schrijvers}, {Sweijen}, {Siewert},
  {Timmerman}, {Vaccari}, {Vink}, {West}, {Wo{\l}owska}, {Zhang}, \&
  {Zheng}}]{Shimwell2022}
{Shimwell}, T.~W., {Hardcastle}, M.~J., {Tasse}, C., {et~al.}
  2022{\natexlab{a}}, \aap, 659, A1

\bibitem[{{Shimwell} {et~al.}(2022{\natexlab{b}}){Shimwell}, {Hardcastle},
  {Tasse}, {Best}, {R{\"o}ttgering}, {Williams}, {Botteon}, {Drabent},
  {Mechev}, {Shulevski}, {van Weeren}, {Bester}, {Br{\"u}ggen}, {Brunetti},
  {Callingham}, {Chy{\.z}y}, {Conway}, {Dijkema}, {Duncan}, {de Gasperin},
  {Hale}, {Haverkorn}, {Hugo}, {Jackson}, {Mevius}, {Miley}, {Morabito},
  {Morganti}, {Offringa}, {Oonk}, {Rafferty}, {Sabater}, {Smith}, {Schwarz},
  {Smirnov}, {O'Sullivan}, {Vedantham}, {White}, {Albert}, {Alegre}, {Asabere},
  {Bacon}, {Bonafede}, {Bonnassieux}, {Brienza}, {Bilicki}, {Bonato}, {Calistro
  Rivera}, {Cassano}, {Cochrane}, {Croston}, {Cuciti}, {Dallacasa}, {Danezi},
  {Dettmar}, {Di Gennaro}, {Edler}, {En{\ss}lin}, {Emig}, {Franzen},
  {Garc{\'\i}a-Vergara}, {Grange}, {G{\"u}rkan}, {Hajduk}, {Heald}, {Heesen},
  {Hoang}, {Hoeft}, {Horellou}, {Iacobelli}, {Jamrozy}, {Jeli{\'c}},
  {Kondapally}, {Kukreti}, {Kunert-Bajraszewska}, {Magliocchetti}, {Mahatma},
  {Ma{\l}ek}, {Mandal}, {Massaro}, {Meyer-Zhao}, {Mingo}, {Mostert}, {Nair},
  {Nakoneczny}, {Nikiel-Wroczy{\'n}ski}, {Orr{\'u}}, {Pajdosz-{\'S}mierciak},
  {Pasini}, {Prandoni}, {van Piggelen}, {Rajpurohit}, {Retana-Montenegro},
  {Riseley}, {Rowlinson}, {Saxena}, {Schrijvers}, {Sweijen}, {Siewert},
  {Timmerman}, {Vaccari}, {Vink}, {West}, {Wo{\l}owska}, {Zhang}, \&
  {Zheng}}]{LoTSSDR2}
{Shimwell}, T.~W., {Hardcastle}, M.~J., {Tasse}, C., {et~al.}
  2022{\natexlab{b}}, \aap, 659, A1

\bibitem[{{Shimwell} {et~al.}(2017){Shimwell}, {R{\"o}ttgering}, {Best},
  {Williams}, {Dijkema}, {de Gasperin}, {Hardcastle}, {Heald}, {Hoang},
  {Horneffer}, {Intema}, {Mahony}, {Mandal}, {Mechev}, {Morabito}, {Oonk},
  {Rafferty}, {Retana-Montenegro}, {Sabater}, {Tasse}, {van Weeren},
  {Br{\"u}ggen}, {Brunetti}, {Chy{\.z}y}, {Conway}, {Haverkorn}, {Jackson},
  {Jarvis}, {McKean}, {Miley}, {Morganti}, {White}, {Wise}, {van Bemmel},
  {Beck}, {Brienza}, {Bonafede}, {Calistro Rivera}, {Cassano}, {Clarke},
  {Cseh}, {Deller}, {Drabent}, {van Driel}, {Engels}, {Falcke}, {Ferrari},
  {Fr{\"o}hlich}, {Garrett}, {Harwood}, {Heesen}, {Hoeft}, {Horellou},
  {Israel}, {Kapi{\'n}ska}, {Kunert-Bajraszewska}, {McKay}, {Mohan},
  {Orr{\'u}}, {Pizzo}, {Prandoni}, {Schwarz}, {Shulevski}, {Sipior}, {Smith},
  {Sridhar}, {Steinmetz}, {Stroe}, {Varenius}, {van der Werf}, {Zensus}, \&
  {Zwart}}]{Shimwell2017}
{Shimwell}, T.~W., {R{\"o}ttgering}, H.~J.~A., {Best}, P.~N., {et~al.} 2017,
  \aap, 598, A104

\bibitem[{{Shimwell} {et~al.}(2019){Shimwell}, {Tasse}, {Hardcastle}, {Mechev},
  {Williams}, {Best}, {R{\"o}ttgering}, {Callingham}, {Dijkema}, {de Gasperin},
  {Hoang}, {Hugo}, {Mirmont}, {Oonk}, {Prandoni}, {Rafferty}, {Sabater},
  {Smirnov}, {van Weeren}, {White}, {Atemkeng}, {Bester}, {Bonnassieux},
  {Br{\"u}ggen}, {Brunetti}, {Chy{\.z}y}, {Cochrane}, {Conway}, {Croston},
  {Danezi}, {Duncan}, {Haverkorn}, {Heald}, {Iacobelli}, {Intema}, {Jackson},
  {Jamrozy}, {Jarvis}, {Lakhoo}, {Mevius}, {Miley}, {Morabito}, {Morganti},
  {Nisbet}, {Orr{\'u}}, {Perkins}, {Pizzo}, {Schrijvers}, {Smith}, {Vermeulen},
  {Wise}, {Alegre}, {Bacon}, {van Bemmel}, {Beswick}, {Bonafede}, {Botteon},
  {Bourke}, {Brienza}, {Calistro Rivera}, {Cassano}, {Clarke}, {Conselice},
  {Dettmar}, {Drabent}, {Dumba}, {Emig}, {En{\ss}lin}, {Ferrari}, {Garrett},
  {G{\'e}nova-Santos}, {Goyal}, {G{\"u}rkan}, {Hale}, {Harwood}, {Heesen},
  {Hoeft}, {Horellou}, {Jackson}, {Kokotanekov}, {Kondapally},
  {Kunert-Bajraszewska}, {Mahatma}, {Mahony}, {Mandal}, {McKean}, {Merloni},
  {Mingo}, {Miskolczi}, {Mooney}, {Nikiel-Wroczy{\'n}ski}, {O'Sullivan},
  {Quinn}, {Reich}, {Roskowi{\'n}ski}, {Rowlinson}, {Savini}, {Saxena},
  {Schwarz}, {Shulevski}, {Sridhar}, {Stacey}, {Urquhart}, {van der Wiel},
  {Varenius}, {Webster}, \& {Wilber}}]{Shimwell2019}
{Shimwell}, T.~W., {Tasse}, C., {Hardcastle}, M.~J., {et~al.} 2019, \aap, 622,
  A1

\bibitem[{{Skrutskie} {et~al.}(2006){Skrutskie}, {Cutri}, {Stiening},
  {Weinberg}, {Schneider}, {Carpenter}, {Beichman}, {Capps}, {Chester},
  {Elias}, {Huchra}, {Liebert}, {Lonsdale}, {Monet}, {Price}, {Seitzer},
  {Jarrett}, {Kirkpatrick}, {Gizis}, {Howard}, {Evans}, {Fowler}, {Fullmer},
  {Hurt}, {Light}, {Kopan}, {Marsh}, {McCallon}, {Tam}, {Van Dyk}, \&
  {Wheelock}}]{skrutskie2006a}
{Skrutskie}, M.~F., {Cutri}, R.~M., {Stiening}, R., {et~al.} 2006, \aj, 131,
  1163

\bibitem[{{Slijepcevic} {et~al.}(2022){Slijepcevic}, {Scaife}, {Walmsley},
  {Bowles}, {Wong}, {Shabala}, \& {Tang}}]{Slijepcevic2022shift}
{Slijepcevic}, I.~V., {Scaife}, A. M.~M., {Walmsley}, M., {et~al.} 2022,
  \mnras, 514, 2599

\bibitem[{{Smith} {et~al.}(2021){Smith}, {Haskell}, {G{\"u}rkan}, {Best},
  {Hardcastle}, {Kondapally}, {Williams}, {Duncan}, {Cochrane}, {McCheyne},
  {R{\"o}ttgering}, {Sabater}, {Shimwell}, {Tasse}, {Bonato}, {Bondi},
  {Jarvis}, {Leslie}, {Prandoni}, \& {Wang}}]{Smith2021}
{Smith}, D.~J.~B., {Haskell}, P., {G{\"u}rkan}, G., {et~al.} 2021, \aap, 648,
  A6

\bibitem[{Strobl {et~al.}(2007)Strobl, Boulesteix, Zeileis, \&
  Hothorn}]{strobl2007bias}
Strobl, C., Boulesteix, A.-L., Zeileis, A., \& Hothorn, T. 2007, BMC
  bioinformatics, 8, 1

\bibitem[{{Tang} {et~al.}(2019){Tang}, {Scaife}, \& {Leahy}}]{Tang2019transfer}
{Tang}, H., {Scaife}, A.~M.~M., \& {Leahy}, J.~P. 2019, \mnras, 488, 3358

\bibitem[{{The Astropy Collaboration} {et~al.}(2018){The Astropy
  Collaboration}, {Price-Whelan}, {Sip{\H o}cz}, {G{\"u}nther}, {Lim},
  {Crawford}, {Conseil}, {Shupe}, {Craig}, \& {Dencheva}}]{astropy}
{The Astropy Collaboration}, {Price-Whelan}, A.~M.~{Price-Whelan}, A.~M.,
  {Sip{\H o}cz}, B.~M., {et~al.} 2018, \aj, 156, 123

\bibitem[{{van Haarlem} {et~al.}(2013){van Haarlem}, {Wise}, {Gunst}, {Heald},
  {McKean}, {Hessels}, {de Bruyn}, {Nijboer}, {Swinbank}, {Fallows},
  {Brentjens}, {Nelles}, {Beck}, {Falcke}, {Fender}, {H{\"o}randel},
  {Koopmans}, {Mann}, {Miley}, {R{\"o}ttgering}, {Stappers}, {Wijers},
  {Zaroubi}, {van den Akker}, {Alexov}, {Anderson}, {Anderson}, {van Ardenne},
  {Arts}, {Asgekar}, {Avruch}, {Batejat}, {B{\"a}hren}, {Bell}, {Bell}, {van
  Bemmel}, {Bennema}, {Bentum}, {Bernardi}, {Best}, {B{\^\i}rzan}, {Bonafede},
  {Boonstra}, {Braun}, {Bregman}, {Breitling}, {van de Brink}, {Broderick},
  {Broekema}, {Brouw}, {Br{\"u}ggen}, {Butcher}, {van Cappellen}, {Ciardi},
  {Coenen}, {Conway}, {Coolen}, {Corstanje}, {Damstra}, {Davies}, {Deller},
  {Dettmar}, {van Diepen}, {Dijkstra}, {Donker}, {Doorduin}, {Dromer}, {Drost},
  {van Duin}, {Eisl{\"o}ffel}, {van Enst}, {Ferrari}, {Frieswijk}, {Gankema},
  {Garrett}, {de Gasperin}, {Gerbers}, {de Geus}, {Grie{\ss}meier}, {Grit},
  {Gruppen}, {Hamaker}, {Hassall}, {Hoeft}, {Holties}, {Horneffer}, {van der
  Horst}, {van Houwelingen}, {Huijgen}, {Iacobelli}, {Intema}, {Jackson},
  {Jelic}, {de Jong}, {Juette}, {Kant}, {Karastergiou}, {Koers}, {Kollen},
  {Kondratiev}, {Kooistra}, {Koopman}, {Koster}, {Kuniyoshi}, {Kramer},
  {Kuper}, {Lambropoulos}, {Law}, {van Leeuwen}, {Lemaitre}, {Loose}, {Maat},
  {Macario}, {Markoff}, {Masters}, {McFadden}, {McKay-Bukowski}, {Meijering},
  {Meulman}, {Mevius}, {Middelberg}, {Millenaar}, {Miller-Jones}, {Mohan},
  {Mol}, {Morawietz}, {Morganti}, {Mulcahy}, {Mulder}, {Munk}, {Nieuwenhuis},
  {van Nieuwpoort}, {Noordam}, {Norden}, {Noutsos}, {Offringa}, {Olofsson},
  {Omar}, {Orr{\'u}}, {Overeem}, {Paas}, {Pandey-Pommier}, {Pandey}, {Pizzo},
  {Polatidis}, {Rafferty}, {Rawlings}, {Reich}, {de Reijer}, {Reitsma},
  {Renting}, {Riemers}, {Rol}, {Romein}, {Roosjen}, {Ruiter}, {Scaife}, {van
  der Schaaf}, {Scheers}, {Schellart}, {Schoenmakers}, {Schoonderbeek},
  {Serylak}, {Shulevski}, {Sluman}, {Smirnov}, {Sobey}, {Spreeuw}, {Steinmetz},
  {Sterks}, {Stiepel}, {Stuurwold}, {Tagger}, {Tang}, {Tasse}, {Thomas},
  {Thoudam}, {Toribio}, {van der Tol}, {Usov}, {van Veelen}, {van der Veen},
  {ter Veen}, {Verbiest}, {Vermeulen}, {Vermaas}, {Vocks}, {Vogt}, {de Vos},
  {van der Wal}, {van Weeren}, {Weggemans}, {Weltevrede}, {White}, {Wijnholds},
  {Wilhelmsson}, {Wucknitz}, {Yatawatta}, {Zarka}, {Zensus}, \& {van
  Zwieten}}]{vanHaarlem2013}
{van Haarlem}, M.~P., {Wise}, M.~W., {Gunst}, A.~W., {et~al.} 2013, \aap, 556,
  A2

\bibitem[{Villmann {et~al.}(1994)Villmann, Der, \& Martinetz}]{Villmann1994}
Villmann, T., Der, R., \& Martinetz, T. 1994, in Proceedings of the IEEE
  International Conference on Neural Networks (ICNN-94), Orlando, Vol.~II,
  645--648

\bibitem[{{Walmsley} {et~al.}(2022{\natexlab{a}}){Walmsley}, {Scaife},
  {Lintott}, {Lochner}, {Etsebeth}, {G{\'e}ron}, {Dickinson}, {Fortson},
  {Kruk}, {Masters}, {Mantha}, \& {Simmons}}]{Walmsley2022zoobot}
{Walmsley}, M., {Scaife}, A. M.~M., {Lintott}, C., {et~al.} 2022{\natexlab{a}},
  \mnras, 513, 1581

\bibitem[{{Walmsley} {et~al.}(2022{\natexlab{b}}){Walmsley}, {Slijepcevic},
  {Bowles}, \& {Scaife}}]{Walmsley2022hybrid}
{Walmsley}, M., {Slijepcevic}, I.~V., {Bowles}, M., \& {Scaife}, A. M.~M.
  2022{\natexlab{b}}, arXiv e-prints, arXiv:2206.11927

\bibitem[{{Walmsley} {et~al.}(2020){Walmsley}, {Smith}, {Lintott}, {Gal},
  {Bamford}, {Dickinson}, {Fortson}, {Kruk}, {Masters}, {Scarlata}, {Simmons},
  {Smethurst}, \& {Wright}}]{Walmsley2020}
{Walmsley}, M., {Smith}, L., {Lintott}, C., {et~al.} 2020, \mnras, 491, 1554

\bibitem[{{Williams} {et~al.}(2019){Williams}, {Hardcastle}, {Best}, {Sabater},
  {Croston}, {Duncan}, {Shimwell}, {R{\"o}ttgering}, {Nisbet}, {G{\"u}rkan},
  {Alegre}, {Cochrane}, {Goyal}, {Hale}, {Jackson}, {Jamrozy}, {Kondapally},
  {Kunert-Bajraszewska}, {Mahatma}, {Mingo}, {Morabito}, {Prandoni},
  {Roskowinski}, {Shulevski}, {Smith}, {Tasse}, {Urquhart}, {Webster}, {White},
  {Beswick}, {Callingham}, {Chy{\.z}y}, {de Gasperin}, {Harwood}, {Hoeft},
  {Iacobelli}, {McKean}, {Mechev}, {Miley}, {Schwarz}, \& {van
  Weeren}}]{Williams2019}
{Williams}, W.~L., {Hardcastle}, M.~J., {Best}, P.~N., {et~al.} 2019, \aap,
  622, A2

\bibitem[{{Zbontar} {et~al.}(2021){Zbontar}, {Jing}, {Misra}, {LeCun}, \&
  {Deny}}]{Jure2021}
{Zbontar}, J., {Jing}, L., {Misra}, I., {LeCun}, Y., \& {Deny}, S. 2021, arXiv
  e-prints, arXiv:2103.03230

\end{thebibliography}

\begin{appendix}
\section{Examples of radio sources for varying concentration index, clumpiness index and Gini coefficient values}
\label{app:optical_morpho}
\begin{figure}\begin{center}
\includegraphics[width=0.9\columnwidth]{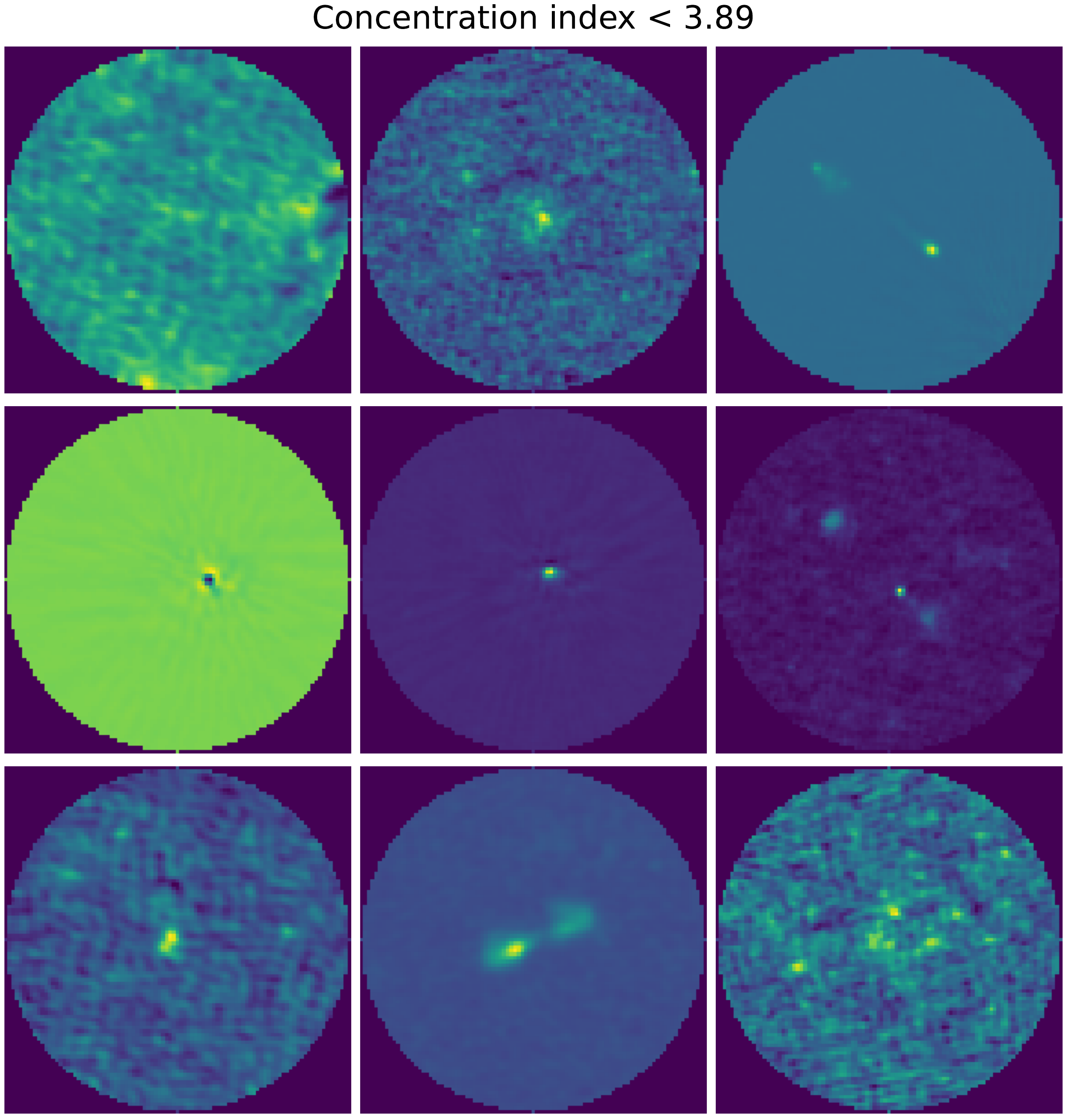}
\includegraphics[width=0.9\columnwidth]{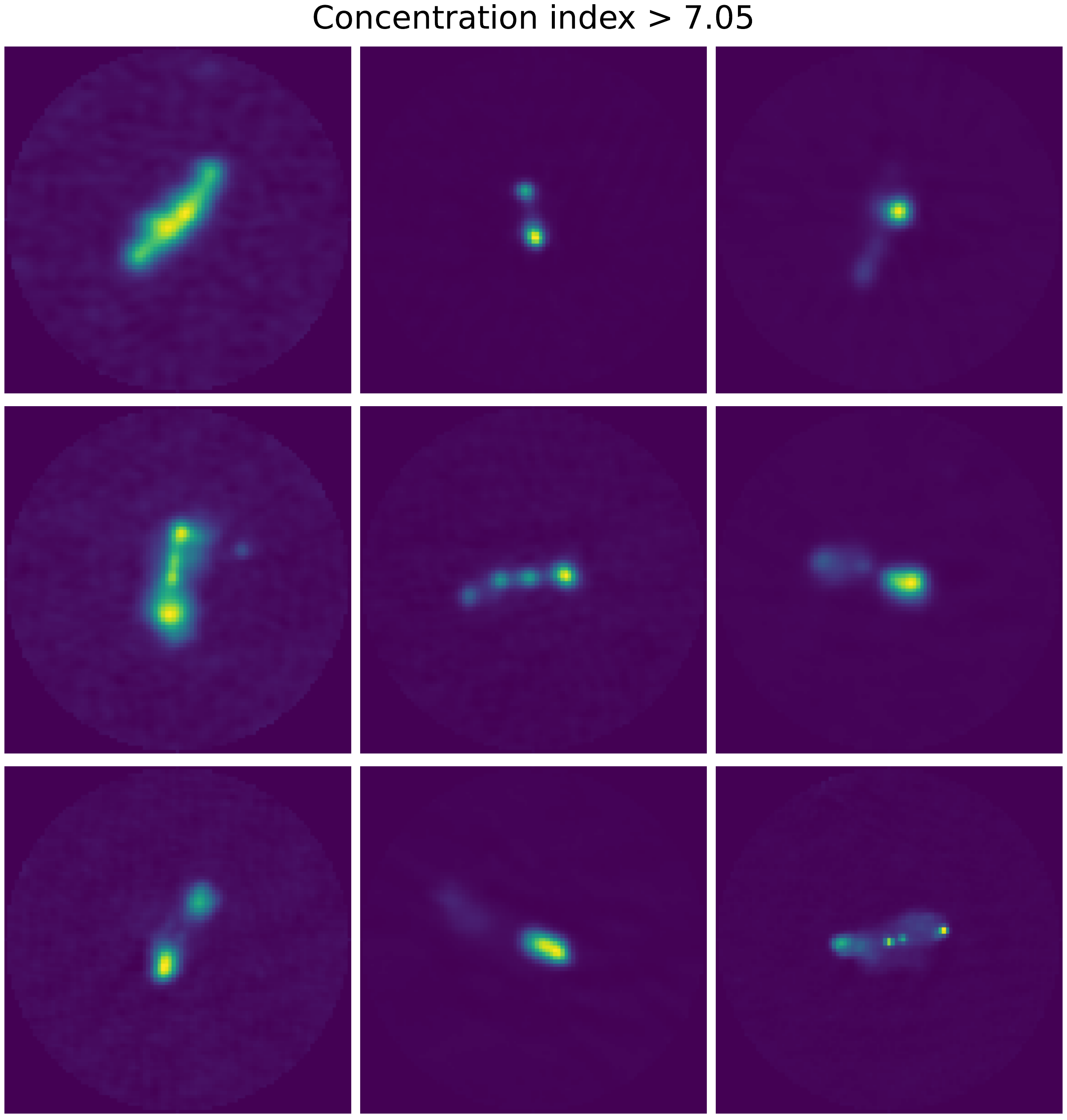}
\caption{Randomly picked sources from the training set that showcase concentration index values below the $20$th and above the $80$th percentile.} 
\label{fig:concentration}
\end{center}\end{figure}

\begin{figure}\begin{center}
\includegraphics[width=0.9\columnwidth]{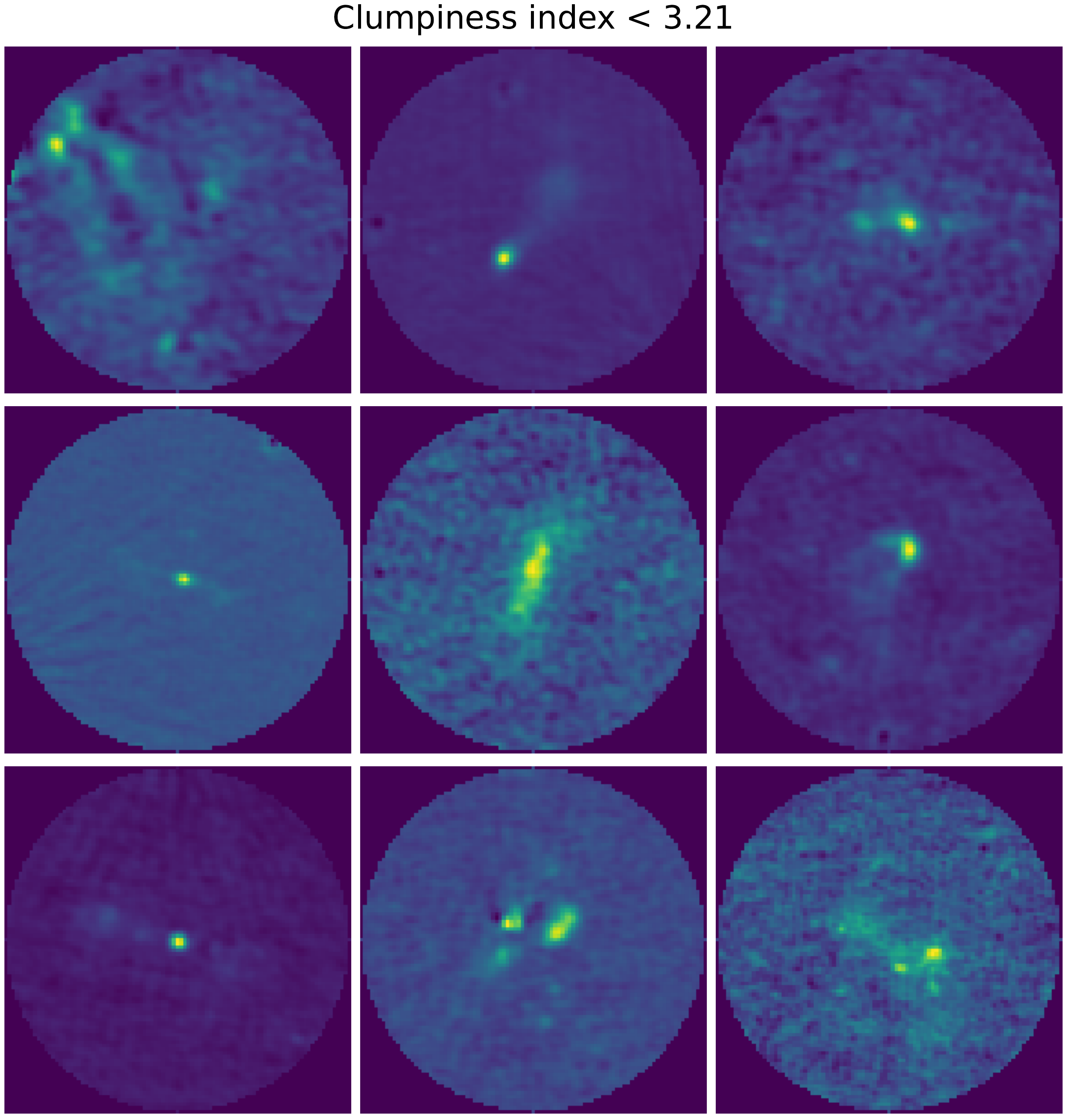}
\includegraphics[width=0.9\columnwidth]{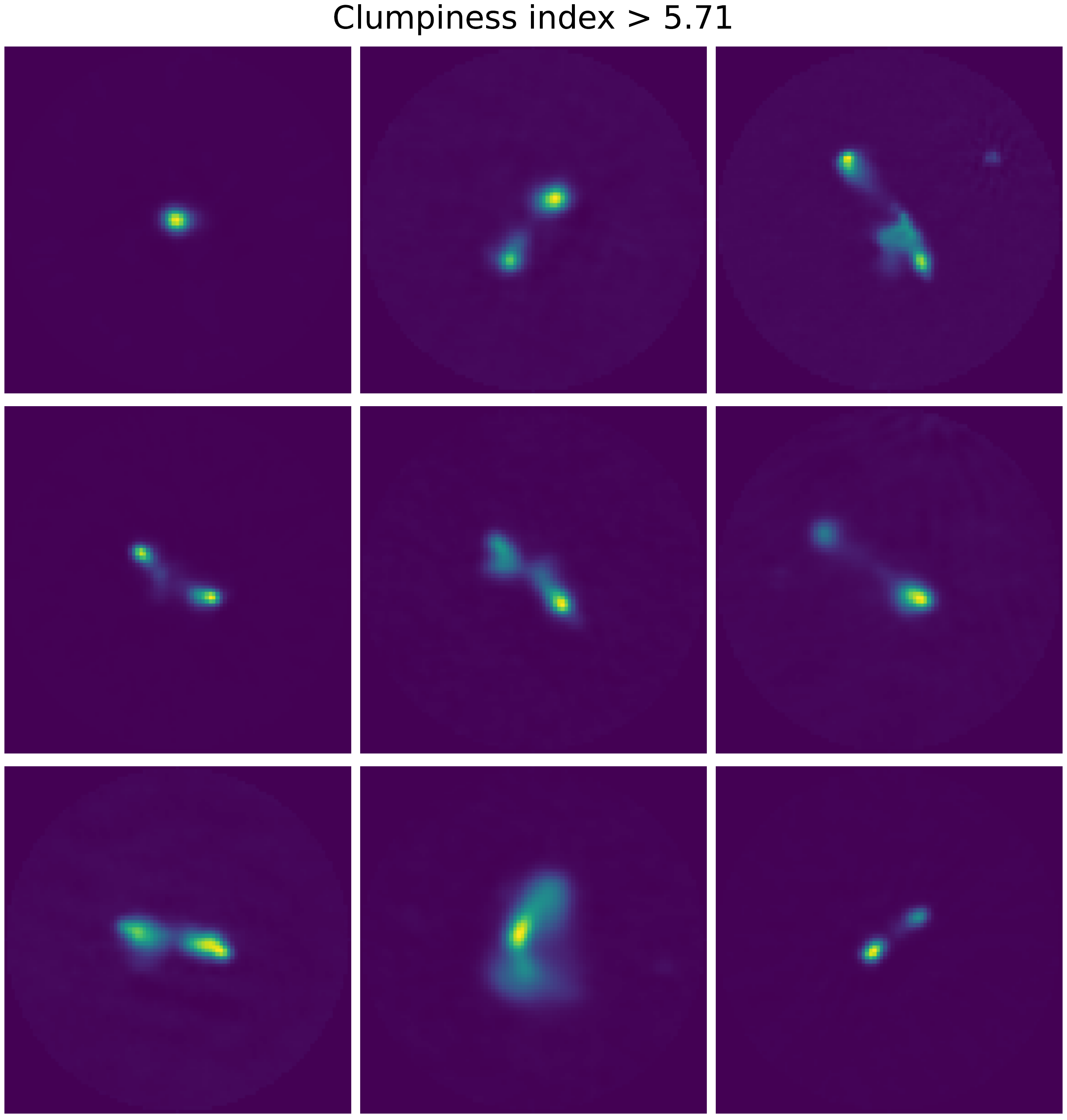}
\caption{Randomly picked sources from the training set that showcase clumpiness index values below the $20$th and above the $80$th percentile.} 
\label{fig:clumpi}
\end{center}\end{figure}

\begin{figure}\begin{center}
\includegraphics[width=0.9\columnwidth]{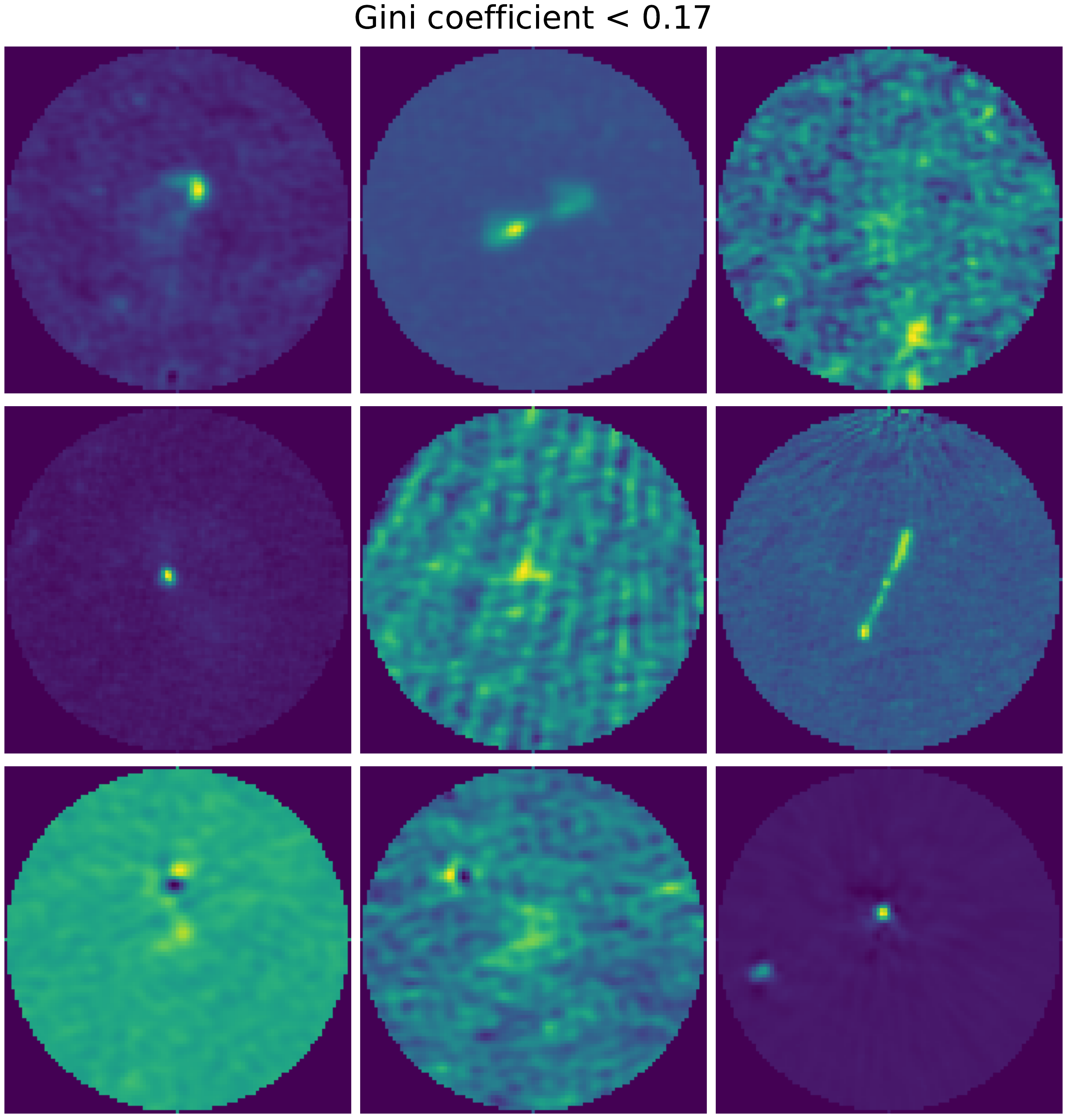}
\includegraphics[width=0.9\columnwidth]{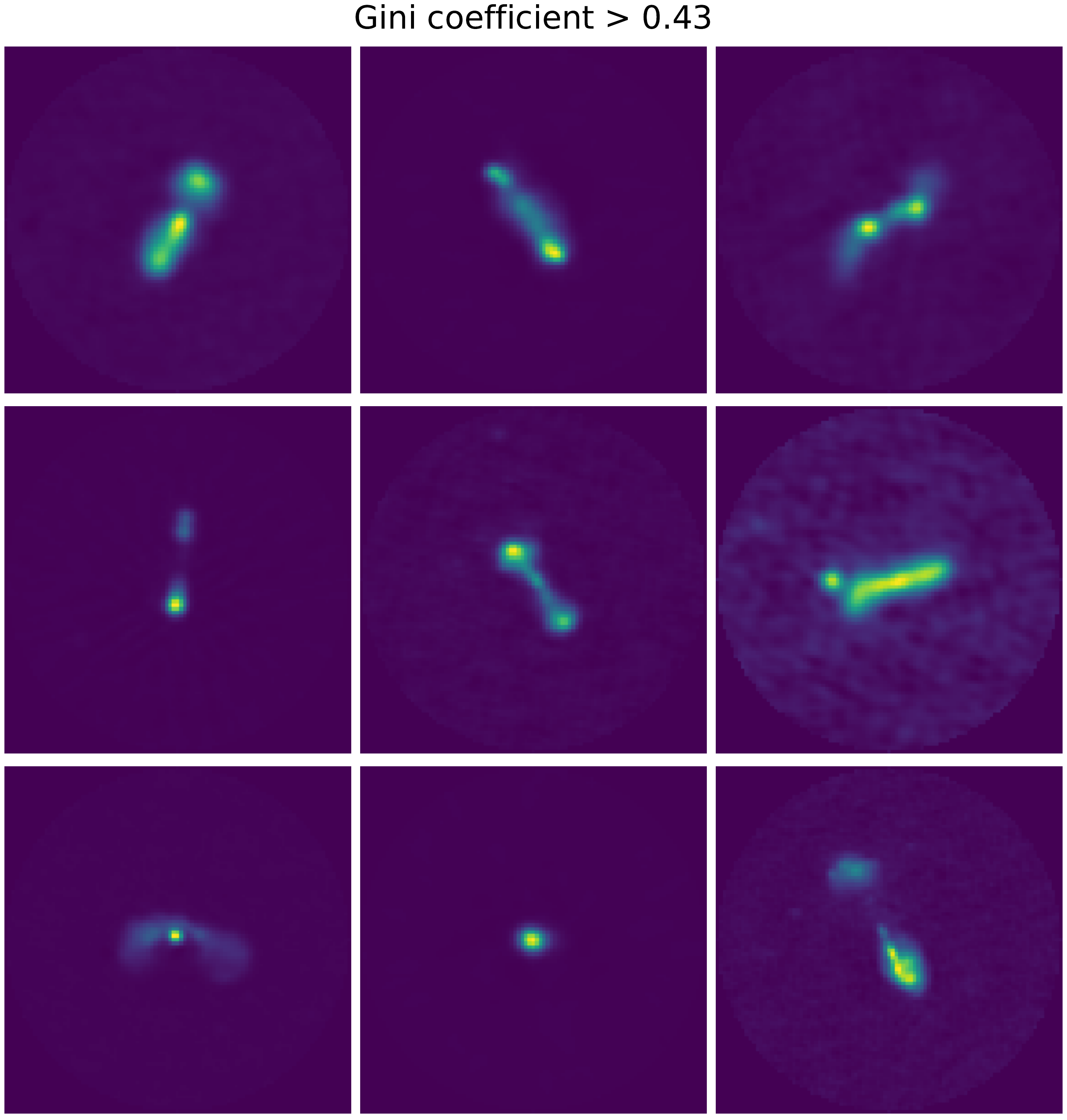}
\caption{Randomly picked sources from the training set that showcase Gini coefficient values below the $20$th and above the $80$th percentile.} 
\label{fig:gini}
\end{center}\end{figure}

To show what features of a radio image are picked up by three morphology metrics normally used to assess optical galaxy morphology, we show nine sources in two parts of the parameter space for the concentration index (Fig. \ref{fig:concentration}), the clumpiness index (Fig. \ref{fig:clumpi}), and the Gini coefficient (Fig. \ref{fig:gini}).
It is interesting to see that, given our dataset of $>60$ arcsec radio sources, four of the nine sources with high concentration values that we randomly picked (Fig. \ref{fig:concentration}) show a double-double radio galaxy morphology \citep[DDRG;][]{Schoenmakers2000}, a sign of restarted AGN activity.
\FloatBarrier

\FloatBarrier

\section{Examples of sources from different Haralick clusters}
\label{app:hara}
To see what features of a radio image are picked up by the clustered Haralick features, Figs. \ref{fig:hara1} and \ref{fig:hara2} show nine randomly sampled sources for each of the four clusters, plus the one `noise'-cluster.

\begin{figure}[H]\begin{center}
\includegraphics[width=0.7\columnwidth]{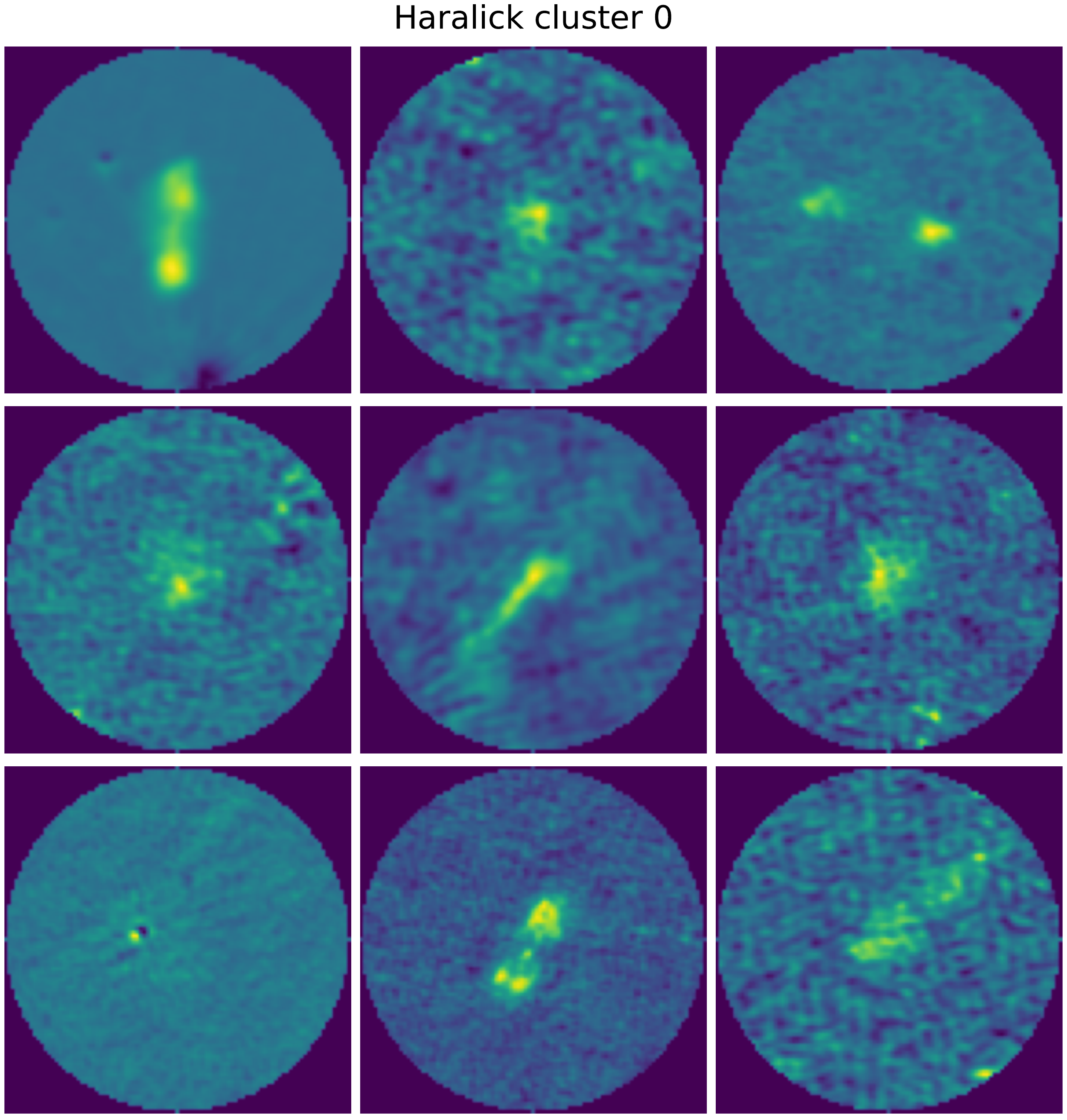}
\includegraphics[width=0.7\columnwidth]{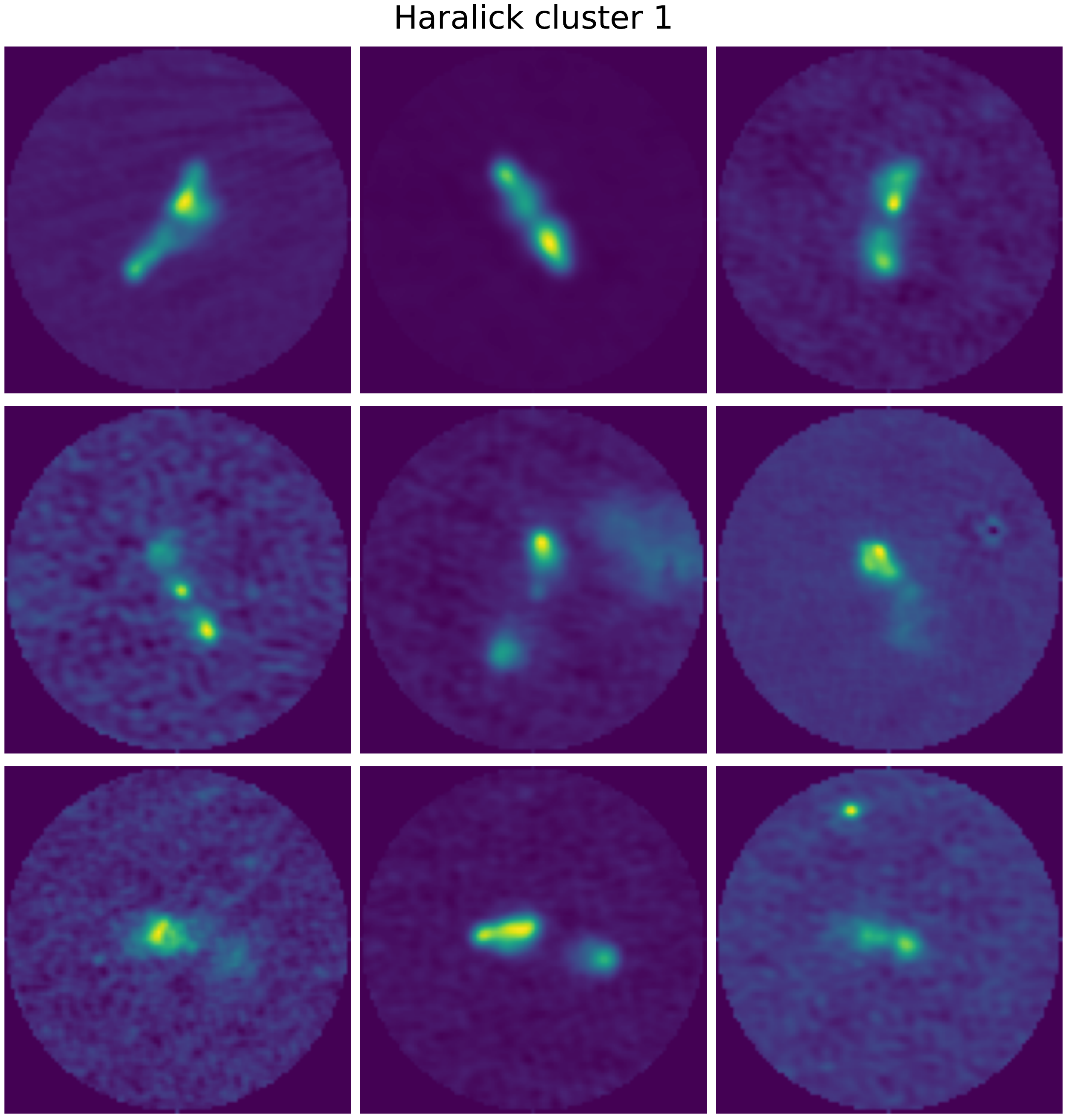}
\caption{Nine randomly picked sources from the training set for each Haralick cluster.} 
\label{fig:hara1}
\end{center}\end{figure}
\begin{figure}\begin{center}
\includegraphics[width=0.7\columnwidth]{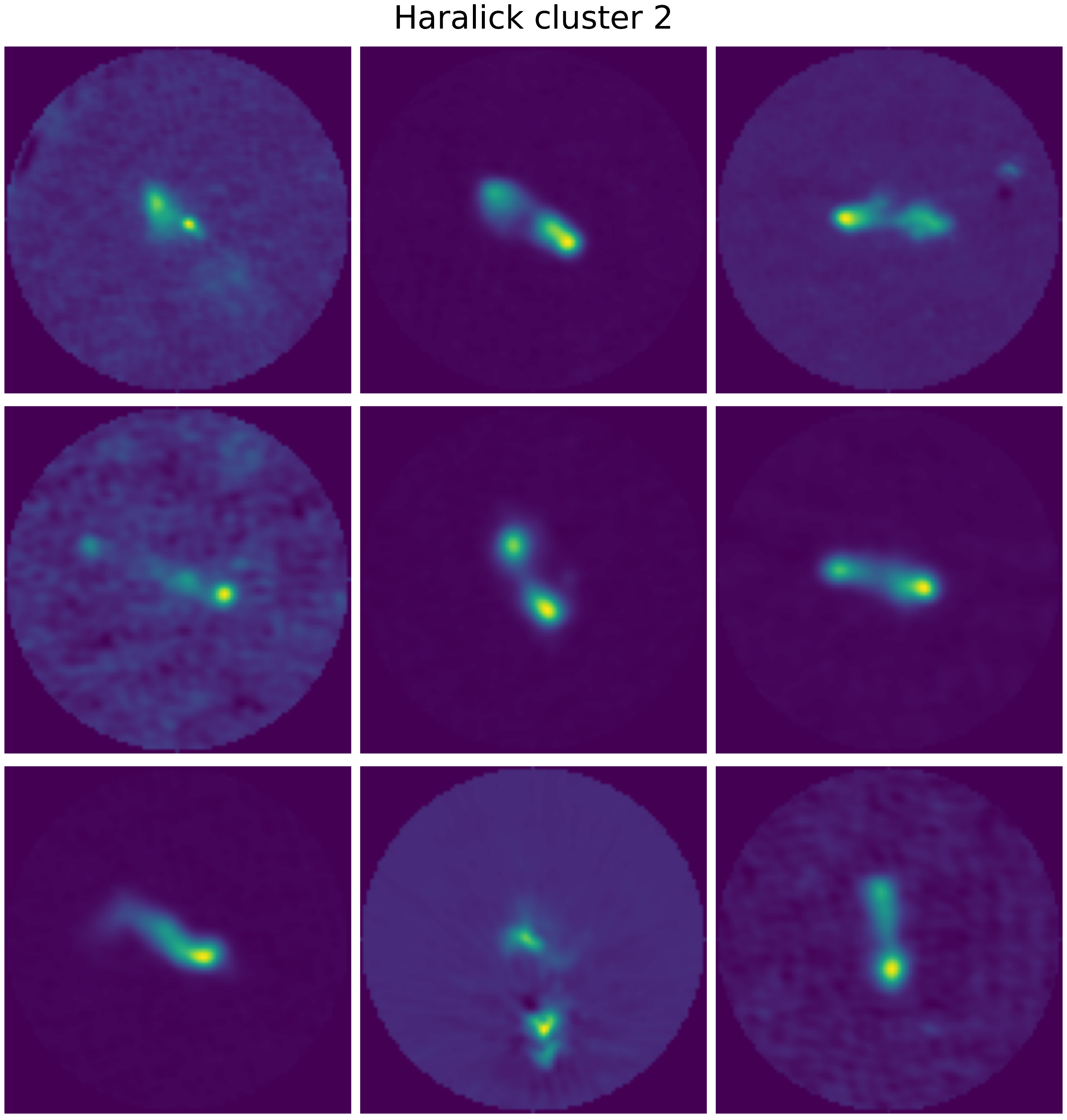}
\includegraphics[width=0.7\columnwidth]{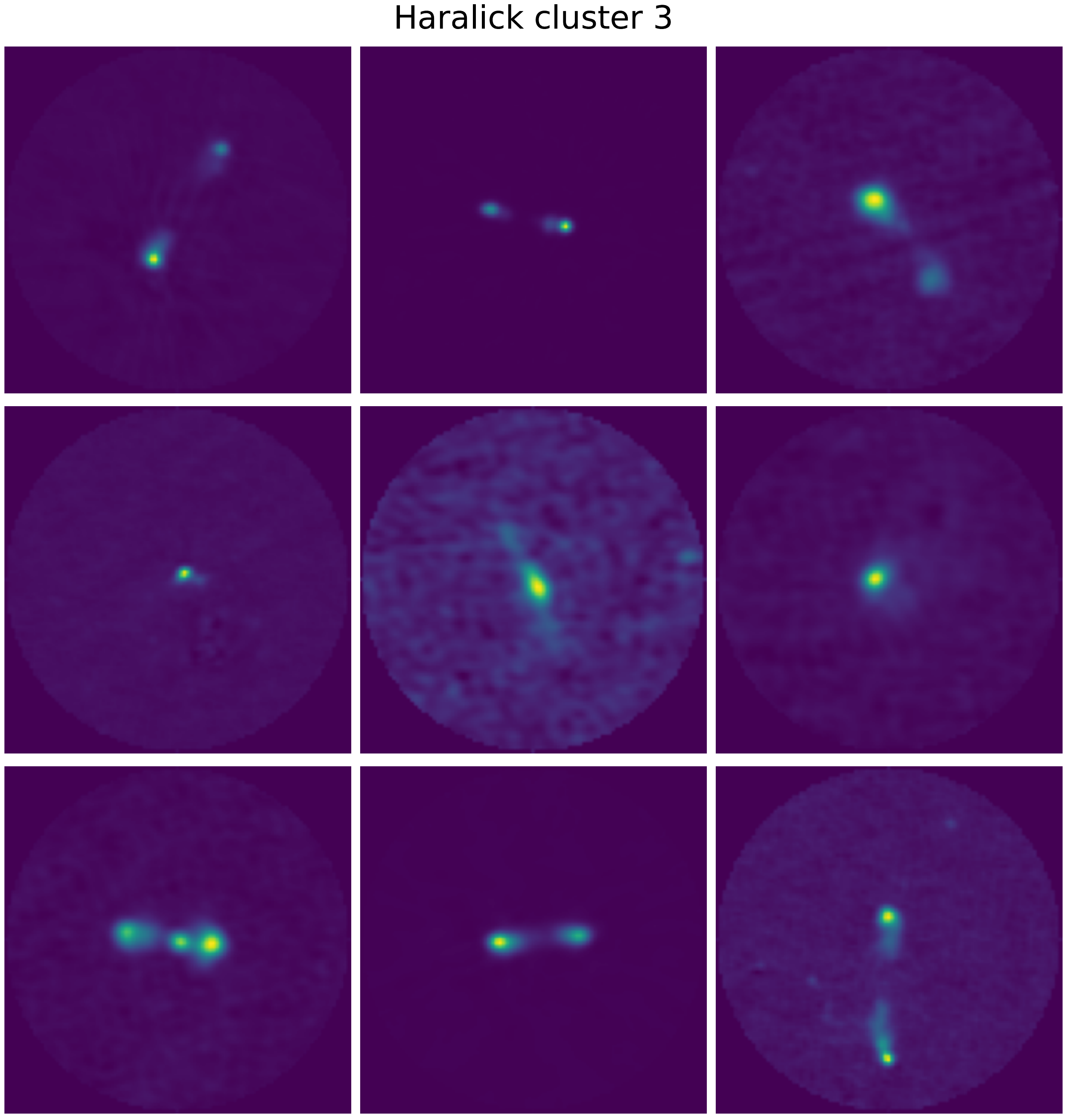}
\includegraphics[width=0.7\columnwidth]{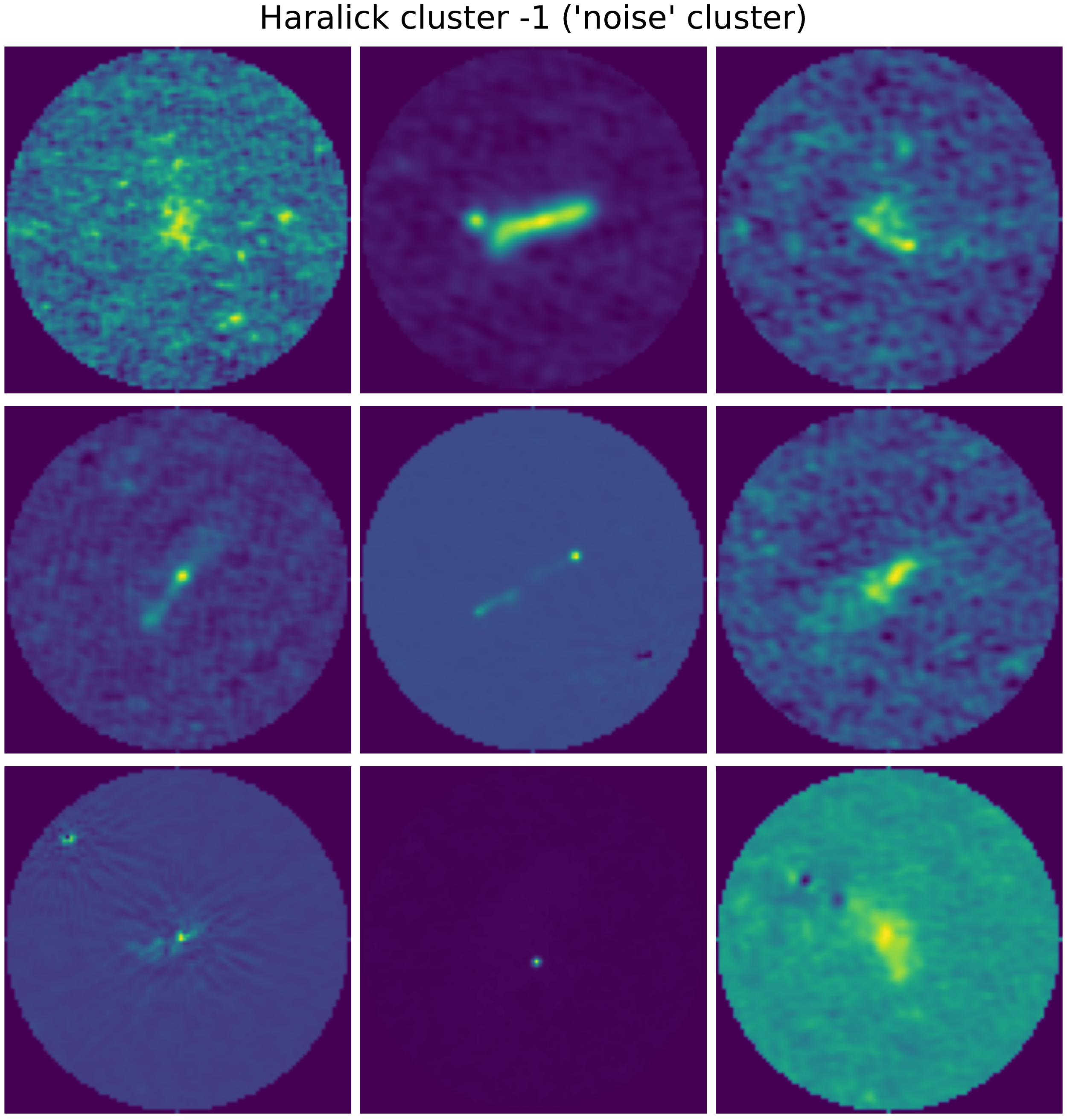}
\caption{continued.} 
\label{fig:hara2}
\end{center}\end{figure}

\FloatBarrier
\section{Derivation of statistical estimates}
\label{app:derivation}
We trained our RF classifier to separate the objects with 
class label `AGN remnant candidate' from the other objects for which we do not yet 
know their class label.
Our test set (containing $N=1173$ sources), can be 
divided into two subsets.
Dataset 1 comprises all $N_1=313$ positive predictions (sources in the test set that the 
model predicts to be `AGN remnant candidate'),
and dataset 2 comprises all $N_2=860$ negative predictions 
(sources in the test set that the 
model predicts to be `noncandidate').
Dataset 1 can be further split into three subsets: 
the $K_1^A=45$ sources with known `AGN remnant candidate' class label,
the unknown number ($U_1^A$) of sources that would have `AGN remnant candidate' as label 
upon visual inspection, and the unknown number ($U_1^N$) of sources that 
would have `noncandidate' as label upon visual inspection.
We know that $U_1^A + U_1^N = N_1-K_1^A = 268$.

We want to know the probability that visual inspection reveals `AGN remnant candidate' 
upon sampling a source from dataset 1:
\begin{equation}
\label{C1}
p = (K_1^A + U_1^A) / N_1 = K_1^A / N_1 + U_1^A / N_1.
\end{equation}
Because $N_1$, $K_1^A$, and $U_1^A$ are constants and not random variables, $p$ is a constant too.
To estimate this probability, we randomly sample (without replacement) and visually 
inspect $N_s=36$ 
sources from dataset 1.
The number of sampled sources that turn out to be `AGN remnant candidate' after visual inspection will be called $x$.
In this case $x \sim \mathrm{Hypergeometric}(U_1^A + U_1^N, U_1^A, N_s)$,
yielding the expected value:
\begin{equation}
\mathbf{E}[x] = N_s \cdot U_1^A / (U_1^A + U_1^N),
\end{equation}
which can be rewritten to:
\begin{equation}
\label{exp}
\mathbf{E}[x / N_s] = U_1^A / (U_1^A + U_1^N),
\end{equation}
and to:
\begin{equation}
\label{C4}
\mathbf{E}[x / N_s \cdot  (U_1^A + U_1^N) / N_1] =  U_1^A / N_1.
\end{equation}
Combining Eqs. \ref{C1} and \ref{C4}, we find that:
\begin{equation}
p = K_1^A / N_1 + \mathbf{E}[x / N_s \cdot  (U_1^A + U_1^N) / N_1],
\end{equation}
and it follows that the random variable
\begin{equation}
p* := K_1^A / N_1 + x / N_s \cdot  (U_1^A + U_1^N) / N_1
\end{equation}
is an unbiased estimator of $p$.

The variance of $p*$ is
\begin{multline}
\mathbf{V}[p*] = \mathbf{V}[K_1^A / N_1 + x / N_s \cdot  (U_1^A + U_1^N) / N_1] \\
= \mathbf{V}[x / N_s \cdot  (U_1^A + U_1^N) / N_1],
\end{multline}

as the variance of constants is zero.
Furthermore,
\begin{equation}
\mathbf{V}[p*] = (U_1^A + U_1^N)^2 / (N_1^2 \cdot N_s^2) \mathbf{V}[x],
\end{equation}
since $ \mathbf{V}[aX]=a^2\mathbf{V}[X]$ if $a$ is a constant and $X$ a random variable.

As we still consider the case $x \sim \mathrm{Hypergeometric}(U_1^A + U_1^N, U_1^A, N_s)$,
we know that:
\begin{multline}
\mathbf{V}[x] = N_s \cdot U_1^A / (U_1^A + U_1^N) \cdot U_1^N / (U_1^A + U_1^N) \\
\cdot (U_1^A + U_1^N - N_s) / (U_1^A + U_1^N - 1).
\end{multline}

We know from Eq. \ref{exp} that $x / N_s$ is an unbiased estimator of $U_1^A / (U_1^A + U_1^N)$.
Similarly $1 - x / N_s $ is an unbiased estimator of  $U_1^N / (U_1^A + U_1^N)$.
We can thus estimate $\mathbf{V}[p*]$ using random variable $W$:
\begin{equation}
W := (U_1^A + U_1^N)^2 / (N_1^2 \cdot N_s^2) V,
\end{equation}
where 
\begin{equation}
V := N_s \cdot  x / N_s \cdot (1 - x / N_s) \cdot (U_1^A + U_1^N - N_s) / (U_1^A + U_1^N - 1).
\end{equation}
We remark that $W$ is not an \textit{unbiased} estimator of $\mathbf{V}[p*]$.

\end{appendix}

\end{document}